\documentclass[12pt,a4paper,openany]{book}
\usepackage{anysize}
\usepackage[latin1]{inputenc}
\usepackage[ruled,chapter]{algorithm}
\usepackage{makeidx}
\usepackage{subfig}
\usepackage[intlimits,fixamsmath]{mathtools}
\usepackage{amssymb}
\usepackage{amsthm}
\usepackage{mathrsfs}
\usepackage{graphicx}
\usepackage{fancyhdr}
\usepackage{array}
\usepackage{geometry}
\usepackage{simplewick}
\usepackage{latexsym}
\usepackage[all]{xy}
\usepackage{euscript}
\usepackage{enumerate}
\usepackage{dsfont}
\usepackage{slashed}
\usepackage[plainpages=false,pagebackref=false,pdftex]{hyperref}
\usepackage{pdfpages}
\usepackage{graphicx}
\everymath{\displaystyle}

\let\sin\undefined
\DeclareMathOperator{\sin}{sen}

\DeclareMathOperator{\sech}{sech}

\renewcommand{\vec}[1]{\boldsymbol #1}								

\newcommand{\dpar}[2]{\frac{\partial#1}{\partial#2}}				
\newcommand{\diff}[1]{\,\mathrm{d}#1}								
\newcommand{\dtot}[2]{\frac{\mathrm{d}#1}{\mathrm{d}#2}}			

\let\oldsqrt\sqrt
\def\sqrt{\mathpalette\DHLhksqrt}
\def\DHLhksqrt#1#2{%
\setbox0=\hbox{$#1\oldsqrt{#2\,}$}\dimen0=\ht0
\advance\dimen0-0.2\ht0
\setbox2=\hbox{\vrule height\ht0 depth -\dimen0}%
{\box0\lower0.4pt\box2}}

\makeindex

\marginsize{25mm}{25mm}{25mm}{25mm}

\hypersetup{
colorlinks=true,breaklinks=true,
linkcolor=black,citecolor=black,
urlcolor=black}

\begin{document}

\setlength{\abovecaptionskip}{0.0cm}
\setlength{\belowcaptionskip}{0.0cm}
\setlength{\baselineskip}{24pt}
\setlength{\headheight}{15pt}

\pagestyle{fancy}
\lhead{}
\chead{}
\rhead{\thepage}
\lfoot{}
\cfoot{}
\rfoot{}

\fancypagestyle{plain}
{
	\fancyhf{}
	\lhead{}
	\chead{}
	\rhead{\thepage}
	\lfoot{}
	\cfoot{}
	\rfoot{}
}

\renewcommand{\headrulewidth}{0pt}


\frontmatter 

\thispagestyle{empty}

\begin{figure}[h]
	\includegraphics[scale=0.8]{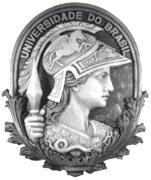}
\end{figure}

\vspace{15pt}

\begin{center}

\textbf{UNIVERSIDADE FEDERAL DO RIO DE JANEIRO}

\textbf{INSTITUTO DE F\'ISICA}

\vspace{30pt}

{\Large \bf Cosmic Censorship Conjecture violation: \\ A semiclassical approach.}

\vspace{25pt}

{\large \bf Rodrigo Lipparelli Fernandez}

\vspace{35pt}

\begin{flushright}
\parbox{10.3cm}{Tese de Doutorado apresentada ao Programa de P\'os-Gradua\c{c}\~ao em F\'isica do Instituto de F\'isica da Universidade Federal do Rio de Janeiro -- UFRJ, como parte dos requisitos necess\'arios \`a obten\c{c}\~ao do t\'itulo de Doutor em Ci\^encias (F\'isica).

\vspace{18pt}

{\large \bf Orientador: Ribamar Rondon de Rezende dos Reis}

\vspace{12pt}

{\large \bf Coorientador: Sergio Eduardo de Carvalho Eyer Jor\'as}}
\end{flushright}

\vspace{80pt}

\textbf{Rio de Janeiro}

\textbf{Mar\c{c}o de 2020}

\end{center}



\newpage
\includepdf{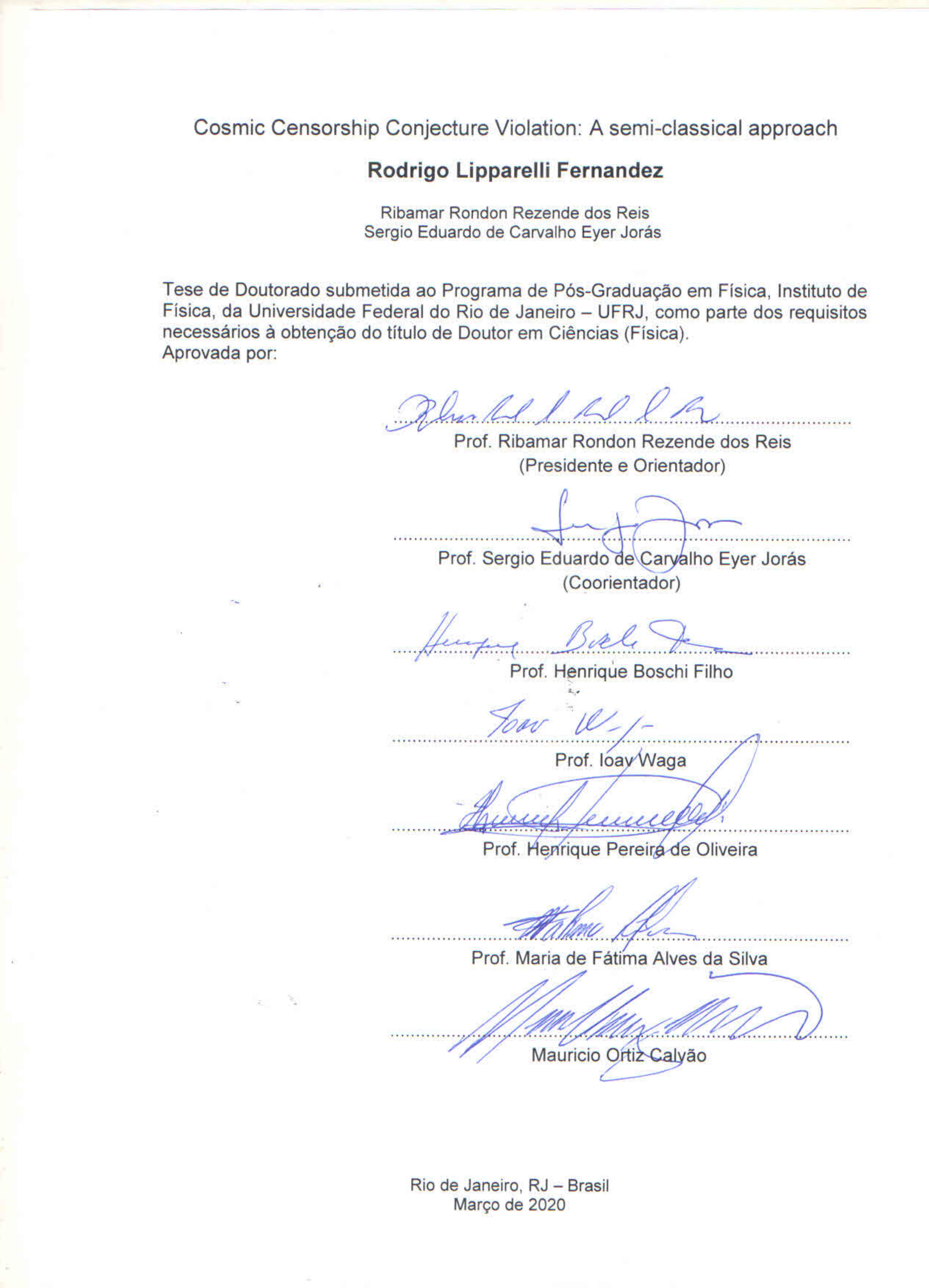}







\newpage
\includepdf{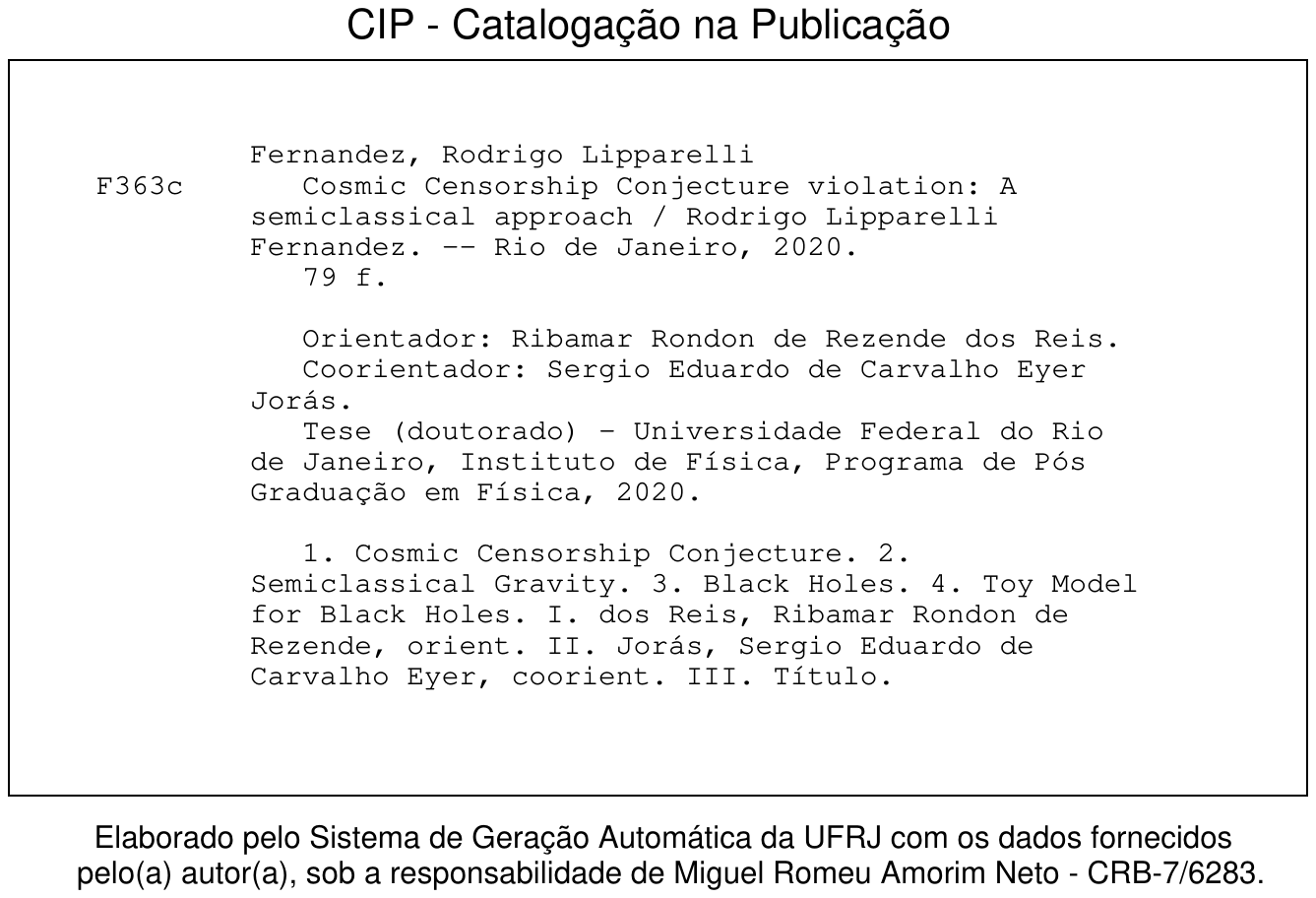}



\newpage

\noindent

\vspace*{20pt}
\begin{center}
{\LARGE\bf Resumo}\\
\vspace{15pt}
{\Large\bf Cosmic Censorship Conjecture violation: \\ A semi-classical approach}\\
\vspace{6pt}
{\bf Rodrigo Lipparelli Fernandez}\\
\vspace{12pt}
{\bf Orientador: Ribamar Rondon de Rezende dos Reis}\\
{\bf Coorientador: Sergio Eduardo de Carvalho Eyer Jor\'as}\\
\vspace{20pt}
\parbox{14cm}{Resumo da Tese de Doutorado apresentada ao Programa de P\'os-Gradua\c{c}\~ao em F\'isica do Instituto de F\'isica da Universidade Federal do Rio de Janeiro -- UFRJ, como parte dos requisitos necess\'arios \`a obten\c{c}\~ao do t\'itulo de Doutor em Ci\^encias (F\'isica).}
\end{center}
\vspace*{35pt}

A Conjectura da Censura C\'osmica (CCC) nos diz que toda sigularidade (exceto pela singularidade cosmol\'ogica) deve aparecer ``vestida'' no universo. Essa afirma\c{c}\~ao foi introduzida por Roger Penrose (Penrose, 1969), dizendo que toda singularidade (exceto o Big Bang) no universo deve estar escondida atr\'as de um Horizonte de Eventos. Matematicamente, essa afirma\c{c}\~ao pode ser posta na desigualdade $M^2 \geqslant Q^2 + a^2$ (em unidades geom\'etricas), com $M$ sendo a massa do buraco negro, $Q$ sua carga e $a := J/M$ seu momento angular por unidade de massa. Essencialmente, estas tr\^es quantidades determinam unicamente um buraco negro, no que diz respeito ao teorema da calv\'icie.

N\'os estudamos a probabilidade da emiss\~ao de um pacote de ondas massiva ($m_w$) e sem carga, uma representa\c{c}\~ao semicl\'assica da part\'icula, por um buraco negro est\'atico e carregado. N\'os mostramos que para alguns valores da massa $\mathcal{M}:=M+\delta M$ (onde $M$ \'e o valor fixo da massa e $\delta M$ uma pequena varia\c{c}\~ao de $M$ na ordem de $m_w$) com diferentes valores de $\delta M$ e carga $Q$ fixa para o buraco negro, que a probabilidade de emiss\~ao tende a zero uma vez que a Conjectura da Censura C\'osmica est\'a pr\'oxima de ser violada, ou seja, quando o pacote emitido \'e tal que a nova quantidade $\mathcal{M}':=\mathcal{M}-m_w$ violaria a desigualdade $\mathcal{M}' > Q$.

\vspace{15pt}

\textbf{Palavras-chave:} Conjectura da censura c\'osmica, Aproxima\c{c}\~ao semicl\'assica, Toy model.



\newpage

\noindent

\vspace*{20pt}
\begin{center}
{\LARGE\bf Abstract}\\
\vspace{15pt}
{\Large\bf Cosmic Censorship Cojecture violation: \\ A semi-classical approach}\\
\vspace{6pt}
{\bf Rodrigo Lipparelli Fernandez}\\
\vspace{12pt}
{\bf Orientador: Ribamar Rondon de Rezende dos Reis}\\
{\bf Coorientador: Sergio Eduardo de Carvalho Eyer Jor\'as}\\
\vspace{20pt}
\parbox{14cm}{\emph{Abstract} da Tese de Doutorado apresentada ao Programa de P\'os-Gradua\c{c}\~ao em F\'isica do Instituto de F\'isica da Universidade Federal do Rio de Janeiro -- UFRJ, como parte dos requisitos necess\'arios \`a obten\c{c}\~ao do t\'itulo de Doutor em Ci\^encias (F\'isica).} 
\end{center}
\vspace*{35pt}

The Cosmic Censorship Conjecture (CCC) states that every singularity (except the cosmological one) must appear ``dressed'' in the universe. This statement was introduced by Roger Penrose (Penrose, 1969), meaning that every singularity (except the Big Bang) in the universe must be hidden inside an Event Horizon. Mathematically, this is described by the inequality $M^2 \geqslant Q^2 + a^2$ (in geometrized unit system), with $M$ being the mass of the black hole, $Q$ its charge and $a := J/M$ its specific angular momentum. Essentially, this three quantities determines uniquely a black hole, as stated by the no-hair theorem.

We study the emission probability of a massive ($m_w$) uncharged scalar wave packet, a semi-classical approximation for a particle, by a static, charged black hole. We show that for a few values of the mass $\mathcal{M} := M+\delta M$ (where $M$ is the fixed value for the mass and $\delta M$ being a small variation to $M$ in the order of $m_w$) with different values for $\delta M$ and fixed charge $Q$ for the black hole, the emission probability tends to zero once the Cosmic Censorship Conjecture is close to be violated, that is, when the emitted packet is such that the new quantity $\mathcal{M}' := \mathcal{M}-m_w$ would violate the inequality $\mathcal{M}' > Q$.

\vspace{15pt}




\newpage

\noindent

\vspace*{20pt}

\begin{center}

{\LARGE\bf Acknowledgements}

\end{center}

\vspace*{40pt}

I'd like to thank both of my parents, Gil Ferreira Fernandez and Ivanira Lipparelli Fernandez, who made it all possible ever since. I also would like to thank my family who have always been by my side throughout my academic career, specially my brothers Daniel Lipparelli Fernandez and Eduardo Lipparelli Fernandez.

My advisors, Ribamar Reis and Sergio Jor\'as, who taught me almost everything I learned through this four years and always guided me to become a better person and an even better scientist. Thank you for making the thesis process enjoyable and by inspiring me the whole time.

I also would like to thank CAPES for the financial support during the elaboration of this thesis under the grant process number 88882.331077/2019-01.



\newpage
\phantomsection
\addcontentsline{toc}{chapter}{Summary}
\tableofcontents

\newpage
\phantomsection
\addcontentsline{toc}{chapter}{List of Figures}
\listoffigures


\mainmatter
\begin{chapter}{Introduction}
\label{intro}

\hspace{5 mm} The \emph{Cosmic Censorship Conjecture} (CCC) states that every singularity (except the cosmological one) must appear ``dressed'' in the universe. This statement was introduced by Roger Penrose~\cite{Penrose2002}, meaning that every singularity (except the Big Bang) in the universe must be hidden inside an Event Horizon (EH) --- that is, if the \emph{hoop conjecture}~\cite{Klauder:1972lsv,PhysRevLett.66.994} is satisfied. The hoop conjecture states that in order for an object to produce a ``dressed'' singularity under gravitational collapse, it must do so below the radius of a hoop defined by the circumference $\mathcal{C} = 2\pi r_s$ in every spatial dimension, where $r_s := 2GM/c^2$ is the Schwarzschild radius and $M$ the mass of said object. That way, if a nonspherical, oblate object collapses in every direction but one under the Schwarzschild radius, it may produce a naked singularity.

For black holes, the CCC is described by the inequality $M^2 \geqslant Q^2 + a^2$ (in geometrized unit system), with $M$ being its mass, $Q$ its charge and $a := J/M$ its specific angular momentum (where $J$ is the total angular momentum). Essentially, this three quantities uniquely defines a black hole, as stated by the \emph{no-hair theorem}~\cite{MTW}. Therefore, the CCC tells us that a black hole will only absorb a particle of mass $m$, charge $q$ and specific angular momentum $j/m$ if the new quantities $M':=M+m$, $Q':=Q+q$, $J':=J+j$ obey $M'^2 \geqslant Q'^2 + a'^2$, and any other particle who would came to violate this condition would be scattered, keeping the singularity dressed.

In the pioneer work of Wald~\cite{1974AnPhy..82..548W}, considering an extreme black hole ($M^2=Q^2+a^2$), he showed that it is classically impossible to obtain a violation of the CCC, for the electromagnetic field and the centrifugal barriers would scatter the test particle, keeping the singularity hidden inside the EH. Nonetheless, in 1979, Hiscock~\cite{Hiscock1979} said that if the CCC would upheld, then ``strange'' results would appear for a Schwarzschild black hole. For instance, there would be a stable circular orbit in a radius $2M+\epsilon$, $\epsilon\ll M$ outside the EH for a massive test particle. In response to Hiscock's work, Needham~\cite{PhysRevD.22.791} states that the interaction particle-black hole had been ignored when Hiscock obtained the area theorem, leading to the wrong conclusion.

About 20 years later, Hubeny~\cite{PhysRevD.59.064013} saw that on taking a nearly-extreme static and charged black hole ($M \gtrsim Q$ and $a=0$) it was possible to violate the CCC through an overcharge of the black hole by absorbing a test particle where $q \ll Q$ and $m \ll M$, although effects like backreaction and superradiance have been ignored through her work and that those effects could prevent the CCC from being violated. In 1999 Penrose~\cite{Penrose1999} publish another article claiming that previous works were not conclusive.

Felice and Yunqiang~\cite{0264-9381-18-7-307} took into account the effects of backreaction and superradiance on turning a Reissner-Nordstr\"om black hole ($Q\neq0$ and $a=0$) into a Kerr naked singularity ($Q=0$ and $a\neq0$) and confirmed that these effects were not sufficient to uphold the CCC. But Hod~\cite{PhysRevD.66.024016} states that not only those effects would have come into account, but also the particle self-energy, saying that wouldn't be possible to obtain a violation to the CCC.

With the work of Matsas and da Silva~\cite{PhysRevLett.99.181301} it was brought to light the possibility of a plane wave tunneling the potential barrier (a quantum only effect) of the black hole and overspin it. In their calculations, they obtained a non-null probability of a plane wave with high angular momentum to tunnelate a static and charged black hole with energy lower than the potential barrier, keeping its mass almost the same while increasing its angular momentum to the point of violate the CCC by a quantum process. This time, Hod~\cite{PhysRevLett.100.121101} stated that not only backreaction and superradiance were not considered, but also for an ensemble those effects would be extremely important. Matsas \emph{et al.}~\cite{PhysRevD.79.101502} replied that those effects were not sufficient to abide the CCC, but also when considering an ensemble the classical results were recovered.

Not long after, Richartz and Saa~\cite{PhysRevD.84.104021} obtained their first interesting results regarding the particle spin, where they considered particles with spin $0$ and $1/2$. For spin $0$ they showed that the effects of backreaction must be taken into account in the absorption of the particle --- actually, a plane wave --- by the black hole, while those effects are negligible for a plane wave with spin $1/2$, concluding that the backreaction is irrelevant for fermions. Saa and Santarelli~\cite{PhysRevD.84.027501} went far beyond and obtained the optimal results for the particle's characteristics being absorbed by the black hole, minimizing the backreaction. Koray and Semiz~\cite{PhysRevD.88.064043} obtained general results taking into account integer spin fields, leading to the violation of the cosmic censorship conjecture, but still disregarding the backreaction and superradiance, limiting their studies considering plane waves in the low frequency regime.

Recently, Leite \emph{et al}~\cite{PhysRevD.96.044043} calculated numerically the absorption (scattering) of a massive scalar field by a charged, rotating black hole. The results include an analysis of the absorption cross section and the incidence angle, introducing a dependence of the absorption rate regarding those quantities.

Classically, the scientific community common sense regarding the CCC is that it is impossible for an extreme black hole to become a naked singularity, whereas for a near-extreme black hole we may have basically two lines of thought: those who believe that backreaction, superradiance and the particle self-energy take place preventing the CCC to be violated, and those who believe that these effects are not sufficient to sustain the CCC, and thus its violation is possible. Many works were and are being done regarding the possibility of a CCC violation through quantum tunneling --- that is, a particle with energy lower than the black hole potential barrier to be absorbed/emitted and leave the black hole with mass $M'$ such that $M'^2 < Q'^2 + a'^2$ (where the prime quantities are the black hole's new quantities after the absorption/emission). But still there is a lot of divergence where the aforementioned effects may prevent that to happen, where it is believed that the definite answer relies on a quantum theory of gravity.

Basically, we have three lines of thought regarding a quantum violation of the CCC. In no particular order, the first one is the overspin of a black hole, where a particle with mass and angular momentum gets absorbed by it and adds more angular momentum than mass, violating the CCC. Matsas and da Silva~\cite{PhysRevLett.99.181301} were the pioneers in the framework of overspinning a black hole through quantum tunneling using plane waves.

The second consideration is to overcharge a black hole due to the absorption of a charged particle with very low mass. In this case, the absorbed particle adds more charge than mass, thus overcharging the black hole and turning it into a naked singularity. In particular, this was carried out by Richartz and Saa~\cite{PhysRevD.78.081503}, where they overcharged a static and charged black hole through the absorption of spin$-0$ and spin$-\tfrac{1}{2}$ plane waves using the low frequency regime.

In the last case scenario, we may consider the idea of a particle escaping the black hole, that is, a particle being emitted by the black hole. Even though this is an unorthodox approach, it is possible due to the many mechanisms discussed in the literature, such as Hawking radiation~\cite{Hawking:1974rv,hawking1975}, for example. In this case, instead of overspinning or overcharging the black hole, the emitted particle may subtract mass, charge and angular momentum from it, where the case of interest is the one in which the particle carries out more mass than charge and/or angular momentum, violating the CCC and exposing the singularity.

In this work we take a step further into the last case described above. To be more specific, we will determine a semi-analytic function that describes the emission probability of an uncharged scalar particle by a static and charged black hole. Using the same procedure present in the literature, we first consider a massive scalar field being absorbed by it. On solving the Klein-Gordon equation for the given scalar field, we find a Schr\"odinger-like equation for the radial part when changing to the tortoise coordinate that has an actual effective potential with no analytical solution. To circumvent this problem, instead of a numerical approach or work in a low (or high) frequency regime, we propose a toy model that is as close as possible to the actual effective potential and has an analytical solution. In the asymptotic limit, we recover plane waves that allow us to define the reflection and transmission rates, and the latter may be identified as the absorption probability due to the mapping from the physical coordinate system to the tortoise coordinate.

Up to this point this is the standard procedure described in the literature (except for the toy model, which we will discuss below). From here on, we propose a new approach by building a gaussian wave packet from the incoming plane waves as a semiclassical representation of a particle, where we will define the absorption probability for this packet. Since a wave packet is a localized object in space --- contrary to the plane waves approach which consists in an infinite beam of particles --- we can define the absorption probability for this packet at each instant of time as it travels through space. Everything that follows in this work is new using this approach.

So far we have considered the absorption of a particle by the black hole, but we are interested in the case of particle emission by it. Fortunately, this is easily achieved by using the symmetry of the problem and relating the transmission rate from the absorption case with the emission probability (that being allowed due the symmetry of the reflection and transmission rates, which we will explore in detail in appendix~\ref{apendiceb}) and study the probability for the black hole to emit a particle with mass $m_w$. If the particle emitted is such that the black hole's new mass $M'\equiv M-m_w < Q$, then we have a violation of the CCC, that is, a naked singularity.

Even though toy models have been used before in another context~\cite{Skakala2010,Batic:2012rm,Maldacena:2016hyu,Maldacena:2016upp,Maldacena:2017axo}, the novelty about our proposition is that it will be described by a piecewise function that provides an analytical solution without considering any frequency regime. Being so, our toy model can describe an asymmetric potential, which is the case of the actual effective potential.

We used the Pr\"ufer method for calculating the phase-shift, as it is show in appendix~\ref{apendicea}, to validate our method, by writting a C code to calculate the phase-shift from the incoming packet to the absorbed one (remembering that the symmetry of the problem allows us to mirror them to become the outgoing packet from the black hole and the emitted packet) and proceeding to obtain the reflection and transmission rates. The Pr\"ufer method is a numerical tool, thus we must choose a set of parameters and compare both results, where if our method is in agreement with the results from the Pr\"ufer method, it will give us the confidence to proceed in our calculations.

Throughout this work we use the signature $(-,+,+,+)$ for all metrics and the \emph{geometrized unit system} where $G=c=1$, meaning that mass and time have the same unit as length, and so on. For example, the mass of the sun is $M_\odot \approx 10^{30}\,\mbox{kg}$, and in the geometrized unit system it is $M_\odot\approx 1.5\times 10^3\,\mbox{m}$, where the conversion factor to the geometrized unit system is $G/c^2$ in this case. With that in mind, all other conversions are easily done once restored the values for $G$ and $c$ and combining them to get the desired quantity. Also, for simplicity and without loss of generality, we will set $M=1\,\mbox{m}$, unless specified otherwise, and every quantity will be given as a function of the mass $M$ of the black hole, thus any quantity, if wanted to describe a more or less massive black hole depends only on the parameters rescale.

\end{chapter}

\begin{chapter}{The Cosmic Censorship Conjecture for Black Holes}
\label{cap2}

\hspace{5 mm} To have a glimpse on what the CCC is for black holes, we first need to describe the most general case of a black hole. Regarding the no-hair theorem, this case is described by a massive, charged and rotating black hole. Second, we are interested in the outside region of such object. Last and most important, we need a metric to describe such spacetime, which for this case is the Kerr-Newman metric, given as~\cite{doi:10.1063/1.1704351}
\begin{equation}\label{eq:mKerrNewman}
	\diff{s}^2 = -\frac{\Delta^2}{\rho^2}\Big(\diff{t} - a\sin^2\theta\diff\varphi\Big)^2 
	+\frac{\sin^2\theta}{\rho^2}\Big[\Big(r^2+a^2\Big)\diff\varphi - a\diff{t}\Big]^2 + \frac{\rho^2}{\Delta^2}\diff{r}^2 +	 \rho^2\diff\theta^2,
\end{equation}
with
\begin{equation}\label{eq:mKNdefs}
	\Delta^2 := r^2 - 2Mr + Q^2 + a^2, \quad \rho^2 := r^2 + a^2\cos^2\theta, \quad a := \frac{J}{M}
\end{equation}
where $M$ is mass of the black hole, $Q$ its charge and $a$ its specific angular momentum (notice that charge, specific angular momentum and mass has the same unit as length in this unit system). For any given particle, the metric in equation~\eqref{eq:mKerrNewman} will be used to describe its motion in the surrounding region of the black hole. Also, notice that this spacetime is asymptotically flat, that is, by rearranging the metric in equation~\eqref{eq:mKerrNewman},
\begin{multline}
    \diff{s}^2 = 
    \Bigg(\frac{-\Delta^2 + a^2\sin^2\theta}{\rho^2}\Bigg)\diff{t}^2 
    + \frac{\rho^2}{\Delta^2}\diff{r}^2 + \rho^2\diff\theta^2 
    + \left\{\frac{2a\sin^2\theta\Big[\Delta^2 - (r^2+a^2)\Big]}{\rho^2}\right\}\diff{t}\diff\varphi \\[.5em]
    + \Bigg[\frac{(r^2+a^2)^2 - a^2\Delta^2\sin^2}{\rho^2}\theta\Bigg]\sin^2\diff\varphi^2
\end{multline}
and expanding at $r\to+\infty$ disregarding terms of order $\mathcal{O}(r^{-1})$, using the definitions for $\Delta$ and $\rho$ in equation~\eqref{eq:mKNdefs},
\begin{equation}
    \diff{s}^2 \approx -\diff{t}^2 + \diff{r}^2 + r^2\diff\theta^2 + r^2\sin^2\theta\,\diff\varphi^2
\end{equation}
which is the (Minkowski) flat spacetime. This is a requirement in the original formulation of the CCC by Penrose (cf.~\cite{Penrose2002}).

We see that for the quantity $\Delta$ there are exactly two non-trivial roots, given by
\begin{equation}\label{eq:CCC}
    \Delta^2 = 0 \iff r_{H\pm} := M \pm \sqrt{M^2 - Q^2 - a^2}.
\end{equation}
where we define the Event Horizon radius, $r_{H+}$, and the Cauchy Horizon radius, $r_{H-}$. These horizons are key on defining what we call the \emph{weak} and \emph{strong} formulations of the CCC. Before we describe each one, we will briefly describe each horizon and their implications.

\textbf{The Event Horizon (EH):} The EH defines the region around the black hole where the escape velocity from inside it is the speed of light in vaccum $c$ for any given massive and/or massless particle. This means that not even a photon --- the particles of which light is made of, and thus, light itself --- can escape from inside this region, meaning no event can be causally connected to an outside event, i.e., cannot be observed by any outside observer. 

\textbf{The Cauchy Horizon (CH):} The CH defines the region where the flux of gravitational and/or electromagnetic radiation diverges at the crossing~\cite{1982RSPSA.384..301C,Costa:2019uny}. This statement poses a serious problem to the deterministic equations in general relativity, where a problem of initial values cannot be uniquely defined, since in one side of the CH we have space-like ($\diff{s}^2>0$) geodesics while on the other side we have time-like ($\diff{s}^2<0$) geodesics. 

At this point it is interesting to notice the following: once a particle crosses the EH radius, $t$ and $r$ interchange signs, and the metric with signature $(-,+,+,+)$ becomes $(+,-,+,+)$ in a discontinuous way (for the sake of argument, consider here $a=Q=0$). This causes the particle's light cone to tilt 90$^\circ$ and its world line will be directed towards the center of the singularity at $r=0$. Before that happens (now considering $Q\neq0$ and/or $a\neq0$), there is also the CH radius, where once again we have a change in the metric signature from $(+,-,+,+)$ to $(-,+,+,+)$, which means that once inside the region delimited by the CH the particle will have a world line just like outside the EH. This subtle change to the metric signature causes the particle to access a classical forbidden region and thus violate the deterministic equations of motion, suggesting that such black holes may be doorways to a universe besides our own. This is, of course, an extraordinary claim, since its implications ranges from parallel worlds to violation of energy conservation --- and most important, it endangers the deterministic equations of motion, since the problem of initial values is not unique, as stated before. That is the reason why Penrose stated that there must be some kind of mechanism, a \emph{``censor''} that prevents that to happen, thus the name ``Cosmic Censorship Conjecture'', which we now give a brief description in its strong and weak formulations.

\textbf{Strong Cosmic Censorship Conjecture (sCCC):} The sCCC states that the \emph{maximal Cauchy problem} should be completely determined by the initial values. This means that if any singularity should arise in the universe (which poses the aforementioned problem to the deterministic equations of motion), it will happen in such a way that this singularity and the region around it cannot cause any influence whatsoever to an outside observer. Interestingly, this hypothesis seems to have been already disproved~\cite{Dafermos:2017dbw}. 

\textbf{Weak Cosmic Censorship Conjecture (wCCC):} The wCCC states that if any singularity should occur in the universe, it will always happen in a way that it will be behind an EH. This means that any outside observer in the past null infinity is completely oblivious to the existence of this singularity.

Both formulations have the same goal: to forbid access to the singularity by some unknown mechanism. But they are independent formulations, which means that there are cases in which the sCCC holds while the wCCC fails, and \emph{vice-versa}.

Even though we are mainly using the wCCC to work with, we are actually interested in what happens after the collapse, that is, after the black hole is already formed (or it may be considered eternal, as well), as we will reason now.

We can see that both roots $r_{H\pm}$ are real if, and only if, $M^2 \geqslant Q^2 + a^2$, as it is shown in equation~\eqref{eq:CCC}. If a black hole absorbs a particle with quantities $m$, $q$ and $j$ (the latter being the particle total angular momentum), then the black hole's new quantities are $M'=M+m$, $Q'=Q+q$ and $a'=a+j/m$, and the ``censor'' will guarantee that $M'^2 \geqslant Q'^2 + a'^2$. If the new quantities violate this inequality in any way, the ``censor'' would reject that particle and the black hole will not absorb it, keeping the singularity hidden.

Now, say that this ``censor'' does not see a particle coming towards the black hole, and this particle has quantities $m$, $q$ and $j$ that will accrete more charge and/or angular momentum than mass, meaning that the inequality may be violated and the black hole will end up with $M'^2 < Q'^2 + a'^2$. This means that both $r_{H\pm}$ will get imaginary parts and cease to be real. In this scenario, both horizons will cease to exist (since there are no imaginary quantities in the physical world), exposing the singularity. 

The process described in the last paragraph is the idea behind the quantum tunneling of a particle beyond the potential barrier of the black hole, in which a particle with energy lower than the potential barrier gets absorbed by the black hole instead of being scattered (rejected). We will actually do the other way around. The black hole will emit a particle of mass $m_w$ such that the new black hole $M'=M-m_w$ violates the CCC.

\end{chapter}

\begin{chapter}{Scattering by a Schwarzschild Black Hole}
\label{cap3}

\hspace{5 mm} In this chapter we will discuss how the method is applied. Since there is no violation of the CCC for the Schwarzschild black hole when it emits (or absorbs) a particle (the EH only diminishes to zero in the emission or increases in the absorption, while the CH is inexistent), this is done for learning purposes to see if the method is well behaved.

\section{Classical Approach}
Following the classical problem of orbits around a Schwarzschild black hole using the prescription by Schutz~\cite{schutz_2009}, the metric describing the empty spacetime outside a Schwarzschild black hole is
\begin{equation}\label{eq:mSchwarzschild}
    \diff{s}^2 = -\Bigg(1-\frac{r_s}{r}\Bigg)\diff{t}^2 + \Bigg(1 - \frac{r_s}{r}\Bigg)^{-1}\diff{r}^2
		+ r^2\diff\theta^2 + r^2\sin^2\theta\diff\varphi^2
\end{equation}
where 
\begin{equation}
    r_s := \frac{2GM}{c^2}
\end{equation}
is the well known \emph{Schwarzschild radius} and $M$ is the black hole's mass. In the geometrized unit system, we have $r_s:=2M$. We may obtain the Scwharzschild metric from equation~\eqref{eq:mKerrNewman} by setting $Q=a=0$, where in equation~\eqref{eq:CCC} we get $r_{H+} = 2M$ and $r_{H-}=0$, so that the Scwharzschild black hole has an EH with radius $r_{H+}=2M$ and a CH with radius $r_{H-}=0$.

The Lagrangian associated with a free particle of mass $m$ in the spacetime outside a Scwharzschild black hole is~\cite{lemos2007mecanica}
\begin{equation}
    \mathcal{L} = \frac{m}{2}g_{\mu\nu}\dot{x}^\mu\dot{x}^\nu
\end{equation}
where $g_{\mu\nu}$ is the metric tensor given by equation~\eqref{eq:mSchwarzschild} and $\dot{x}^\mu$ is the particle $4-$velocity. The equations of motion are given by the Euler-Lagrange equations,
\begin{equation}
    \dtot{}{\tau}\dpar{\mathcal{L}}{\dot{x}^\mu} = \dpar{\mathcal{L}}{x^\mu}.
\end{equation}
It is not hard to see that if the partial derivative of the Lagrangian with respect to $x^\mu$ is null, then the momentum $p_\mu$ is conserved, that is,
\begin{equation}
    \dpar{\mathcal{L}}{x^\mu} = 0 
    \implies \dtot{}{\tau}\dpar{\mathcal{L}}{\dot{x}^\mu} = 0 
    \implies p_\mu := \dpar{\mathcal{L}}{\dot{x}^\mu} = \mathrm{const.}
\end{equation}
This conserved momentum tells us that some quantities are kept constant throughout the particle motion. For instance, since the metric in equation~\eqref{eq:mSchwarzschild} shows no explicit dependence of the time and $\varphi$ coordinates, we may as well define the conserved quantities
\begin{equation}
    \tilde{E}:=-\frac{p_t}{m}, \qquad \tilde{L}:=\frac{p_\varphi}{m}
\end{equation}
being the particle energy and angular momentum per unit mass. Given the spherical symmetry this means that the movement is confined to a plane, and we might as well choose it as the equatorial plane $\theta=\pi/2$, without any loss of generality. For the $p_\theta$ component, then,
\begin{equation}
    p_\theta = \dpar{\mathcal{L}}{\dot{\theta}} = mr^2\dot{\theta}
\end{equation}
but since $\theta=\pi/2$, it implies that we have $\dot{\theta}=0$, meaning that $p_\theta$ vanishes. Thus, we have the components of the $4-$momentum to be
\begin{equation}
    p^t := g^{tt}p_t = \frac{m\tilde{E}}{\displaystyle 1-\frac{r_s}{r}}, \quad p^r := g^{rr}p_r = m\dot{r}, \quad 
    p^\theta := g^{\theta\theta}p_\theta = 0, \quad p^\varphi := g^{\varphi\varphi}p_\varphi = \frac{m\tilde{L}}{r^2}
\end{equation}
and using the relation $p^\mu p_\mu = -m^2$,
\begin{equation}
    p^\mu p_\mu = -\Bigg(1-\frac{r_s}{r}\Bigg)^{-1}m^2\tilde{E}^2 + m^2\Bigg(1-\frac{r_s}{r}\Bigg)^{-1}\dot{r}^2
    + \frac{m^2\tilde{L}^2}{r^2} = -m^2
\end{equation}
we find the orbit equation for a massive particle outside a Schwarzschild black hole,
\begin{equation}
    \dot{r}^2 = \tilde{E}^2 - \Bigg(1 - \frac{r_s}{r}\Bigg)\Bigg(\frac{\tilde{L}^2}{r^2} + 1\Bigg)
\end{equation}
where we define the effective potential
\begin{equation}
    V_{massive}(r) := \Bigg(1-\frac{r_s}{r}\Bigg)\Bigg(\frac{\tilde{L}^2}{r^2} + 1\Bigg).
\end{equation}
The same procedure may be applied to a massless particle, where the derivatives are taken with respect to some affine parameter $\gamma$, leading to the effective potential
\begin{equation}
    V_{massless}(r) := \Bigg(1-\frac{r_s}{r}\Bigg)\frac{p_\varphi^2}{r^2}.
\end{equation}

In general, the effective potential may be written as
\begin{equation}
    V(r) = \Bigg(1 - \frac{r_s}{r}\Bigg)\Bigg(\frac{L^2}{r^2} + \delta_{p}\Bigg)
\end{equation}
where $\delta_p = 1$ for a massive particle (and $L=p_\varphi/m$) or $\delta_p=0$ (and $L=p_\varphi$) for a massless particle. If then we would like to know about the extrema of the potential to seek for circular orbits, we must set $V'(r)=0$ and solve for $r$. Thus, for a massive particle,
\begin{equation}
    V'_{massive}(r) = \frac{(3r_s - 2r)L^2 + r_s r^2}{r^4} = 0 \iff
    r_\pm = \frac{L^2}{r_s}\left[1 \pm \sqrt{1 - \Bigg(\frac{\sqrt{3}r_s}{L}\Bigg)^2}\right]
\end{equation}
where we have two roots $r_+$ and $r_-$. The root $r_-$ is associated to the maximum of the potential, while $r_+$ is associated to the minimum, as shown in figure~\ref{fig:classical}. If $L=L_c:=\sqrt{3}r_s$, then both maximum and minimum are coincident. If $L$ is chosen to be $L=L_c$, then we have $r_+=r_-=r_c:= 3r_s$, which is known as the \emph{Innermost Stable Circular Orbit} (ISCO) radius, that is the smallest radius around the black hole for a massive particle to have a stable circular orbit, as shown in figure~\ref{fig:ISCO}.

On the other hand, for the massless particle,
\begin{equation}
    V'_{massless}(r) = \frac{(3r_s-2r)L^2}{r^4} = 0 \iff r=r_p := \frac{3r_s}{2}
\end{equation}
regardless of $L$. Note that $r_p$ is an unstable circular orbit (as shown in figure~\ref{fig:classical}) and it is two times smaller than the ISCO.

\begin{figure}
    \centering
    \includegraphics[width=0.95\textwidth]{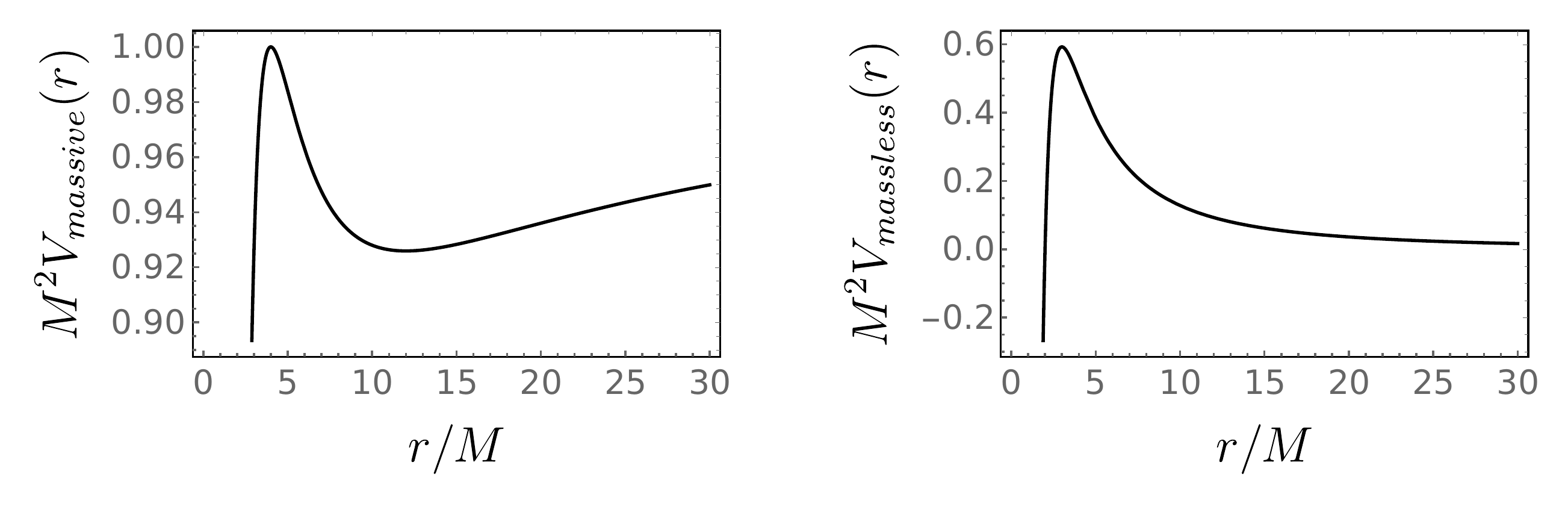}
    \caption[Classical scattering]{Classical effective potential for the scattering of a massive particle (left) and a massless particle (right). The case for a massive particle shows a maximum and a minimum, where the minimum is the stable circular orbit and the maximum is the unstable circular orbit, while for the massless particle we only have the unstable circular orbit (the maximum). Here we have fixed $L = 4$ for both panels.}\label{fig:classical}
\end{figure}

\begin{figure}
    \centering
    \includegraphics[width=0.95\textwidth]{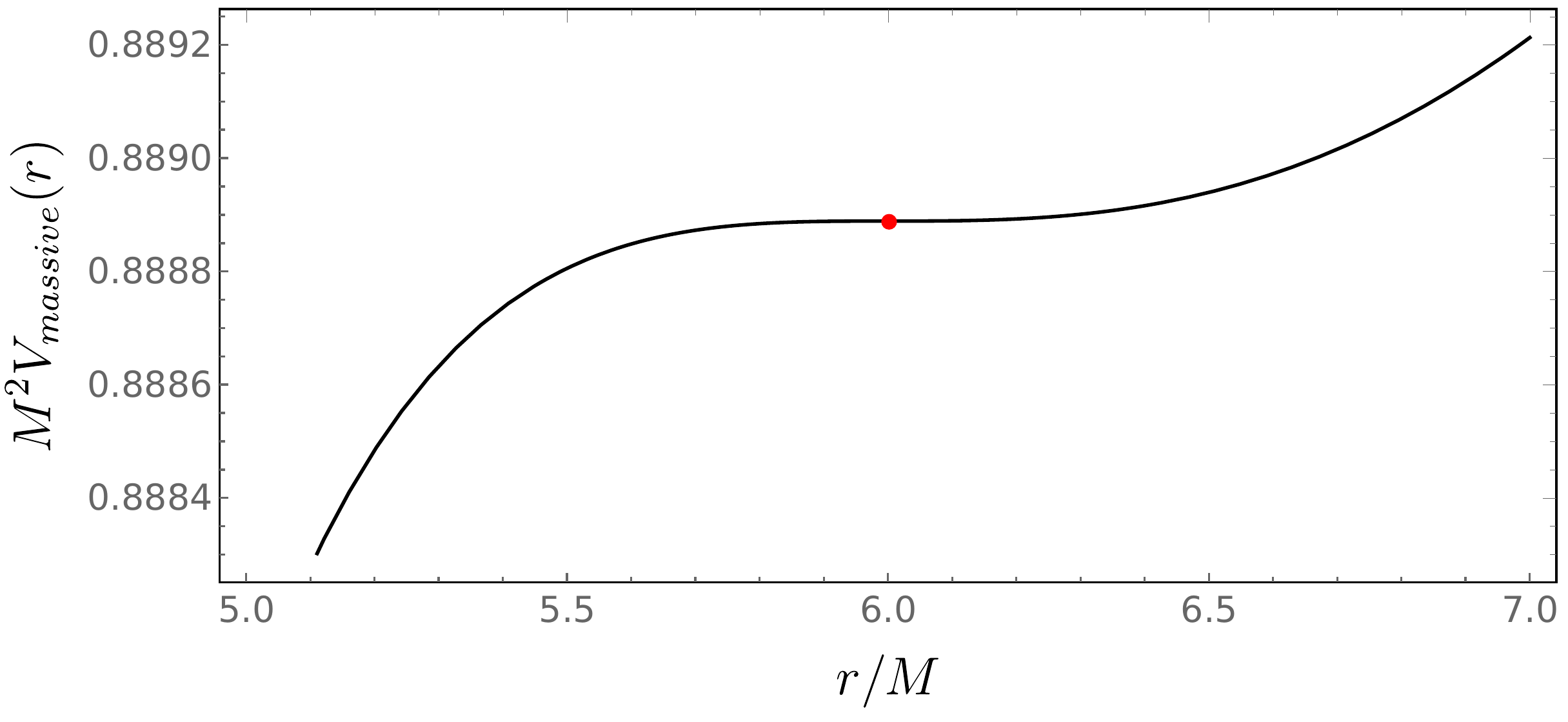}
    \caption[ISCO]{The red dot represents the position of the ISCO at $r=r_c$ for a massive particle with $L=L_c$. Notice that there is no distinction between the maximum and the minimum in this case.}\label{fig:ISCO}
\end{figure}

So it is clear from the classical point of view that there is an unstable circular orbit for both massless and massive cases, but there is only a stable circular orbit for the massive case. Given that, it is to be expected that in the semiclassical approach we must recover that information in the classical limit, that is, considering a wave packet of massive or massless waves representing a massive or a massless particle, in the optical limit, we would get the same results as described here in the classical limit. For that, we begin our next section keeping that in mind for a massive scalar field.

\section{Semiclassical Approach}
The next step is to consider a massive scalar field $\Phi(t,\vec{r})$ outside the black hole and solve for the Klein-Gordon equation~\cite{doi:10.1080/01422418608228771,1926ZPhy...40..117G}
\begin{equation}\label{eq:KG}
    \square\Phi(t,\vec{r}) - \mu^2\Phi(t,\vec{r}) = 0, \quad 
    \square := \frac{1}{\sqrt{-g}}\partial_\mu \sqrt{-g}g^{\mu\nu}\partial_\nu
\end{equation}
where $g_{\mu\nu}$ is the metric in equation~\eqref{eq:mSchwarzschild}, $g$ the determinant of the metric tensor, $\square$ is the D'Alembert operator, and $m_w:=\mu\hbar$ is the scalar field's mass (note that $[\mu] = \mbox{m}^{-1}$). Using separation of variables for the massive scalar field such that $\Phi(t,\vec{r}) = R(r)\Theta(\theta)e^{i(m\varphi - \omega t)}$, where $m\in\mathbb{Z}$ is the projection of the orbital angular momentum along the $z-$axis and $\omega$ is the scalar field frequency, we arrive at the couple of equations
\begin{align}
	r^2\Bigg(1-\frac{r_s}{r}\Bigg)\dtot{^2R}{r^2} + r\Bigg(2-\frac{r_s}{r}\Bigg)\dtot{R}{r} 
		+ \Bigg(\frac{\omega^2r^2 + (r_s - r)\mu^2r}{1-\frac{r_s}{r}} - \mathcal{A}\Bigg)R\label{eq:radial} = 0,\\[.5em]
	\dtot{^2\Theta}{\theta^2} + \cot\theta\dtot{\Theta}{\theta} + \Bigg(\mathcal{A} - \frac{m^2}{\sin^2\theta}\Bigg)\Theta = 0\label{eq:AscLegendre},
\end{align}
with $\mathcal{A}$ being the separation constant. The solutions to equation~\eqref{eq:AscLegendre} are the well known Legendre Polynomials and the separation constant is $\mathcal{A}=\mathcal{A}_\ell=\ell(\ell+1)$ where $|m| \leqslant \ell\in\mathbb{N}$ is the orbital angular momentum. For the radial equation~\eqref{eq:radial} then
\begin{equation}\label{eq:radial2}
	r^2\Bigg(1-\frac{r_s}{r}\Bigg)\dtot{^2R}{r^2} 
	+ r\Bigg(2-\frac{r_s}{r}\Bigg)\dtot{R}{r} 
	+ \Bigg(\frac{\omega^2r^2 + (r_s - r)\mu^2r}{1-\frac{r_s}{r}} -\ell(\ell+1)\Bigg)R = 0.
\end{equation}

We can see that the domain for the $r$-coordinate is $r\in(r_{H+},+\infty)$, but we are only interested in the region outside the black hole, so we proceed to a change of variables to \emph{Eddington-Finkelstein coordinate}, also known as the \emph{tortoise coordinate} $r^\star$ defined by
\begin{equation}\label{eq:tortoise}
    \dtot{r}{r^\star} := 1 - \frac{r_s}{r} 
\end{equation}
so that $r^\star\in(-\infty,+\infty)$, and we ``pushed'' the EH at $r_{H+}=2M$ to $r^\star\to-\infty$. The tortoise coordinate as a function of the $r$-coordinate is easily obtaind by integrating equation~\eqref{eq:tortoise},
\begin{equation}
    r^\star(r) = r + r_s\log\left(\frac{r}{r_s} - 1\right) + C
\end{equation}
where $C$ is the integration constant, and without any loss of generality we might as well set $C=0$. For a discussion on the choice of the boundary conditions for the tortoise coordinate to set the integration constant in such a way that in the asymptotic limit it matches the classical Newtonian phase-shift, we invite the reader to see reference~\cite{Glampedakis:2001cx}.

In the tortoise coordinate system, every $r^\star$-coordinate is outside the black hole, so we don't have to worry about its value to know if we are inside or outside the black hole. Applying equation~\eqref{eq:tortoise} in equation~\eqref{eq:radial2} we get
\begin{equation}\label{eq:Veff}
    \Bigg[\dtot{}{r^{\star^2}} + \omega^2 - V_{eff}(r)\Bigg]u_{\omega\ell m}(r)=0, \quad
    V_{eff}(r) := \Bigg(1 - \frac{r_s}{r}\Bigg)\Bigg(\frac{\ell(\ell+1)}{r^2} + \frac{r_s}{r^3} + \mu^2\Bigg),
\end{equation}
where $u_{\omega\ell m}(r) = rR_{\omega\ell m}(r)$ and $V_{eff}(r)$ is the effective potential for the radial coordinate. Note the mixed ($r$ and $r^\star$) coordinate notation. Equation~\eqref{eq:Veff} is known as the \emph{Schr\"odinger-like} equation because of its clear resemblance to the original time independent Schr\"odinger equation for a particle with energy $E=\hbar\omega$ (but note that the frequency in the Schr\"odinger-like equation is squared, unlike it would be in the original time independent Schr\"odinger equation. Also, the dispersion relation for the Schr\"odinger-like equation gives us $|k|=\omega$). 

Equation~\eqref{eq:Veff} is not analytically integrable for the given potential, and the options to continue would be to make an approximation of the problem for very low (or very high) frequencies, or by proceeding to numerical analysis. Either option limits the problem at hand, while the first option will lead to a very narrow set of solutions, the second one implies that we must choose a particular set of parameters to get the desired information at each time, which can become costly in the computational sense.

To circumvent these issues, we will introduce a \emph{toy model} for the effective potential that is analytically integrable in equation~\eqref{eq:Veff} and is as close as possible to the actual effective potential. Unlike the approximation or numerical methods, this will give us analytical wave functions to the Schr\"odinger-like equation~\eqref{eq:Veff} that will be valid to any frequency value and the deviation from the actual effective potential should be as small as possible, but the method itself has its limitations. Once we aim to exploit the maximum, the minimum and the asymptotic values of the potential, which shows a direct connection to the set of parameters $\{M,\ell,\mu\}$, we cannot randomly pick this set --- the method requires that the aforementioned characteristics to be present for it to work properly.

Nevertheless, we still need to validate the new method. For this, we use the Pr\"ufer method to compare the results with our toy model considering a region of the parameter space in which both are valid. The Pr\"ufer method is a known and well tested numerical tool to calculate the phase-shift, as shown in appendix~\ref{apendicea}. Once our method met in agreement with the Pr\"ufer method, it gave us the confidence to continue with our calculations.

\section{Toy Model}
As discussed before, we need a function for the effective potential in equation~\eqref{eq:Veff} that is as close as possible to the actual one and is analytically integrable to solve for the wave function $u_{\ell m}(r)$. For that matter, we \emph{propose} the following piecewise function
\begin{equation}\label{eq:Vtoy}
    V_{toy}(r^\star) = 
    \begin{cases} 
        V_1(r^\star) = b_1\sech^2[a_1(r^\star-c_1)], & r^\star < r_0^\star \\[.7em]
        V_2(r^\star) = (b_2-\mu^2)\left\{\left[1-e^{-a_2(r^\star-c_2)}\right]^2-1\right\}, & r^\star \geqslant r_0^\star
    \end{cases}
\end{equation}

Both $V_1(r^\star)$ and $V_2(r^\star)$ are analytically integrable once plugged in the Schr\"odinger-like equation~\eqref{eq:Veff}, and are known as the P\"oschl-Teller~\cite{Poschl:1933zz} and Morse~\cite{PhysRev.34.57} potentials, respectively. Figure~\ref{fig:toy} shows a plot of the toy model along the actual effective potential.

\begin{figure}
    \centering
    \includegraphics[width=0.95\textwidth]{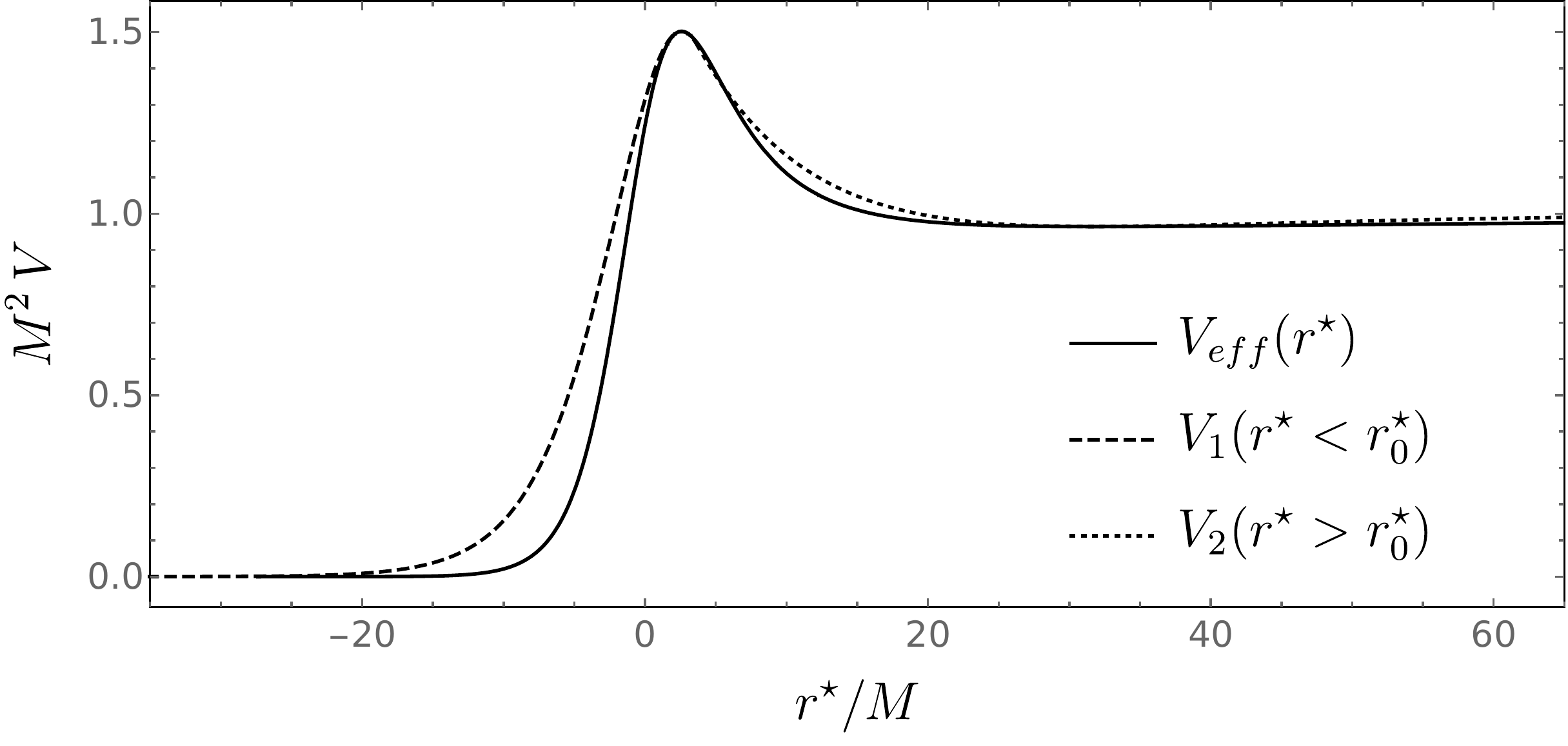}
    
    \flushright
    \includegraphics[width=0.95\textwidth]{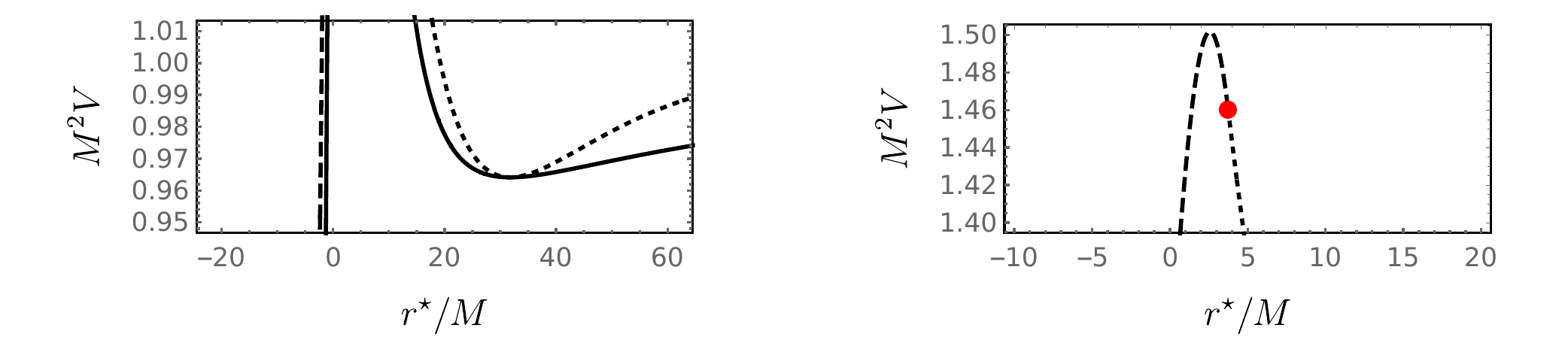}
    \caption[Toy Model for the Effective Potential]{Plot of the actual effective potential (solid line) along with the proposed toy model (dashed and dotted lines) for $\ell=5$ and $M\mu=1.0$ in the above panel. The maximum and the minimum (the latter zoomed in the bottom left panel) of both potentials are exactly the same, and so the asymptotic behavior. In the bottom right panel we show the junction point $r_0^\star$ (red dot), where we can clearly see that $r_0^\star$ lies somewhere in between the maximum and the minimum (in this case, close to the maximum, but not always necessarily so).}\label{fig:toy}
\end{figure}

The asymmetry of the toy model was key on developing the function, since the potential itself is asymmetric. The parameters $a_i,\,b_i,\,c_i\in\mathbb{R}_+$, with $i=1,2$ (which depends on the mass $M$ of the black hole, the angular momentum $\ell$ and mass $\mu$ of the scalar field) control the potential width, height/depth and position of the maximum and the minimum, respectively, in the tortoise coordinate system. Finally, $r_0^\star$ is the junction point for the functions of the toy model.

To correctly set these three parameters, we impose that the extrema of the toy model must match the extrema of the actual effective potential in tortoise coordinate, that is
\begin{equation}
    c_1 = r^\star(r_{\rm max}), \quad c_2 = r^\star(r_{\rm min})
\end{equation}
where $r_{\rm max}$ and $r_{\rm min}$ are the positions of the maximum and minimum of the actual effective potential in the physical $r-$coordinate. The toy model height and depth are given by the height and depth by the actual effective potential (since the tortoise coordinate only changes the horizontal scale, the vertical scale is kept), respectively,
\begin{equation}
    b_1 = V_{eff}(r_{\rm max}), \quad b_2 = V_{eff}(r_{\rm min})
\end{equation}
and finally, for the potential width $a_1$ and $a_2$, we have chosen that $a_1$ is set by taking the second derivatives of the actual effective potential and the toy model at their respective maxima (that is, at $r^\star=c_1$) with respect to $r^\star$ and equating them both, arriving to
\begin{equation}
    V''_{toy}(c_1) = V''_{eff}(c_1) \implies a_1 = \sqrt{-\frac{V''_{eff}(c_1)}{2b_1}}
\end{equation}
where the prime denotes the derivative with respect to $r^\star$. At last, $a_2$ is found together with $r_0^\star$ by imposing the continuity of both the potential and its first derivative at $r^\star=r_0^\star$,
\begin{align}
    \lim_{r^\star \to r_0^{\star -}} V_{toy}(r^\star) &= \lim_{r^\star \to r_0^{\star+}} V_{toy}(r^\star)
    \label{eq:system1} \\[.5em]
    \lim_{r^\star \to r_0^{\star -}} V'_{toy}(r^\star) &= \lim_{r^\star \to r_0^{\star+}} V'_{toy}(r^\star)
    \label{eq:system2}
\end{align}
This condition guarantee that the potential will be smoothly connected at $r_0^\star$. Also, notice that, as shown in figure~\ref{fig:toy}, $c_1 < r_0^\star < c_2$.

The solutions to equation~\eqref{eq:Veff} for a massive scalar field with frequency $\omega$ and the potential given by the toy model, in the tortoise coordinate system, are~\cite{CEVIK20161600,2012arXiv1203.1285P}
\begin{multline}\label{eq:sol1}
    u^{(1)}_{\omega\ell m}(r^\star)
    = \alpha_1\left(\frac{1+y}{1-y}\right)^{\lambda/2}
		{}_2F_1\left(\nu,1-\nu,1+\lambda;\frac{1+y}{2}\right) +\\[.7em]
    + \beta_1 \left[\frac{4}{(1+y)(1-y)}\right]^{\lambda/2}  
		{}_2F_1\left(\nu-\lambda,1-\nu-\lambda,1-\lambda;\frac{1+y}{2}\right)
\end{multline}
for $r^\star < r_0^\star$, where
\begin{equation}
    \lambda := \frac{i\omega}{a_1}, \quad \nu := \frac{1}{2}\left(1+\sqrt{1-\frac{4b_1}{a_2^2}}\right), \quad
    y := \tanh[a_1(r^\star-c_1)]
\end{equation}
with $\lambda,\nu\in\mathbb{C}$, and ${}_2F_1(a,b,c;x)$ is the Gaussian hypergeometric function. For $r^\star \geqslant r_0^\star$,
\begin{multline}\label{eq:sol2}
    u^{(2)}_{\omega\ell m}(r^\star) =
    e^{-z/2}\left\{
		\alpha_2 z^\eta {}_1F_1\left(\frac{1}{2}+\eta-\zeta,1+2\eta;z\right) + \right.\\[.7em]
		\left. + \beta_2 z^{-\eta} {}_1F_1\left(\frac{1}{2}-\eta-\zeta,1-2\eta;z\right)
	\right\}
\end{multline}
with
\begin{equation}
    \eta := \frac{i\omega}{a_2}, \quad \zeta := \frac{\sqrt{b_2}}{a_2}, \quad z := 2\zeta e^{-a_2(r^\star-c_2)}
\end{equation}
where $\eta\in\mathbb{C}$, while $\zeta\in\mathbb{R}$, and ${}_1F_1(a,b;x)$ is the Kummer's function. The coefficients $\alpha_1$, $\alpha_2$, $\beta_1$ and $\beta_2$ are integration constants to be numerically determined by the condition that the wave functions and their derivatives are both continuous at the junction point $r_0^\star$. The complete solution to equation~\eqref{eq:Veff} is then
\begin{equation}
    u_{\omega\ell m}(r^\star) =
    \begin{cases} 
        u^{(1)}_{\omega\ell m}(r^\star), &r^\star < r_0^\star \\[.5em]
        u^{(2)}_{\omega\ell m}(r^\star), &r^\star \geqslant r_0^\star
    \end{cases}
\end{equation}
and its not hard to see that in the limit $r^\star\to\pm\infty$, given the behavior of the hypergeometric functions~\cite{Olver:2010:NHM:1830479}, we have plane waves as expected
\begin{equation}\label{eq:asymexp}
    u_{\omega\ell m}(r^\star) \sim
    \begin{cases} 
        t_{\omega\ell m}e^{-i\omega r^\star}, &r^\star \to -\infty \\[.5em]
        k_{\omega\ell m}e^{-i\varpi r^\star} + r_{\omega\ell m}e^{i\varpi r^\star}, &r^\star \to +\infty
    \end{cases}    
\end{equation}
where $\varpi := \sqrt{\omega^2-\mu^2}$ (here and throughout this work, $\omega>\mu$). The coefficients $k$, $r$ and $t$ (that also depend on $M$, $\mu$ and the wave's amplitudes $\alpha_i$ and $\beta_i$) are determined by the asymptotic expansion and the boundary conditions of a plane wave coming from $r^\star\to+\infty$ to $r^\star\to-\infty$. The condition of no outgoing wave from $r^\star\to-\infty$ was already applied, thus the coefficient of the plane wave $e^{i\omega r^\star}$ is set to zero (i.e., there are no outgoing waves from the black hole), as shown in figure~\ref{fig:absorption}.

\begin{figure}[h]
	\centering
    \includegraphics[width=0.95\linewidth]{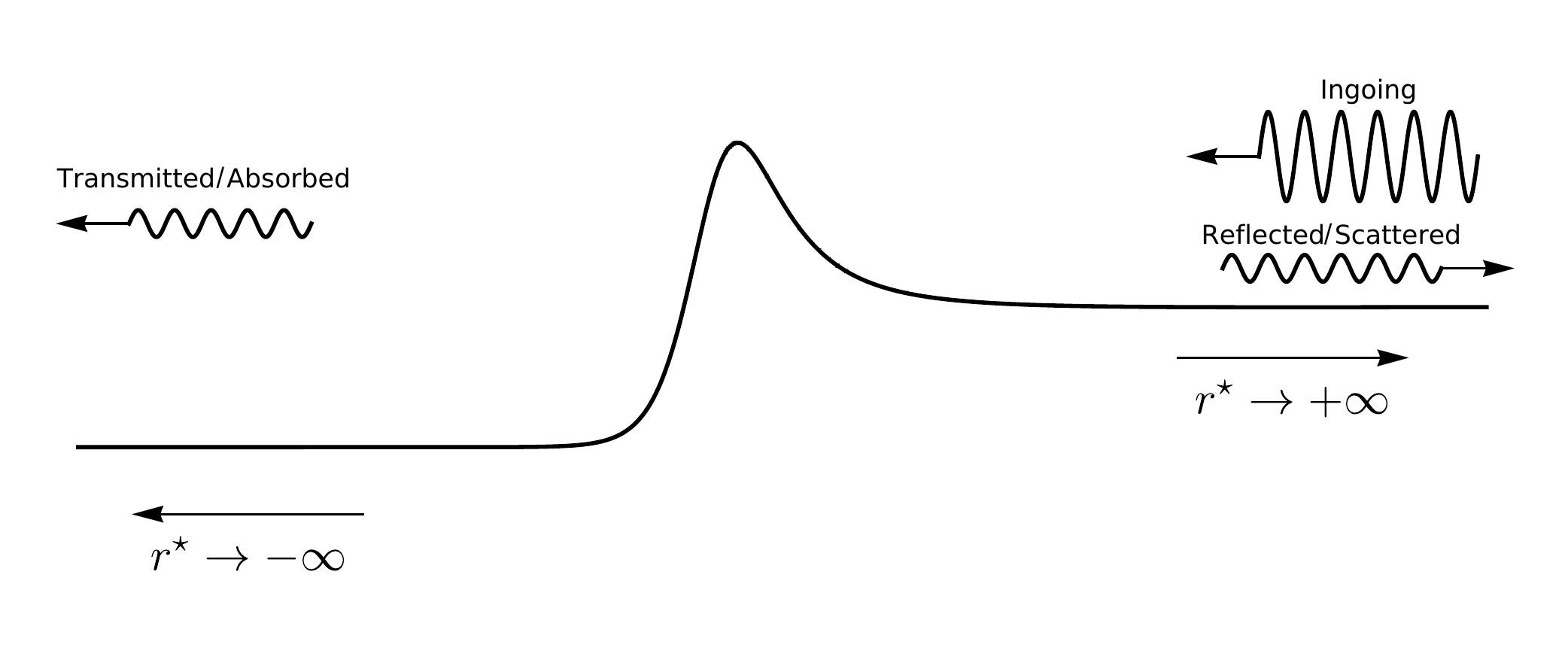}
    \caption[Schematic representation of the scattering/absorption of a plane wave]{Schematic representation of an ingoing plane wave being scattered/absorbed by the potential barrier of the black hole, as represented by the asymptotic expansion in equation~\eqref{eq:asymexp}.}\label{fig:absorption}
\end{figure}

The problem is similar to a quantum-tunneling problem, where we have incoming, reflected and transmitted plane waves (although for the most general the tunneling problem there may be an outgoing mode from the left side, given the appropriate boundary conditions), thus we might as well use that piece of information to give meaning to the quantities
\begin{equation}
    R_w := \Big|\frac{r}{k}\Big|^2, \quad T_w := \Big|\frac{t}{k}\Big|^2
\end{equation}
as the Reflection and Transmission rates for the given plane waves of frequency $\omega$, respectively. With the conservation of flux we get the relation between $T_w$ and $R_w$ to be
\begin{equation}\label{eq:wavecf}
    R_w + \left|\frac{\omega}{\varpi}\right|T_w = 1.
\end{equation}
Also, we may calculate the scattering matrix $S_\ell$ and the phase-shift $\delta_\ell$ defined by
\begin{equation}\label{eq:phaseshift}
    S_\ell := (-1)^{\ell+1}\frac{r}{k} =: e^{2i\delta_\ell}
\end{equation}

We can also define the partial and total absorption cross sections of the massive scalar field via
\begin{equation}\label{eq:partial_and_total_cs}
    \sigma_\ell(\omega) := \frac{\pi}{\varpi^2}(2\ell+1)T_w, \quad
    \sigma_{abs}(\omega) := \sum_{\ell=0}^\infty \sigma_\ell(\omega)
\end{equation}
where some of the partial absorption cross sections are plotted in figure~\ref{fig:pcs} for different values of $\ell$ beginning at $\ell=3$, which are in complete agreement with reference~\cite{PhysRevD.89.104053} that obtained the same results using low and high frequency regimes and also numerical methods for plane waves. Also, in figure~\ref{fig:Twave} we show the transmission rates for two waves with different mass configurations, which we will compare later with the transmission rates for the wave packet. From figure~\ref{fig:Twave} we can see that for high frequency values the transmission rate is maximum, while for low frequency values it is zero. This is to be expected, since the lower $\omega$ is, the larger is the wavelength, and the wave is scattered by the black hole, while for high$-\omega$ the wave gets transmitted, i.e., absorbed by the black hole. This means that almost every particle with low$-\omega$ gets scattered by the black hole, while those who gets absorbed must be done so via quantum tunneling.

\begin{figure}
    \centering
    \includegraphics[width=0.95\textwidth]{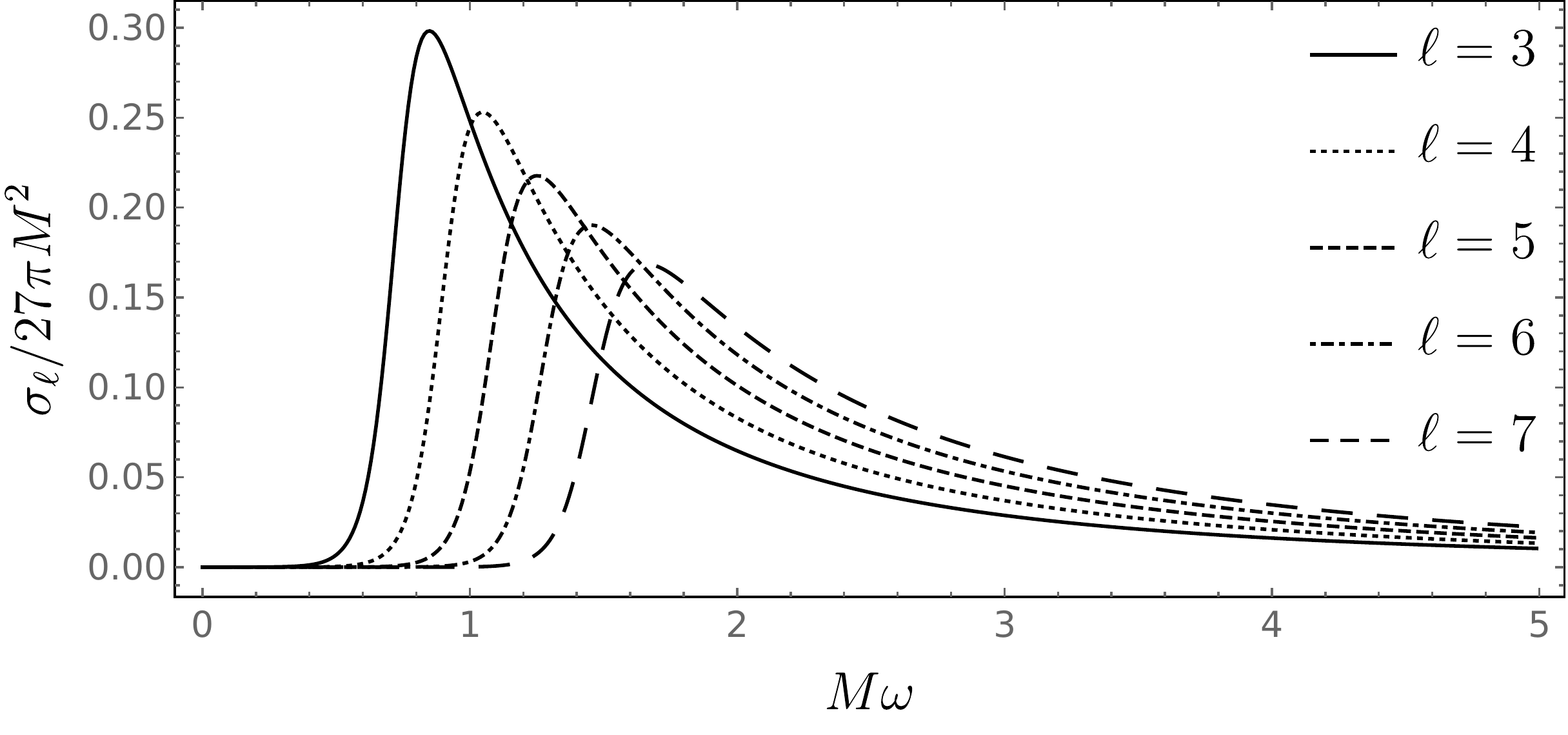}
    \caption[Partial Absorption Cross Section for the Massive Scalar Field]{Partial absorption cross section for different values of the orbital angular momentum $\ell$ with respect to the frequency $\omega$ of the massive scalar field. The normalization constant for the partial absorption cross section was chosen to be $27\pi M^2$ which is the geometric limit for the total absorption cross section. Here, $M\mu=0.5$}\label{fig:pcs}
\end{figure}

\begin{figure}
    \centering
    \includegraphics[width=0.95\textwidth]{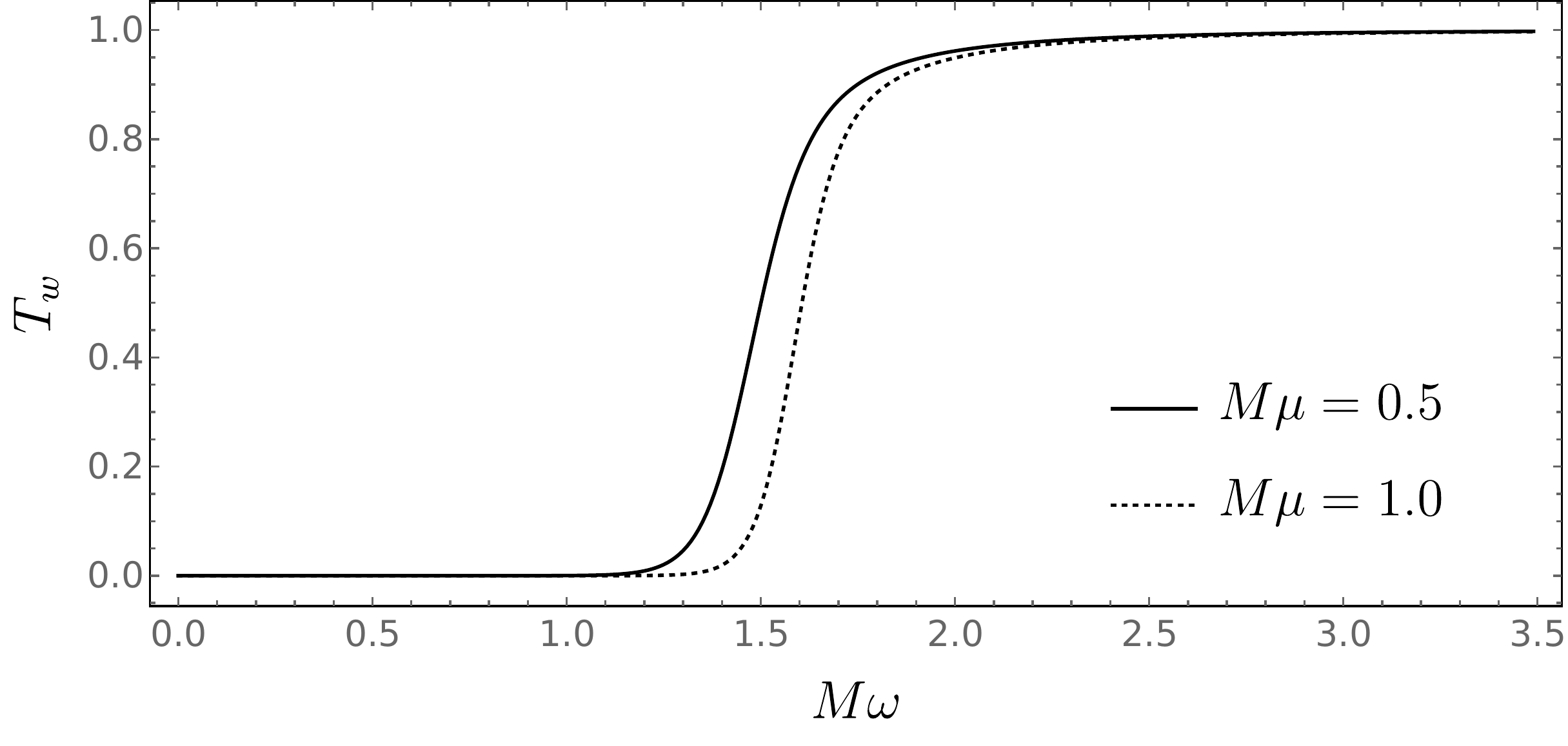}
    \caption[Massive Scalar Field Transmission rate]{Transmission rates for the massive scalar field with respect to its frequency $\omega$. Here, $\ell=7$.}\label{fig:Twave}    
\end{figure}

Matsas and da Silva~\cite{PhysRevLett.99.181301} achieved a violation of the CCC via low$-\omega$ and high$-\ell$ values, thus overspinning a nearly-extreme charged black hole. Using the low$-\omega$ regime, they found an analytical expression for the transmission rate. Using high$-\ell$ values, they found out that, even though small, there is a non-null probability of a particle with low mass and high angular momentum to be absorbed by the black hole, stating that ``the censor may be oblivious to process involving quantum effects''. Two configurations were considered during their calculations, a black hole of mass $M=10^2M_p$ (where $M_p = 2.176\times 10^{-8}\,\mbox{kg}$ is the Planck mass) and $\ell=417$, while the second one was $M=10^5M_\odot$ and $\ell=1141721 \times 10^{58}$. Both configurations considered that the charge of the black hole was $Q = M-e$, where $e$ is the elementary charge, meaning that $M \gtrsim Q$, as stated above.

The problem for the absorption of a scalar field by a black hole has already been discussed for numerous papers throughout the past years, but as for a wave packet is a brand new discussion. Many attempts were proposed in the past 30 years, but very little has been done about the problem. Vishveshwara~\cite{vishveshwara1970scattering} first studied the gravitational radiation of a Schwarzschild black hole carried out by a wave packet, and in his seminal paper, Hawking~\cite{hawking1975} used wave packets to describe the phenomenon now known as \emph{Hawking radiation}, where it is stated that black holes loses mass in the form of thermal radiation.

\section{Wave Packet}

As we have pointed out in the introduction, the systematic approach to the particle absorption problem by a black hole is the use of plane waves, as we also did in the last section. A plane wave represents an \emph{eternal, infinite beam of particles}, and we are calculating the probability (identified as the transmission rate due to the mapping from the physical radial coordinate to the tortoise coordinate) of \emph{one of these particles} to be absorbed by the black hole. The problem in this formulation are the emphasized words.

Since a plane wave is \emph{eternal}, this means that the initial conditions depends on the metric itself. If \emph{one of the particles} in this infinite beam gets absorbed by the black hole, it will change the metric and, consequently, the initial conditions to which this eternal beam of particles was given for. This contradicts the problem itself, where by its own action the plane waves changes its own initial conditions (due to the change in the metric). This is what we call \emph{backreaction}, where the proposed change in the metric changes the initial conditions and we cannot use those plane waves after the absorption because the metric is now different.

To workaround this issue, we propose the use of a \emph{gaussian wave packet} built from the incoming plane waves. Since the wave packet is a localized object in space --- i.e., given an instant in time it has a defined position in space ---, we can define the absorption probability at each instant of time, meaning that the absorption of this packet proposes a local change, contrary to the plane wave, and the problem of backreaction is not an issue once using this approach. This was the idea behind our paper that, up to the publication of this thesis, has been submitted and is in review.

We will build the gaussian wave packet in the position space as the Fourier transform of an also gaussian wave packet in the frequency space, that is
\begin{equation}\label{eq:ftpacket}
    \psi(t,r^\star) = \int_0^\infty \tilde{\psi}(\omega)e^{-i\omega (t+r^\star)} \diff{\omega}, \quad
    \tilde{\psi}(\omega) = \frac{1}{(2\pi\tilde{\xi}^2)^{1/4}}\exp\left[-\frac{(\omega-\omega_0)^2}{4\tilde{\xi}^2}\right]
\end{equation}
notice that $\tilde{\xi}$ is the packet width in the frequency space while $\omega_0$ is its central frequency. The initial condition gives us the desired gaussian wave packet in the position space,
\begin{equation}
	|\psi(0,r^\star)|^2 = \frac{1}{\sqrt{2\pi}}\exp\left(-\frac{r^{\star^2}}{2\xi^2}\right), \quad
	\xi := \tilde{\xi}^{-1}
\end{equation}
where $\xi$ is the packet width in the position space. For the numerical analysis that follows, we will restrict ourselves to the interval $(-2\tilde{\xi},+2\tilde{\xi})$ around the central frequency $\omega_0$, so that the incoming packet is
\begin{equation}\label{eq:incomingpacket}
    \psi_i(t,r^\star) := \int_{\omega_0-2\tilde{\xi}}^{\omega_0+2\tilde{\xi}} 
		k_{\omega\ell m}\,\tilde{\psi}(\omega)e^{-i\varpi(t+r^\star)}\diff{\omega}
\end{equation}
and the same procedure is applied to the transmitted and reflected waves. This was done so that the numerical integration was less time consuming, but the outcome is about the same since we are taking an interval of $4\tilde{\xi}$ for the gaussian wave packet. We choose the values for $\omega_0$ and $\tilde{\xi}$ avoiding negative values for the lower bound in the integration, so we always guarantee that $\omega_0 - 2\tilde{\xi} \geqslant 0$. 


We then proceed on defining the reflection and transmission rates for the wave packet via
\begin{equation}\label{eq:RTwavepacket}
    R:=
    \frac
    {\displaystyle\int_{r_p^\star-\Delta}^{r_p^\star+\Delta} |\psi_r(t,r^\star)|^2\,\diff{r^\star}}
    {\displaystyle\int_{r_p^\star-\Delta}^{r_p^\star+\Delta} |\psi_i(t,r^\star)|^2\,\diff{r^\star}},
    \quad
    T:=
    \frac
    {\displaystyle\int_{r_p^\star-\Delta}^{r_p^\star+\Delta} |\psi_t(t,r^\star)|^2\,\diff{r^\star}}
    {\displaystyle\int_{r_p^\star-\Delta}^{r_p^\star+\Delta} |\psi_i(t,r^\star)|^2\,\diff{r^\star}}    
\end{equation}
with $\psi_i(t,r^\star)$ given in equation~\eqref{eq:incomingpacket} and 
\begin{equation}\label{eq:RTpackets}
    \psi_r(t,r^\star) := \int_{\omega_0-2\tilde{\xi}}^{\omega_0+2\tilde{\xi}}
		r_{\omega\ell m}\, \tilde{\psi}(\omega) e^{i\varpi(t-r^\star)}\,\diff{\omega}, \quad
	\psi_t(t,r^\star) := \int_{\omega_0-2\tilde{\xi}}^{\omega_0+2\tilde{\xi}}
		t_{\omega\ell m}\, \tilde{\psi}(\omega) e^{-i\omega(t+r^\star)}\,\diff{\omega}, \quad
\end{equation}
are the wave packets for the reflected and transmitted waves, respectively. The integration in equation~\eqref{eq:RTwavepacket} is done over a neighborhood $(-\Delta,+\Delta)$ around the central peak $r_p^\star$ of the given incoming packet, just as we did for the wave packet in the frequency space. A good choice for the limits of the integrand would be $\Delta = 2\xi$, which will ensure that the majority of the wave packet will be inside the integration interval around $r_p^\star$.

For the conservation of flux, we have
\begin{equation}
    1 = \frac{|j_r|}{|j_i|} + \frac{|j_t|}{|j_i|}, \quad
    j \propto \psi^\star\dtot{\psi}{r^\star} - \dtot{\psi^\star}{r^\star}\psi
\end{equation}
with $j_i$, $j_r$ and $j_t$ being the probability current density for the incident, reflected and transmitted packets, respectively. Once we are integrating over a narrow region around the central frequency $\omega_0$ of the packet, this means that $\omega \approx \omega_0$, and the prefactor that appears in equation~\eqref{eq:wavecf} is, for the packets,
\begin{equation}\label{eq:fluxcons}
    R + \left|\frac{\omega_0}{\sqrt{\omega_0^2-\mu^2}}\right|T = 1.
\end{equation}

A plot for the transmission rate for the incoming wave packet as a function of the central frequency $\omega_0$ is shown in figure~\ref{fig:Tpacket}.

\begin{figure}
    \centering
    \includegraphics[width=0.95\textwidth]{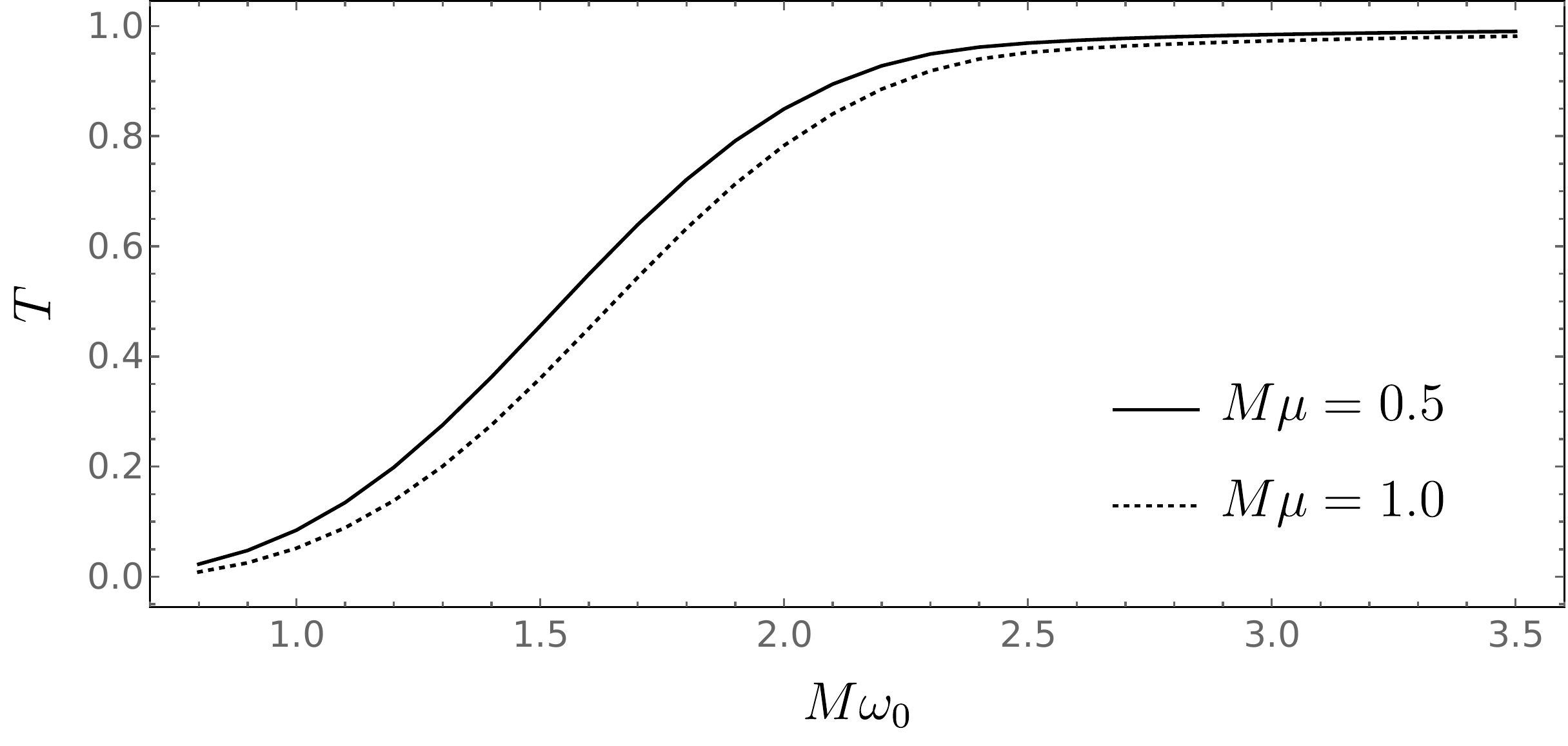}
    \caption[Wave Packet Transmission rate]{Transmission rates for the incoming wave packet with two different mass configurations. The wave packet orbital number is $\ell=7$.}\label{fig:Tpacket}
\end{figure}

Notice how the transmission rate for the wave packet rises more slowly when compared to the transmission rate for the plane wave in figure~\ref{fig:Twave}. Overall, the transmission rate of the packet behaves as the mean of the transmission rates for each individual plane wave contributing to the packet. Consequently, depending on the central frequency and width of the packet, some frequencies may be scattered (low transmission) while others get transmitted, i.e., absorbed. We also show in figure~\ref{fig:comparison} the comparison between the results for the transmission rate for the packet using the Pr\"ufer method and the toy model, showing that both results are in agreement, validating our results. In figure~\ref{fig:fluxerr} we present the error in the flux conservation relative to our method.

\begin{figure}
    \centering
    \includegraphics[width=0.95\linewidth]{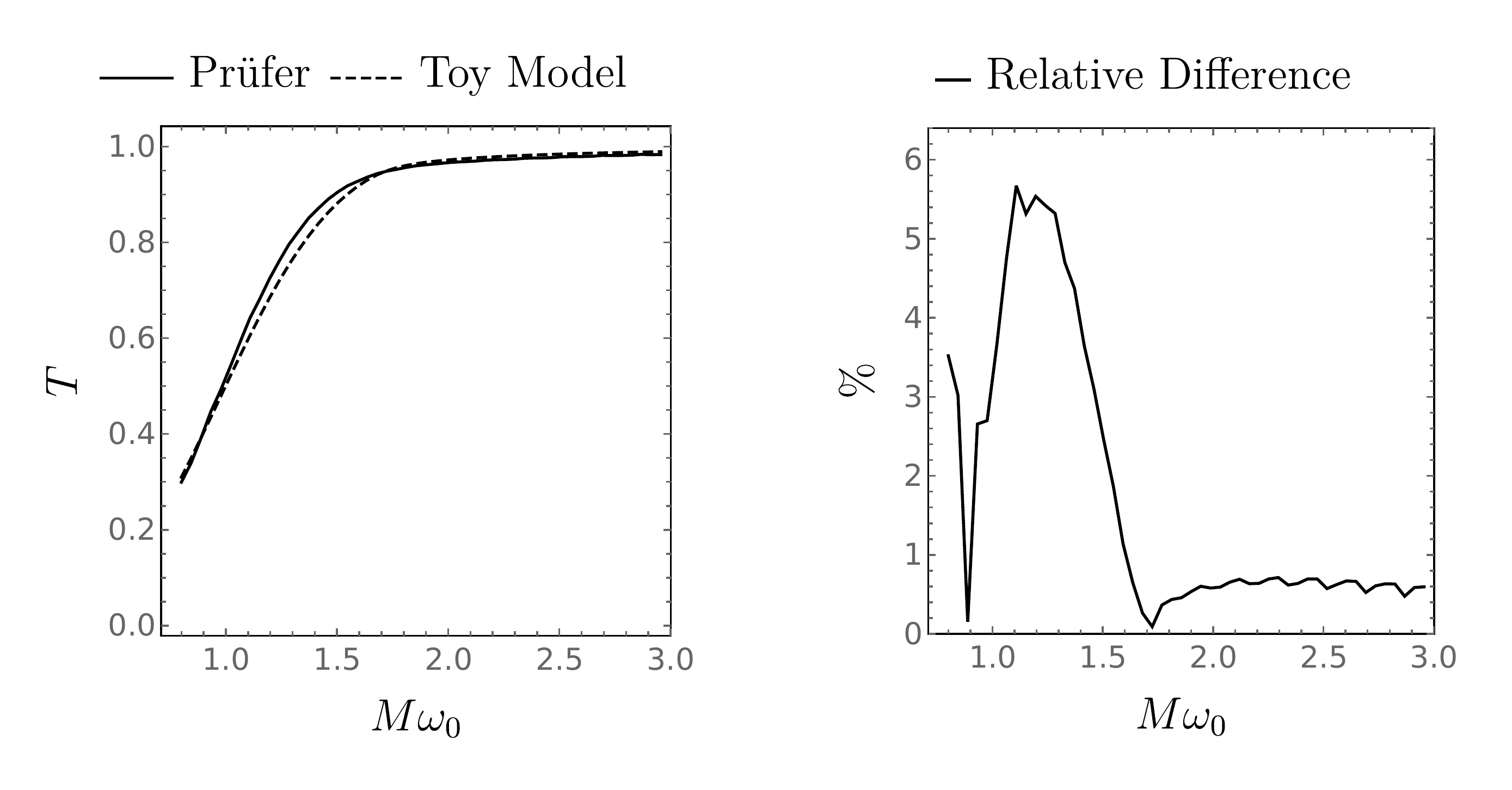}
    \caption[Comparison between the Pr\"ufer method and toy model]{In the left panel we show the comparison of the transmission rate for the packet between the toy model (dashed line) and the Pr\"ufer method (solid line), obtained for a wave packet with $\ell=5$ and $M\mu=0.5$. Notice that the relative difference between toy model and the Pr\"ufer method in the right panel is below 6\% for $\{M,\ell,\mu\}=\{1,5,1\}$, which is one of the set of parameters where both methods are valid.}\label{fig:comparison}
\end{figure}

\begin{figure}
    \centering
    \includegraphics[width=0.95\linewidth]{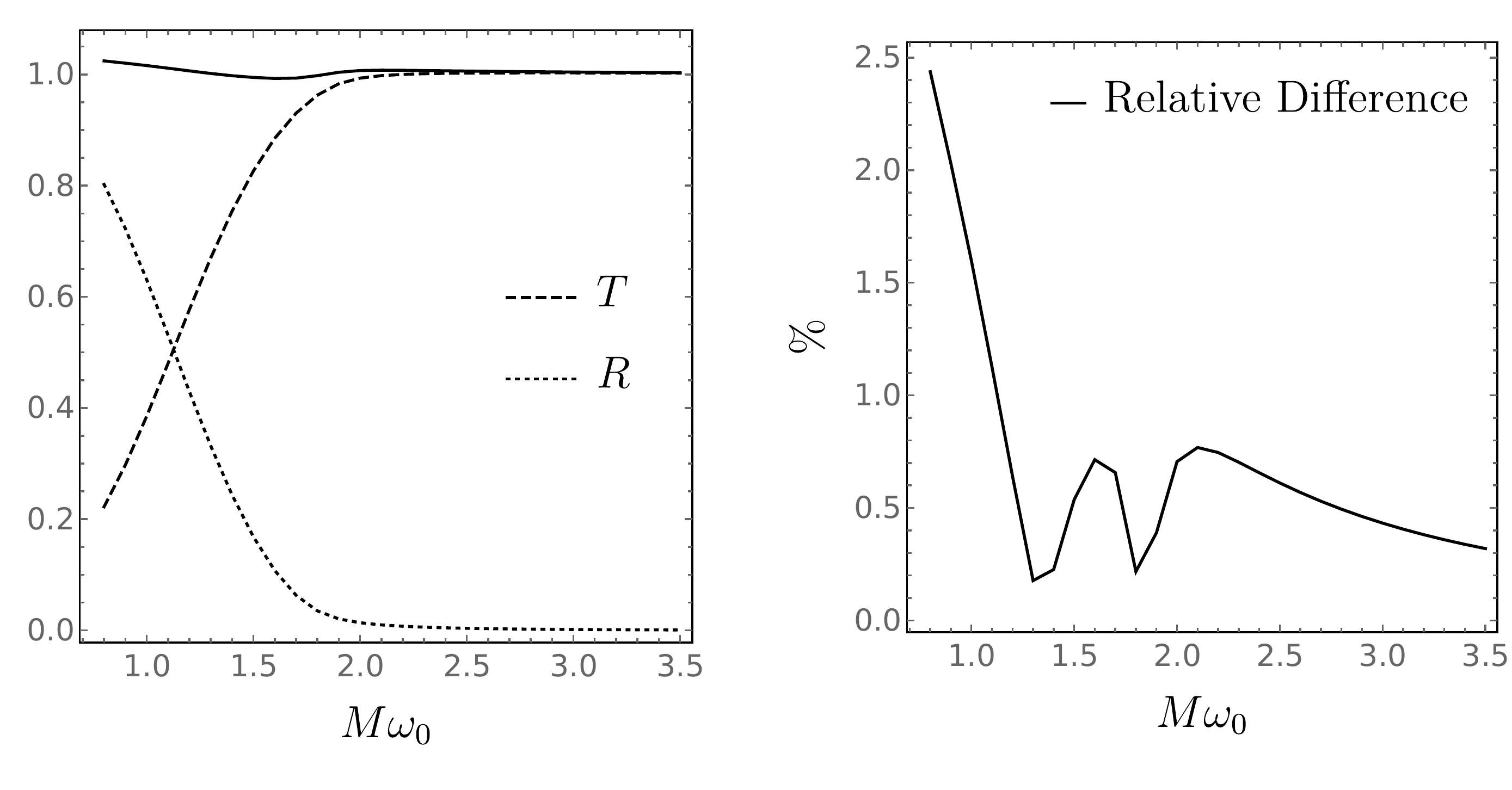}
    \caption[Error in the flux conservation]{The flux conservation for our method with $\ell=5$ and $M\mu=0.5$. The solid line in the left panel is calculated using eq.~\eqref{eq:fluxcons} with the Transmission and Reflections rates obtained with our method, separately. The right panel represents the relative difference of the flux conservation obtained here and compared to its theoretical value (unity), being below $2.5\%$.}\label{fig:fluxerr}
\end{figure}

\section{Integer spin$-s$ Massless Field}
We will end this chapter discussing about integer spin$-s$ for massless fields. In that case, the effective potential in equation~\eqref{eq:Veff} changes qualitatively to~\cite{Sibandze:2016agp}
\begin{equation}
    V_{eff}(r) = \Bigg(1 - \frac{r_s}{r}\Bigg)\Bigg(\frac{\ell(\ell+1)}{r^2} + \frac{(1-s^2)r_s}{r^3}\Bigg)
\end{equation}
where we have introduced the scalar ($s=0$), vector ($s=1$) and tensor ($s=2$) spin contributions for the effective potential. The asymptotic expansion in equation~\eqref{eq:asymexp} for this case is
\begin{equation}
    u_{\omega\ell m}(r^\star) =
    \begin{cases} 
		t_{\omega\ell m}e^{-i\omega r^\star}, &r^\star \to -\infty \\[.5em]
		k_{\omega\ell m}e^{-i\omega r^\star} + r_{\omega\ell m}e^{i\omega r^\star}, &r^\star \to +\infty \\ 
	\end{cases}
\end{equation}
since $\mu=0$. Now not only the coefficients $k$, $r$ and $t$ depends on the black hole's mass, but also on the particle's spin. Though it is not a significant quantitative change in the effective potential at equation~\eqref{eq:Veff}, it is a significant qualitative one. For the effective potential we have just one difference when compared to the massive scalar field which is the lack of the minima, as classical results have already shown for the massless case and as we can see in figure~\ref{fig:massless}

\begin{figure}
    \centering
    \includegraphics[width=0.95\textwidth]{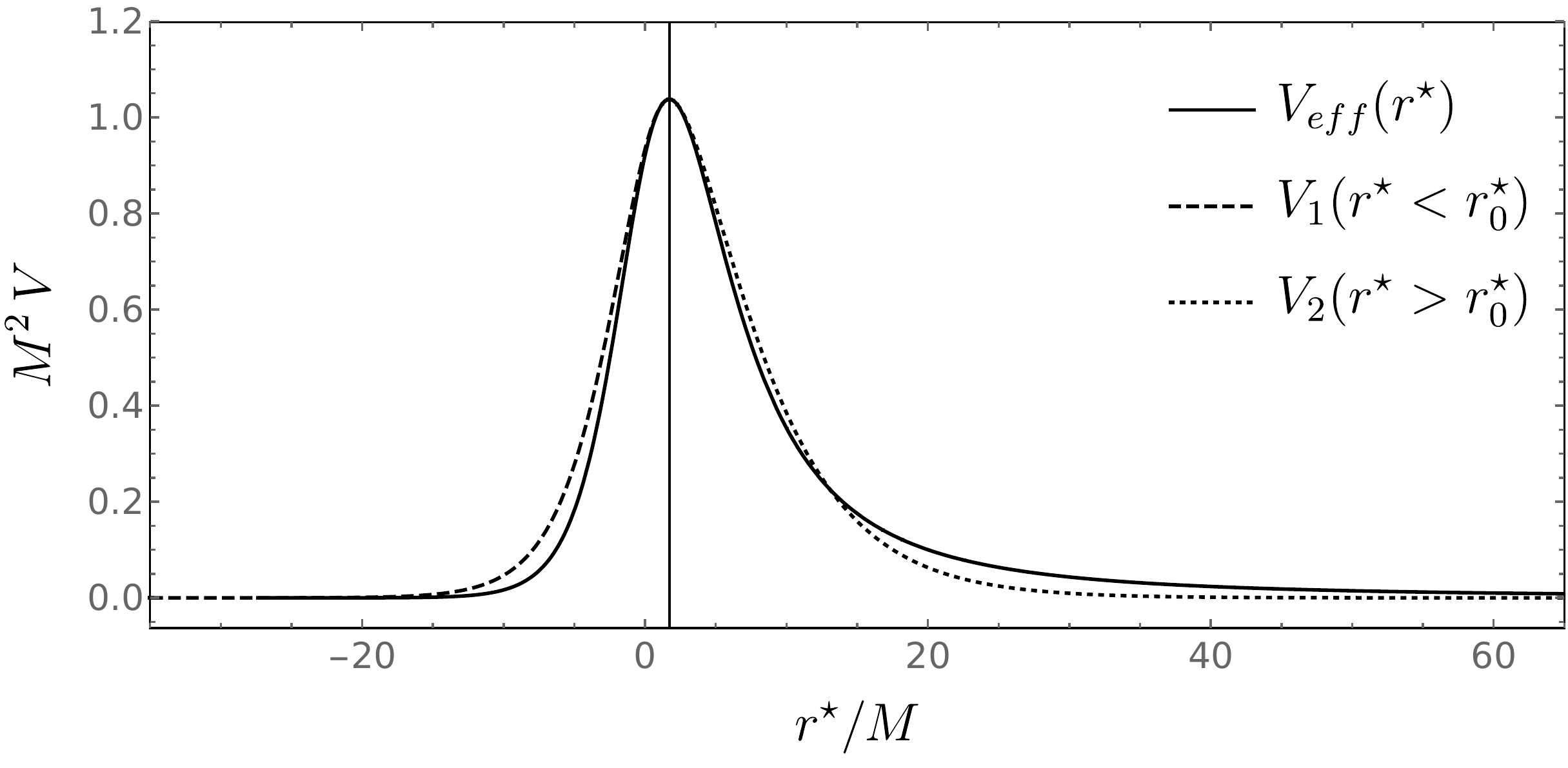}
    \caption[Toy Model for the Effective Potential, with spin]{Effective potential and toy model for a massless spin $s=2$ field. Notice the same behavior as the classical effective potential for a massless particle in figure~\ref{fig:classical}. The field's orbital parameter is $\ell=5$. Also, in this the case of a massless field, the junction point coincides with the maximum, that is, $r_0^\star = c_1$, where the vertical line represents its position $r_0^\star \approx 1.722M$ (which in the physical radial coordinate is $r\approx 3M$).}\label{fig:massless}
\end{figure}

The method itself is easily adapted to consider now the spin of the field, and that allows us to use the same procedure for the massive scalar field to the integer spin$-s$ massless field without having to take a different approach. Thus, in figure~\ref{fig:Twavespin} we show the transmission rate for the plane waves with different values of the spin, and in figure~\ref{fig:Tpacketspin} for the wave packet. 

\begin{figure}
    \centering
    \includegraphics[width=0.95\textwidth]{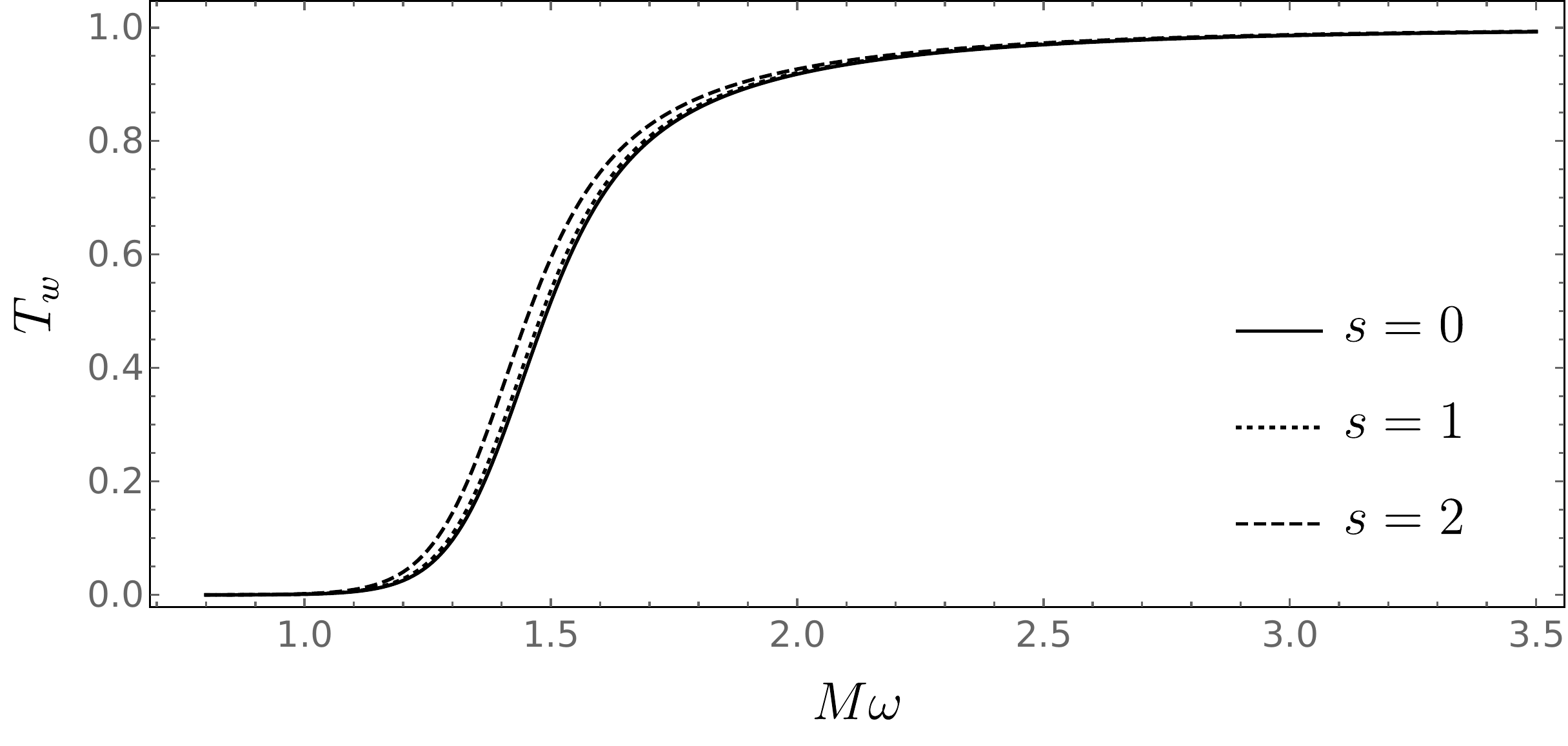}
    \caption[Massless $s$-spin Wave Transmission rate]{Transmission rates for different values of the spin for the plane waves. The difference from $s=0$ (scalar) to $s=1$ (vector) is almost negligible, while it is apparent when $s=2$ (tensor) during the transition phase. The orbital number, as before, is $\ell=7$.}\label{fig:Twavespin}
\end{figure}

\begin{figure}
    \includegraphics[width=0.95\textwidth]{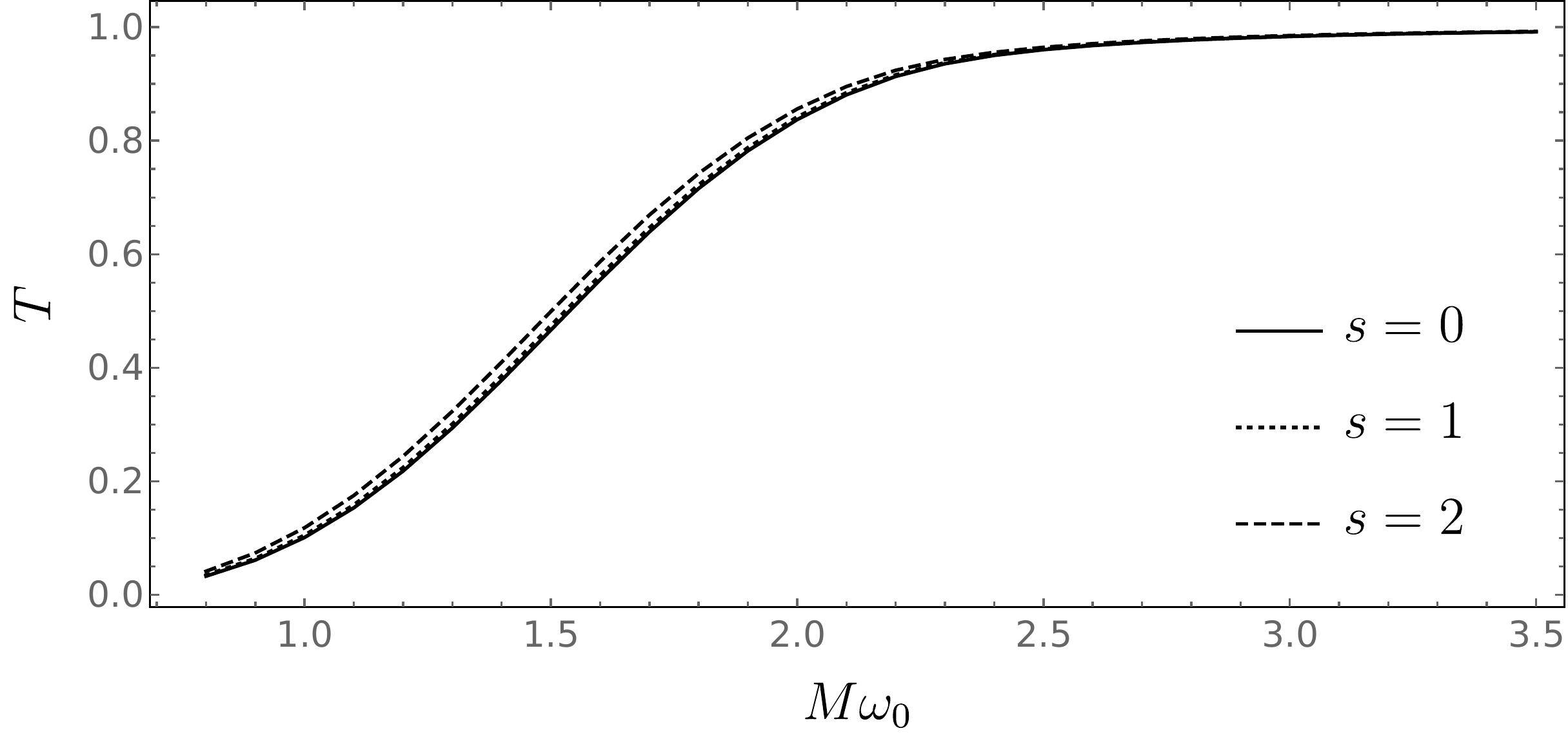}
    \caption[Massless spin$-s$ Wave Packet Transmission rate]{Transmission rates for different values of the spin for the packet. The orbital number, as before, is $\ell=7$.}\label{fig:Tpacketspin}
\end{figure}

A massive integer spin$-s$ field is easily investigated by returning the added term $\mu^2$ in the effective potential and do the same procedure as before. The method handles subtle changes in a very intuitive way and can be adapted without further complications, meaning that we can always try and expand the set of parameters in a convenient way.

From the many studies done in the field here discussed, our method is in complete agreement to the major lines on this topic~\cite{Glampedakis:2001cx,PhysRevD.89.104053,PhysRevD.52.1808,Andersson:2000tf,PhysRevD.18.1030,PhysRevD.91.124030,doi:10.1142/S0218271818430125}. The toy model is a very effective tool to study the problem of the effective potential for the absorption problem of a particle by the black hole. Of course, its not the real problem, but it is a very good approximation to it without having to consider specific values for the plane wave's frequencies and treat the problem as a Taylor expansion up to a given order, nor numerical analysis and have to rerun the code every time we need to change a single or a set of parameters. This approach is completely analytical in every sense of the word without having to restrict ourselves in the cases aforementioned. 

In the next chapter we will consider a subtle modification that will not only change the problem qualitatively, but also quantitatively, by adding charge to the black hole. The addition of a charge brings to existence both the EH and the CH, but also a way to exploit a possible violation of the CCC. 

\end{chapter}

\begin{chapter}{Scattering by a Reissner-Nordstr\"om Black Hole}
\label{cap4}

\hspace{5 mm} The Reissner-Nordstr\"om black hole is a static, charged black hole. Even though observations do not show evidences of its existence (also, its very unlikely to exist a large charged structure in the universe), the reasoning behind it gives us a rich discussion. The change from the Schwarzschild black hole to a Reissner-Nordstr\"om one is a very simple one and the metric itself remains diagonal, then all symmetries involved are kept and the problem remains simple yet complete, and offers us a chance to take a glance at the CCC violation.

\section{Reissner-Nordstr\"om metric}
The region outside a static, charged black hole may be described by the Reissner-Nordstr\"om (RN) metric, given in spherical coordinates as~\cite{doi:10.1002/andp.19163550905,nordstrom1918}
\begin{equation}\label{eq:RNmetric}
    \diff{s}^2 = -\Bigg(1 - \frac{r_s}{r} + \frac{r_q^2}{r^2}\Bigg)\diff{t}^2
		+ \Bigg(1 - \frac{r_s}{r} + \frac{r_q^2}{r^2}\Bigg)^{-1}\diff{r}^2
		+ r^2\diff\theta^2 + r^2\sin^2\theta\diff\varphi^2
\end{equation}
where
\begin{equation}
    r_s :=\frac{2GM}{c^2} = 2M, \quad r_q := \sqrt{\frac{Q^2G}{4\pi\varepsilon_0c^4}} = Q
\end{equation}
We have already defined the Schwarzschild radius $r_s$ and the new \emph{characteristic length} $r_q$, converted to geometrized units.

For a RN black hole, we have the radii $r_{H+}$ and $r_{H-}$ given respectively as
\begin{equation}\label{eq:Hradius}
    r_{H+} = M + \sqrt{M^2 - Q^2}, \quad r_{H-} = M - \sqrt{M^2 - Q^2}
\end{equation}
meaning that we have both an EH radius $r_{H+}$ and a CH radius $r_{H-}$. Unlike the Schwarzschild black hole, there may be now a possibility to violate the CCC by either overcharging the black hole (i.e., absorbing a particle in such a way that it adds more charge than mass) or by emission of an uncharged massive particle by the black hole, meaning that the inequality $M \geqslant Q$ may be violated and both EH and CH become imaginary and cease to exist, exposing the singularity. But, if the CCC is to be upheld, it means that there must be some kind of mechanism (the ``censor'') that will forbid that to happen and will keep the singularity hidden, and that is the discussion that follows.

\section{Massive Wave Packet absorption}

As we have done in chapter~\ref{cap3}, let us consider a massive scalar field for the Klein-Gordon equation, but now using the metric in equation~\eqref{eq:RNmetric}, that is, solve for
\begin{equation}\label{eq:KGRN}
    \square \Phi(t,\vec{r}) - \mu^2\Phi(t,\vec{r}) = 0,
\end{equation}
as we have done before in equation~\eqref{eq:KG}, but now using the metric in equation~\eqref{eq:RNmetric}. Since the difference between equation~\eqref{eq:mSchwarzschild} and equation~\eqref{eq:RNmetric} is a simple addition of $r_q^2/r^2$ in the right places, carrying out the calculations for equation~\eqref{eq:KGRN} using the same separation of variables from the last chapter for the field $\Phi(t,\vec{r})$ gives us a similar pair of equations
\begin{align}
	r^2\Bigg(1-\frac{r_s}{r}+\frac{r_q^2}{r^2}\Bigg)\dtot{^2R}{r^2} + r\Bigg(2-\frac{r_s}{r}\Bigg)\dtot{R}{r} +
		\Bigg(\frac{\omega^2r^2+[(r_s-r)r - r_q^2]\mu^2}{1-\frac{r_s}{r}+\frac{r_q^2}{r^2}} - \mathcal{A}\Bigg)R\label{eq:radialRN} 
		= 0,\\[.5em]
	\dtot{^2\Theta}{\theta^2} + \cot\theta\dtot{\Theta}{\theta} 
		+ \Bigg(\mathcal{A} - \frac{m^2}{\sin^2\theta}\Bigg)\Theta = 0\label{eq:AscLegendreRN},
\end{align}
where, as expected, equation~\eqref{eq:AscLegendreRN} is exactly the same as equation~\eqref{eq:AscLegendre}, since there are no changes in the spherical symmetry of the problem, thus the only change comes from equation~\eqref{eq:radialRN} once compared to equation~\eqref{eq:radial}. Obviously, when $r_q=0$, we recover equation~\eqref{eq:radial}.

We already know that the solutions to equation~\eqref{eq:AscLegendreRN} are the Legendre Polynomials and $\mathcal{A} = \mathcal{A}_\ell = \ell(\ell+1)$ with $|m|\leqslant\ell\in\mathbb{N}$. For the radial equation~\eqref{eq:radialRN}, we are left to
\begin{equation}
    r^2\Bigg(1-\frac{r_s}{r}+\frac{r_q^2}{r^2}\Bigg)\dtot{^2R}{r^2} + r\Bigg(2-\frac{r_s}{r}\Bigg)\dtot{R}{r} 
		+ \Bigg(\frac{\omega^2r^2+[(r_s-r)r - r_q^2]\mu^2}{1-\frac{r_s}{r}+\frac{r_q^2}{r^2}} 
			- \ell(\ell+1)\Bigg)R\label{eq:radial2RN} = 0
\end{equation}
and $r\in(r_{H+},+\infty)$. It comes at hand to, once again, introduce the tortoise coordinate, but this time it is defined as
\begin{equation}\label{eq:tortoiseRN}
    \dtot{r}{r^\star} := 1 - \frac{r_s}{r} + \frac{r_q^2}{r^2}
\end{equation}
and by solving this equation we get $r^\star$ as a function of $r$, that is
\begin{equation}
    r^\star(r) = r + \frac{r_{H+}^2}{r_{H+} - r_{H-}}\log\left(\frac{r}{r_{H+}} - 1\right) 
		- \frac{r_{H-}^2}{r_{H+} - r_{H-}}\log\left(\frac{r}{r_{H-}} - 1\right) + C
\end{equation}
with $r_{H\pm}$ defined in equation~\eqref{eq:Hradius}. Once again, we have $r^\star\in(-\infty,+\infty)$ and both the EH and CH are pushed to $r^\star\to-\infty$, so every $r^\star$ coordinate is guaranteed to be outside the black hole. Applying equation~\eqref{eq:tortoiseRN} in equation~\eqref{eq:radial2RN}, we get the same Schr\"odinger-like as in equation~\eqref{eq:Veff}, but this time with the effective potential given by equation
\begin{equation}\label{eq:VeffRN}
    V_{eff}(r) = \Bigg(1 - \frac{r_s}{r} + \frac{r_q^2}{r^2}\Bigg)
		\Bigg(\frac{\ell(\ell+1)}{r^2} + \frac{r_s}{r^3} - \frac{2r_q^2}{r^4} + \mu^2\Bigg)
\end{equation}
and since the change from the effective potential from equation~\eqref{eq:Veff} is just the addition of the terms relative to $r_q$, we will use the same toy model proposed in equation~\eqref{eq:Vtoy},
\begin{equation}\label{eq:VtoyRN}
    V_{toy}(r^\star) = 
    \begin{cases} 
        V_1(r^\star) = b_1\sech^2[a_1(r^\star-c_1)], & r^\star < r_0^\star \\[.7em]
        V_2(r^\star) = (b_2-\mu^2)\left\{\left[1-e^{-a_2(r^\star-c_2)}\right]^2-1\right\}, & r^\star \geqslant r_0^\star
    \end{cases}
\end{equation}
A plot of the toy model along with the effective potential in equation~\eqref{eq:VeffRN} is shown in figure~\ref{fig:VtoyRN}. This means that equations~\eqref{eq:sol1} and~\eqref{eq:sol2} will be solutions to the Schr\"odinger-like equation~\eqref{eq:Veff} with the effective potential given by equation~\eqref{eq:VeffRN}, keeping in mind that the parameters $a_i$, $b_i$ and $c_i$, $i=1,2$, now will depend on the mass $M$ and charge $Q$ of the black hole, as well as the orbital parameter $\ell$ and mass $\mu$ of the field.

\begin{figure}[!ht]
    \centering
    \includegraphics[width=1\textwidth]{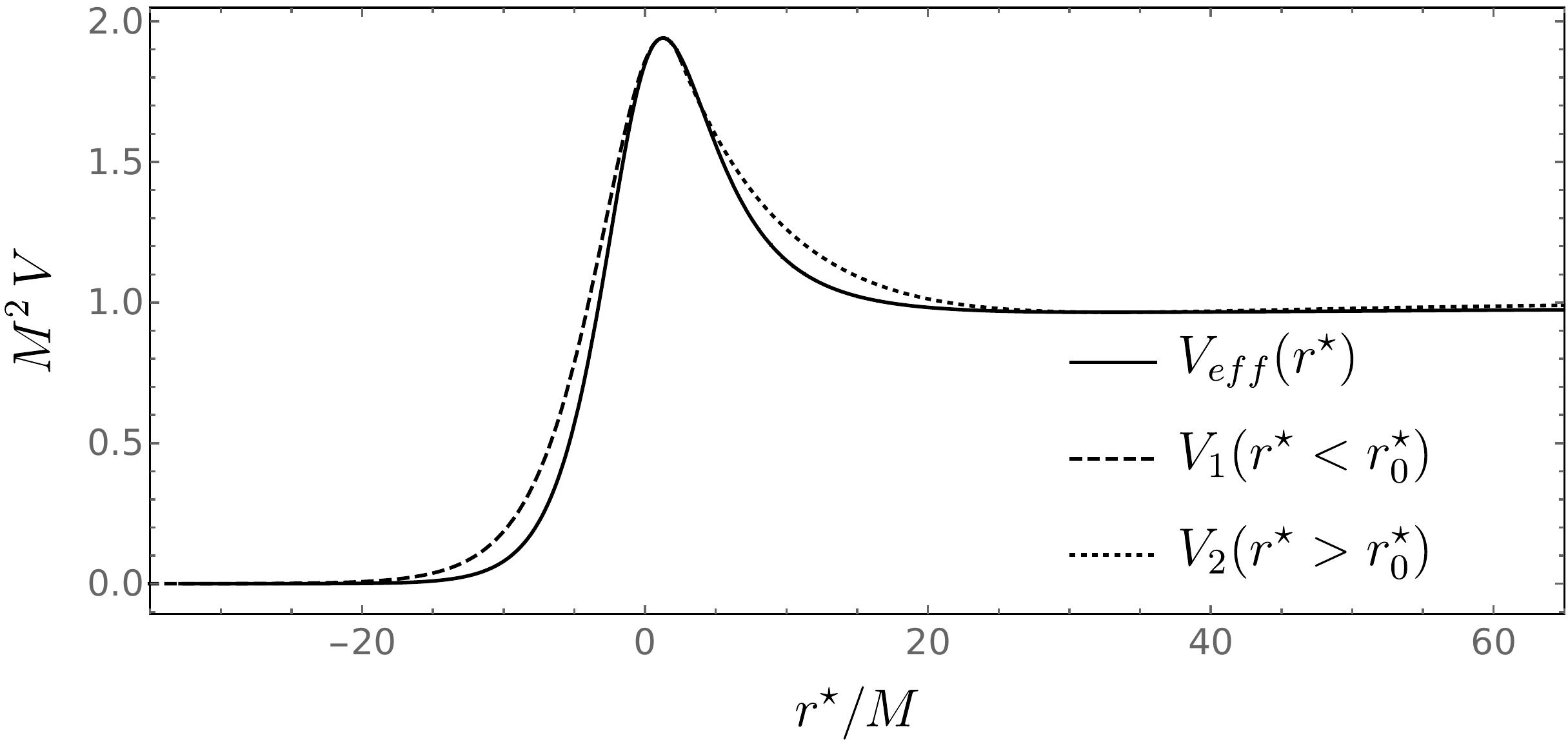}
    \caption[Reissner-Nordstr\"om Effective Potential]{Effective potential (solid) and Toy Model (dashed and dotted) for the static, charged black hole proposed by the RN metric in equation~\eqref{eq:RNmetric}. Here we have set the charge of the black hole to be $Q=0.9M$ (charge has the same unit as length in the geometrized unit system) while the field's orbital number and mass are $\ell=5$ and $M\mu=1.0$, respectively.}\label{fig:VtoyRN}
\end{figure}

So it is quite straightforward to proceed with our method since this new problem is easily treatable within it. Again we build a wave packet for the incoming waves as a semiclassical representation of a particle and then study its absorption by the black hole. We will just show the meaningful quantities here, as we did in chapter~\ref{cap3}.

The transmission rates for the plane waves of the massive scalar field are shown in figure~\ref{fig:TwaveRN} for different values of the charge $Q$ of the black hole, and in figure~\ref{fig:TwaveRN2} for different values of the mass $\mu$ of the field, and also for the partial absorption cross section for fixed values of $\mu$ of the packet and charge $Q$ of the black hole in figure~\ref{fig:pcsRN}.

As for the transmission rates for the wave packet with central frequency $\omega_0$, we show in figure~\ref{fig:TpacketRN} for different values of the charge $Q$ of the black hole, and in figure~\ref{fig:TpacketRN2} for different values of the mass $\mu$ of the plane waves composing the packet.

All other quantities (scattering matrix $S_\ell$, phase-shift $\delta_\ell$, and total absorption cross section $\sigma_{abs}$) are easily obtained within the method itself, since every quantity described and shown here are analytical, as we have mentioned before.

\begin{figure}[!htbp]
    \centering
    \includegraphics[width=0.95\textwidth]{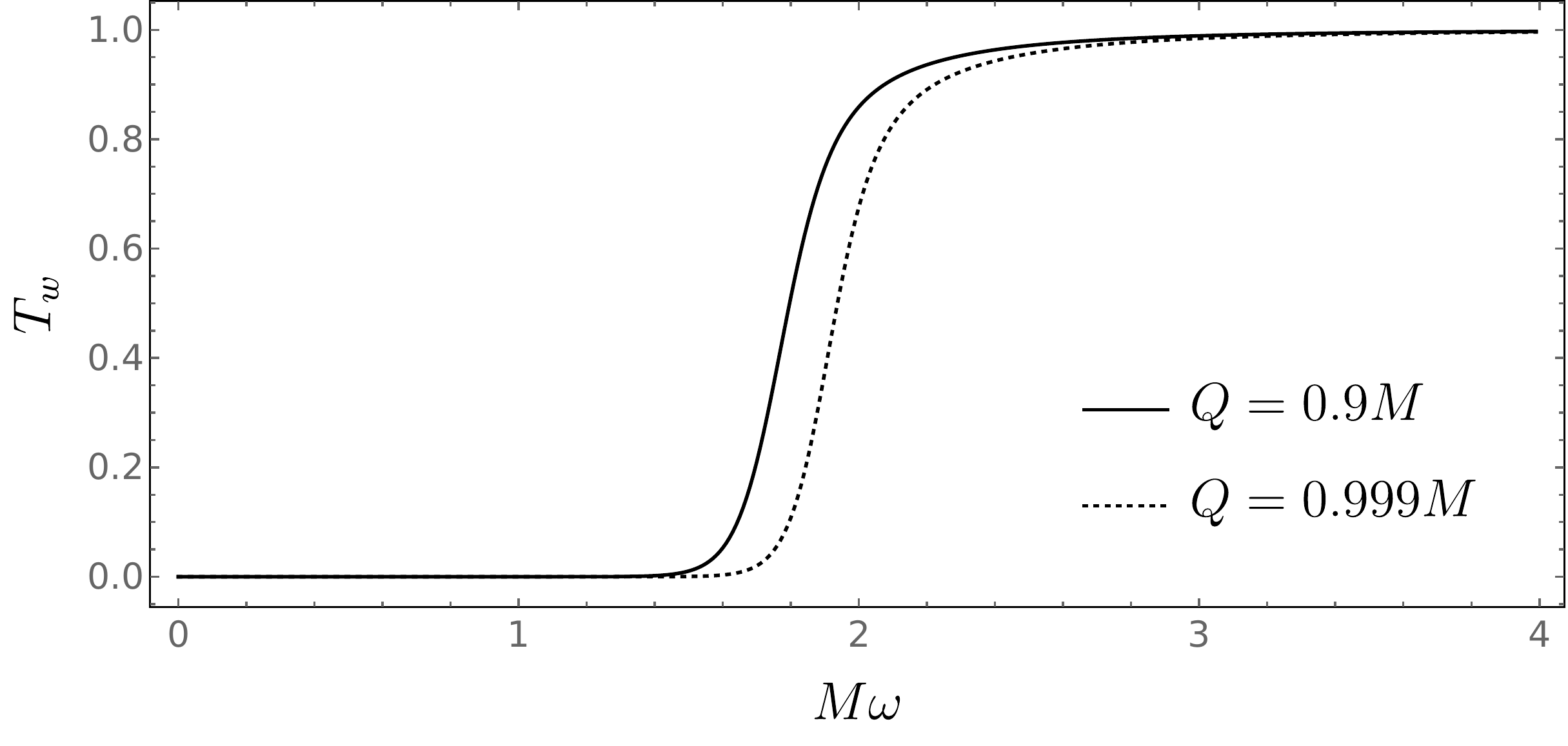}
    \caption[Transmission rates for the Massive Scalar Field, fixed $\mu$]{Transmission rates for the plane waves of the massive scalar field with $M\mu=0.5$ and different values of charge $Q$ of the black hole. For both curves, $\ell=7$}\label{fig:TwaveRN}
\end{figure}

\begin{figure}[!htbp]
    \centering
    \includegraphics[width=0.95\textwidth]{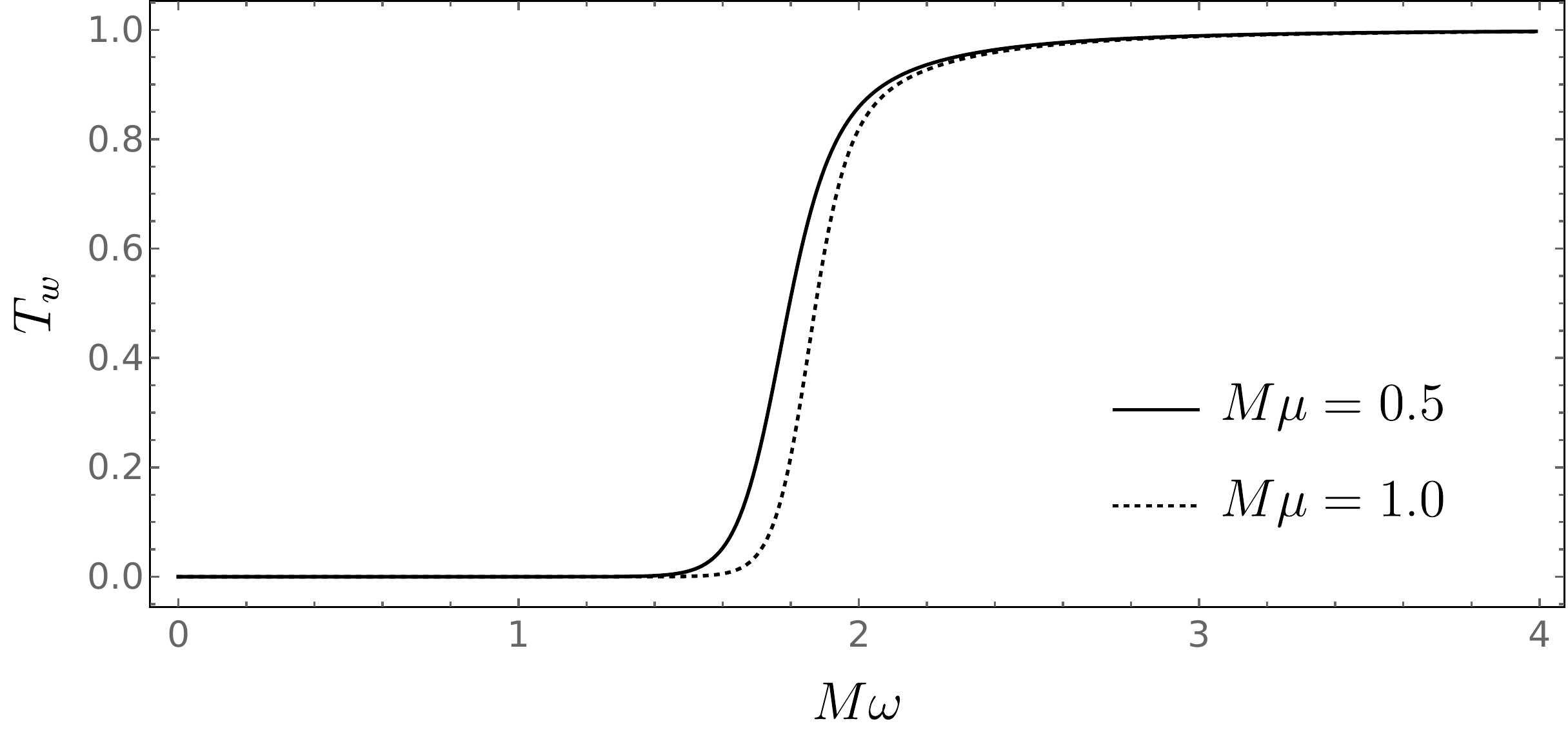}
    \caption[Transmission rates for the Massive Scalar Field, fixed $Q$]{Transmission rates for the plane waves of the massive scalar field with $Q=0.9M$ and different values of the mass $\mu$ of the field and orbital number $\ell=7$.}\label{fig:TwaveRN2}
\end{figure}

\begin{figure}[!htbp]
    \centering
    \includegraphics[width=0.95\textwidth]{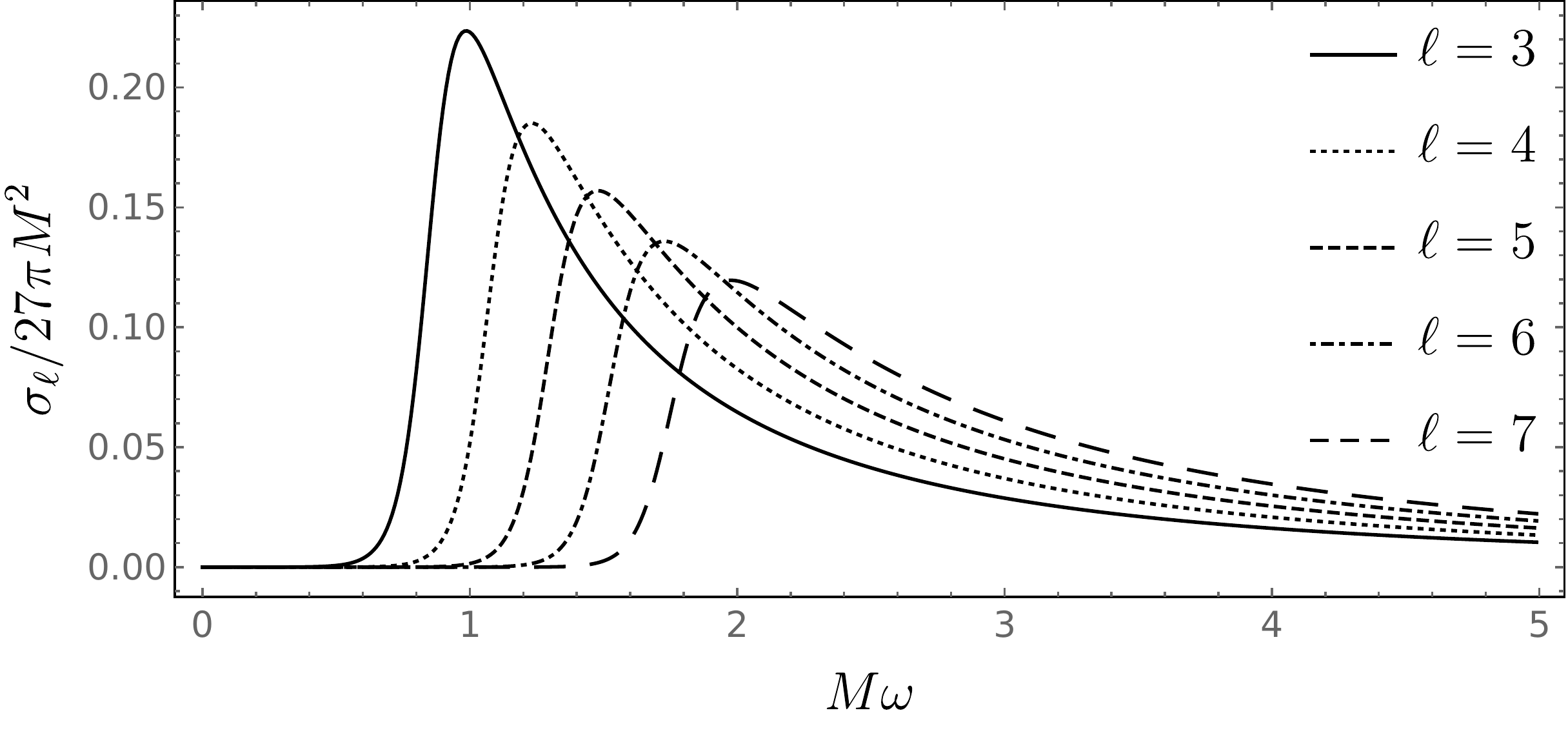}
    \caption[Partial Absorption Cross Section for the Massive Scalar Field]{Partial Absorption cross section for the plane waves of the massive scalar field for different values of the orbital parameter $\ell$ and fixed values of the black hole's charge $Q=0.9M$ and field's mass $M\mu=0.5$. The normalization constant for the vertical axis is chosen to be the optical limit, just as in figure~\ref{fig:pcs}.}\label{fig:pcsRN}
\end{figure}

\begin{figure}[!htbp]
	\centering
	\includegraphics[width=0.95\textwidth]{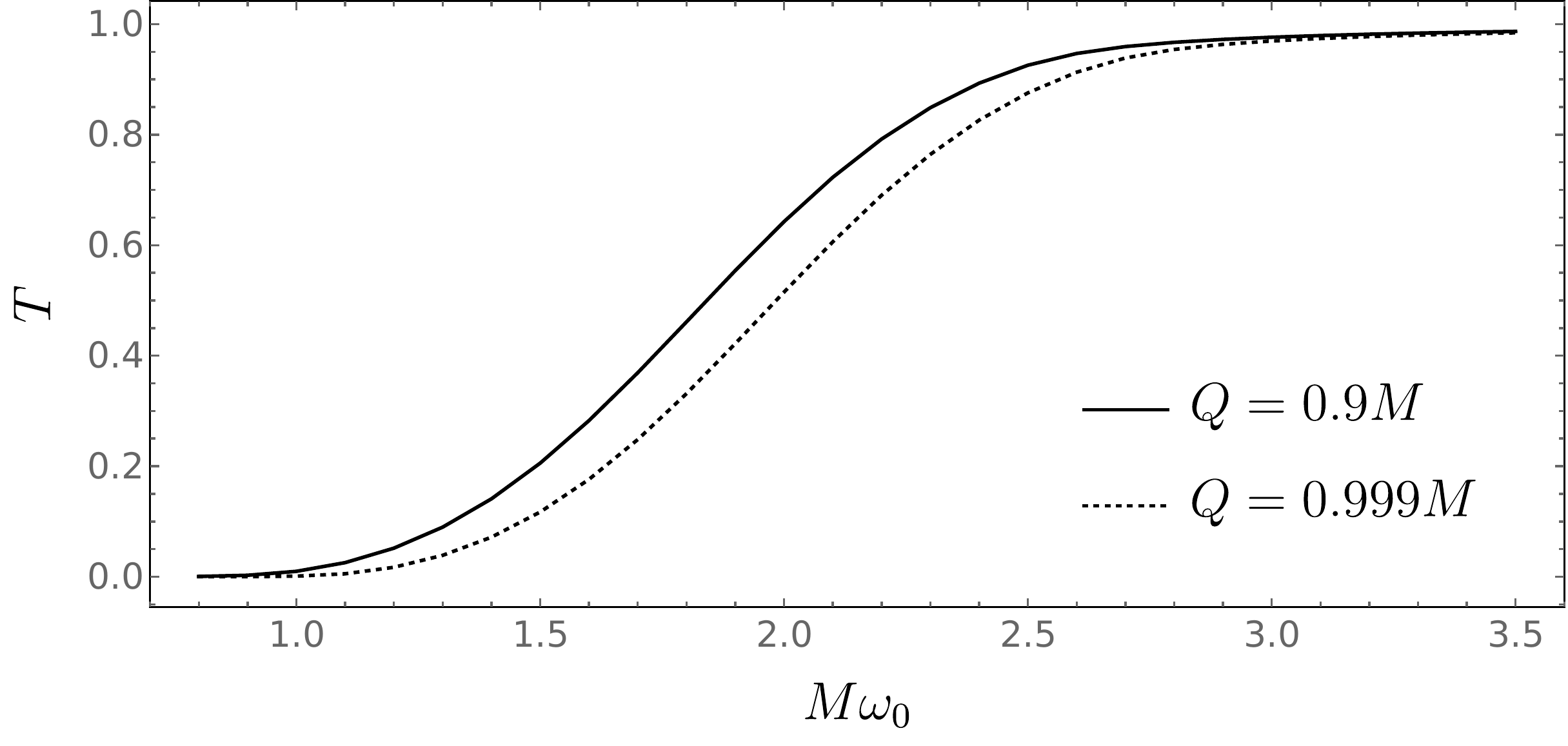}
	\caption[Wave packet Transmission rate for fixed mass $\mu$ of the packet]{Transmission rates for the wave packet with mass $M\mu=0.5$ and central frequency $\omega_0$, with different values for the black hole charge and the packet's orbital parameter as $\ell=5$.}\label{fig:TpacketRN}
\end{figure}

\begin{figure}[!htbp]
	\centering
	\includegraphics[width=0.95\textwidth]{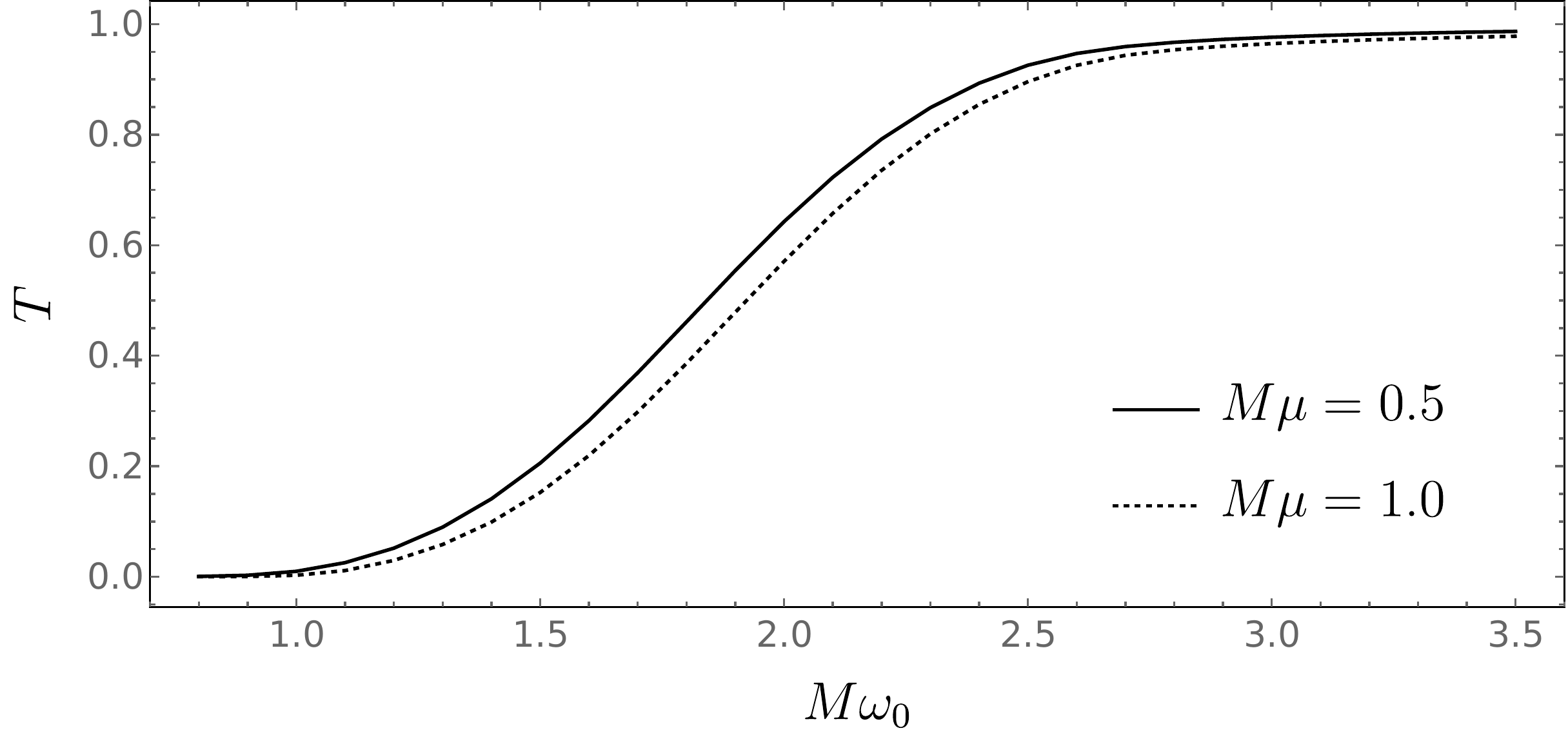}
	\caption[Wave packet Transmission rate for fixed charge $Q$ of the black hole]{Transmission rates for the wave packet with central frequency $\omega_0$. the black hole charge is fixed with $Q=0.9M$ and different values of the mass of the packet and the packet's orbital parameter as $\ell=5$.}\label{fig:TpacketRN2}
\end{figure}

It is also interesting to define some not-so-new quantities, as the packet's partial and total absorption cross section as
\begin{equation}
    \sigma^{(wp)}_\ell(\omega_0) := \int_0^\infty \tilde{\psi}(\omega)\sigma_\ell(\omega)\,\diff\omega, \quad
    \sigma^{(wp)}_{abs}(\omega_0) := \sum_{\ell=0}^\infty \sigma^{(wp)}_\ell(\omega_0)
\end{equation}
with $\tilde{\psi}(\omega)$ as defined in equation~\eqref{eq:ftpacket} and $\sigma_\ell(\omega)$ defined in equation~\eqref{eq:partial_and_total_cs}.

\begin{figure}[!htbp]
	\centering
	\includegraphics[width=1\textwidth]{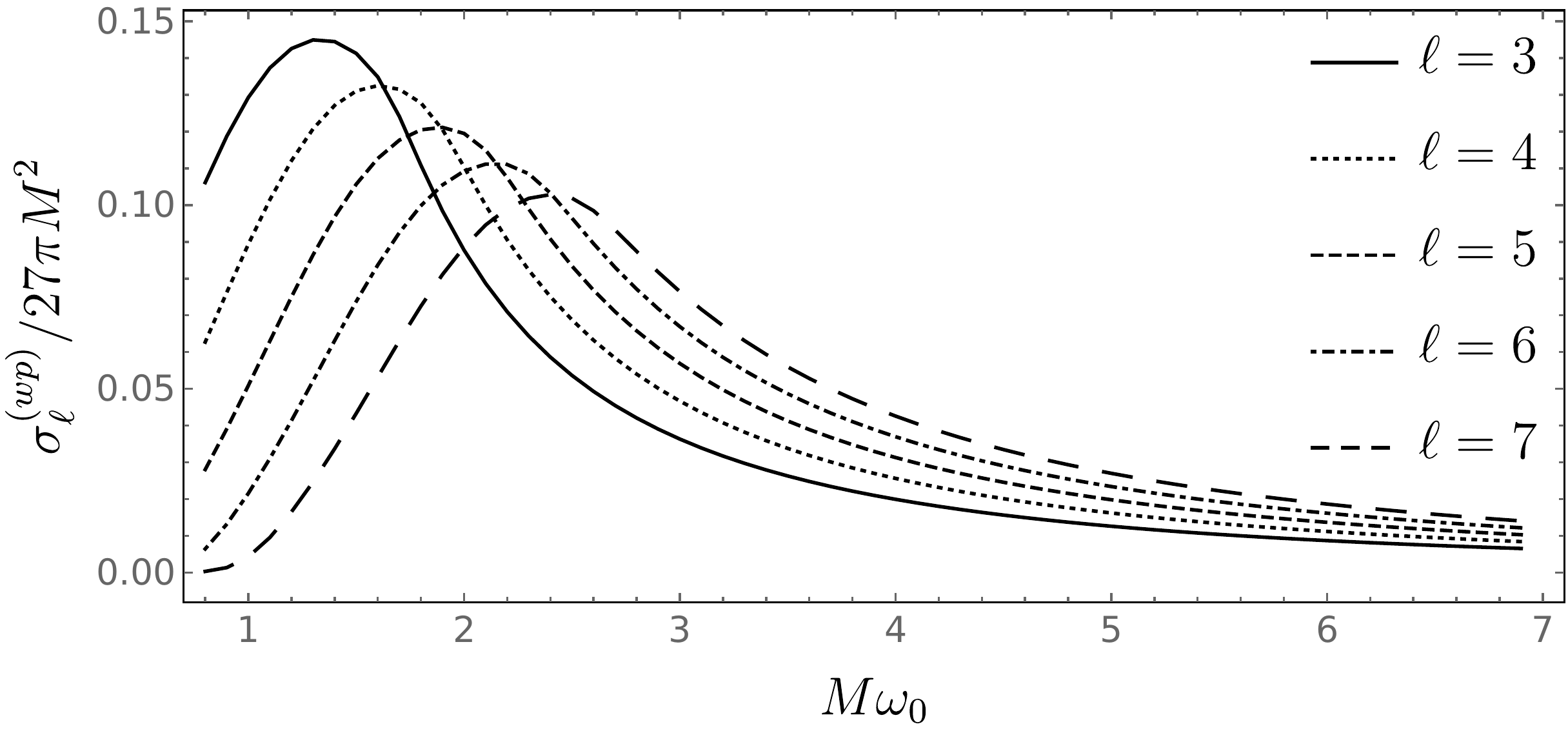}
	\caption[Wave Packet Partial Absorption cross section]{Partial Absorption cross section for the wave packet with central frequency $\omega_0$ and mass $M\mu=0.5$ fixed. The black hole charge set as $Q=0.9M$. The partial absorption cross section is normalized by its value in the optical limit.}\label{fig:packetpcs}
\end{figure}

So far we have been dealing with the problem of an incoming packet of mass $\mu$ from $r^\star\to+\infty$ impinging on the black hole and being absorbed or scattered. If the packet gets absorbed the black hole will accrete its mass and ends up with new mass $M'=M+m_w$, imposing no danger to the CCC, since $M' > Q$. Of course, we could also think of a packet with the same mass $\mu$ but also with charge $q$, then the new quantities of the black hole would be $M'=M+m_w$ and $Q'=Q+q$, and this new quantities must obey $M' > Q'$ or else the packet would just be scattered if the CCC is to be upheld. But, in the works by Matsas~\cite{PhysRevLett.99.181301,PhysRevD.79.101502} it was shown that a charged scalar field could be absorbed by the black hole via quantum tunneling, and the new quantities were in such a way that $M' < Q'$ and the CCC is violated.

The problem in dealing with a charged packet is the appearance of superradiance (when the reflection rate could be greater than unity, meaning that energy was extracted from the black hole) and the complication on the effective potential that the field's frequency $\omega$ would couple with the charge~\cite{PhysRevD.93.024028} (and the effective potential is no longer independent of the frequency of the field) implying that the method we have created would have to suffer major changes, if not rewritten. Besides, the particle's trajectory would also depend on its charge, and not only by the metric, which is crucial here.

We will go on and exploit the symmetry of the problem discussed so far. If we mirror the incoming packet to be coming from $r^\star\to-\infty$, then the absorbed packet becomes the emitted packet. This means that the emitted packet has a mass $\mu$ that will make the black hole to have new mass $M'=M-m_w$. But that emitted packet, according to the CCC, cannot be any --- classically, it must keep the inequality $M' > Q$. Thus, we will focus on a packet that will be emitted by quantum tunneling process, i.e., it will pass through the potential barrier and be emitted, trying to violate the CCC via $M'<Q$.

\section{Analysis of the Transmission rates of the packets}
As said in the end of last section, we will exploit the symmetry of the problem and perform an inversion of the packets, where we had the incoming and reflected packets at $r^\star\to+\infty$, and a transmitted (absorbed) packet at $r^\star\to-\infty$, now it becomes the outcoming and reflected packets at $r^\star\to-\infty$, and a transmitted (now, \emph{emitted}) packet at $r^\star\to+\infty$, as we can see in figure~\ref{fig:emission}. This simple trick changes completely the view of the problem, but none of its consequences, i.e., the procedure is exactly the same we have done so far, and we would get the exact same results. This is only possible due to the symmetry of the reflection and transmission rates which makes no distinction if the plane waves come from $r^\star\to+\infty$ or from $r^\star\to-\infty$. A full discussion on the subject may be seen in appendix~\ref{apendiceb}.

\begin{figure}[h]
	\centering
    \includegraphics[width=0.95\linewidth]{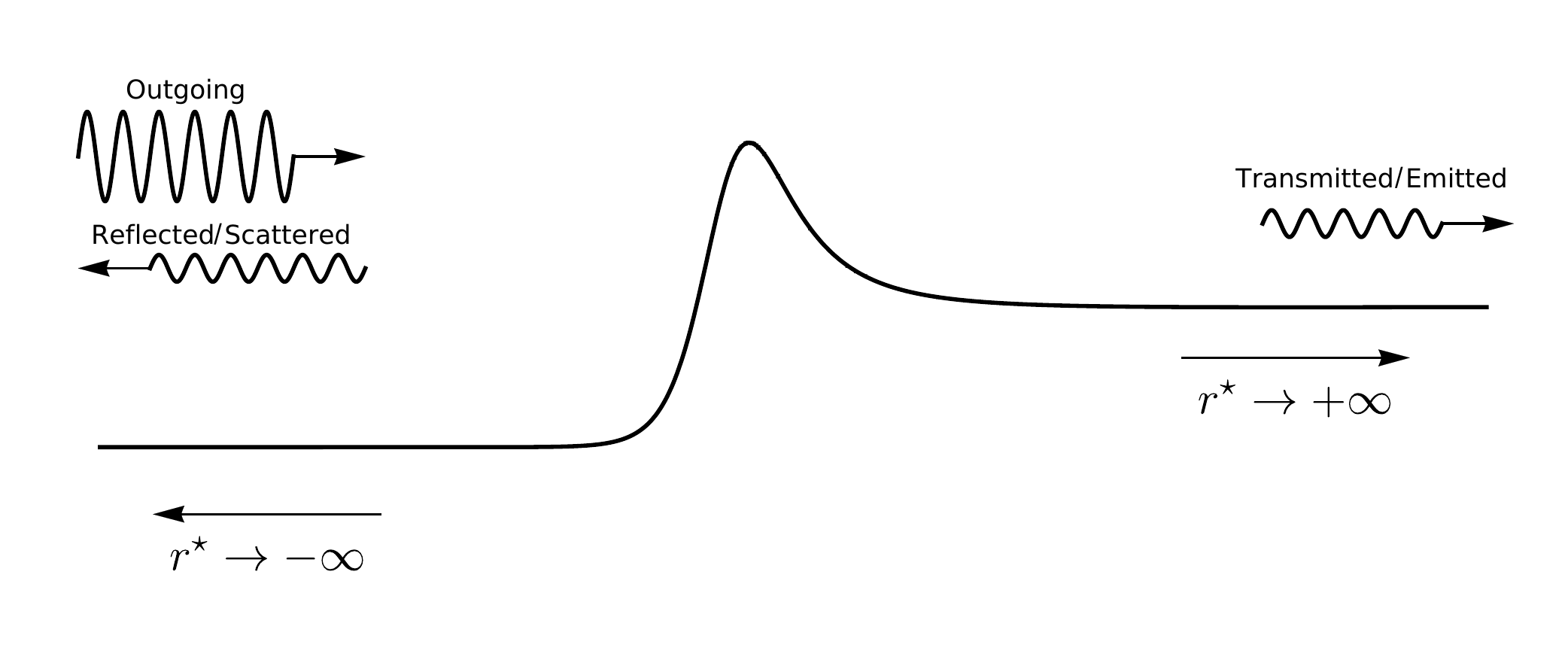}
    \caption[Schematic representation of the scattering/emission of a plane wave]{Schematic representation of an outgoing plane wave being scattered/emitted through the black hole potential barrier. The wave packet is built from those outgoing plane waves.}\label{fig:emission}
\end{figure}

Before we jump in to the next question, let us first get familiar with the quantities we are dealing here. So far we have treated every quantity as being normalized to the black hole's mass $M$. That comes in hand since if we want to do a physical analysis all we need is proceed to a rescale of the parameters. With that in mind, let us discuss about unit conversion for us to get the gist of the numbers. 

The conversion factor for mass from geometrized unit system to S.I. is
\begin{equation}
    M_{SI} = \frac{c^2M_{GSU}}{G}
\end{equation}
where GSU stands for geometrized system unit. For a black hole of $M_{GSU} = 1\,\mbox{m}$ that we have been dealing so far, we have $M_{SI} \approx 10^{27}\,\mbox{kg} \approx 10^{-4}M_\odot$. On the other hand, for the product $M\mu=1$, where $\mu$ we have been calling the particle's mass is, actually, according to the Klein-Gordon equation,
\begin{equation}
    \mu := \frac{m_wc}{\hbar}
\end{equation}
the inverse of the Compton wavelength of a particle with mass $m_w$. Thus, if $M_{GSU}=1\,\mbox{m}$ we have
\begin{equation}
    \mu = 1\,\mbox{m}^{-1} \implies m_w = \frac{\mu\hbar}{c} \approx 10^{-43}\,\mbox{kg} \approx 10^{-7}\,\mbox{eV}/c^2
\end{equation}
is the actual mass in eV$/c^2$ of the given particle. But since choosing $\mu$ is the same as choosing $m_w$, we will keep the idea of $\mu$ being the mass of the packet. Now, if we would want to deal with a more realistic black hole, say $M=10M_\odot$, keeping $M\mu = 1$, we would have a particle with mass $m_w \approx 10^{-11}\,\mbox{eV}/c^2$. Nevertheless, all calculations and graphs would be left unchanged since every quantity is normalized to the black hole's mass. This also shows that the particles we are dealing here are extremely light, axion-like particles~\cite{Ringwald:2014vqa}.

Now, the question that lies at hand is, what is the probability of a black hole of mass $M$ and charge $Q$ to emit a particle of mass $m_w$ in such a way that $M-m_w < Q$, that is, violate the CCC? This analysis will rely on the transmission rate for the packet since its interpretation is exactly what we need to take a glance at this question.

Let us consider two black holes. The first one the mass will be $\mathcal{M}_{1}$ and with charge $\mathcal{Q}$, while the second one the mass will be $\mathcal{M}_2$ and same charge $\mathcal{Q}$. We impose the difference of the masses of both black holes to be as small as possible, that is, $\mathcal{M}_1 - \mathcal{M}_2 = \delta M \ll M$, where $M$ is a base value for the mass. If we choose $\mathcal{M}_1 := M + \delta M$, then we will have $\mathcal{M}_2 := M$. Next, we will impose that the difference\footnote{We have also kept the same mass-charge ratio and the results are qualitatively the same.} between the mass $M$ and charge $\mathcal{Q}$ of both black holes to also be small as possible with $M - \mathcal{Q} = \delta \ll M$, which gives us $\mathcal{Q} := M-\delta$. We introduced the small variations $\delta M$ and $\delta$ so that the CCC inequality 
\begin{equation}
    \mathcal{M} > \mathcal{Q} \implies M + \delta M > M - \delta \implies \delta M + \delta > 0
\end{equation}
meaning that the difference between $\mathcal{M}$ and $\mathcal{Q}$ is as small as possible. For a charged and non-rotating black hole with $\mathcal{M} \gtrsim \mathcal{Q}$ is known as a \emph{near-extreme} Reissner-Nordstr\"om black hole. Thus, if the emitted packet is such that
\begin{equation}
    \mathcal{M}' > \mathcal{Q} \implies \mathcal{M} - m_w > \mathcal{Q} \implies \delta M + \delta > m_w
\end{equation}
which explicitly shows us that if the CCC is to be kept, then the sum of the small variations $\delta M$ and $\delta$ must be greater than the packet's mass. The idea is given in figure~\ref{fig:twolevel}.

\begin{figure}[h]
	\centering
    \includegraphics[width=0.48\linewidth]{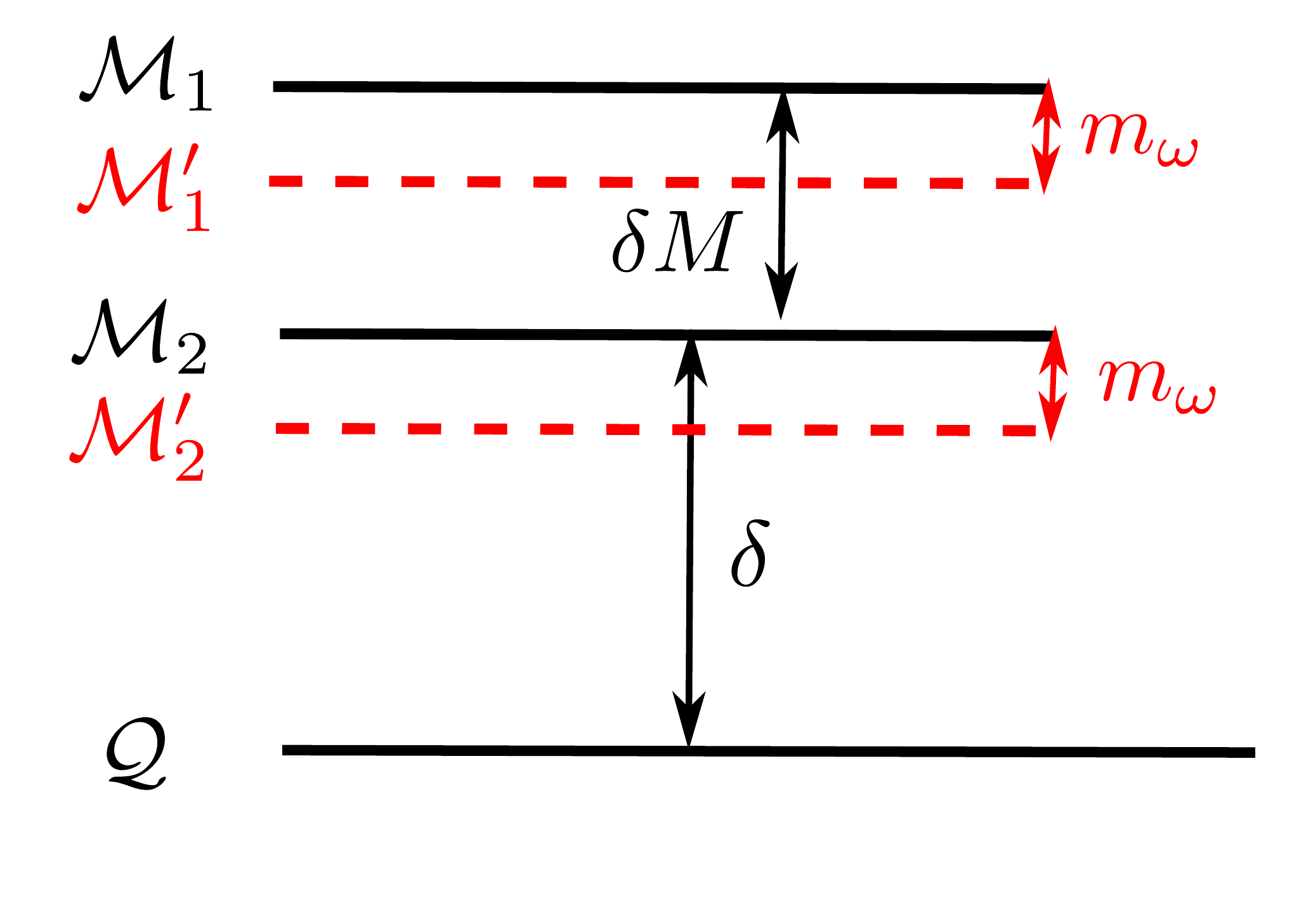}
    \hspace{\fill}
    \includegraphics[width=0.48\linewidth]{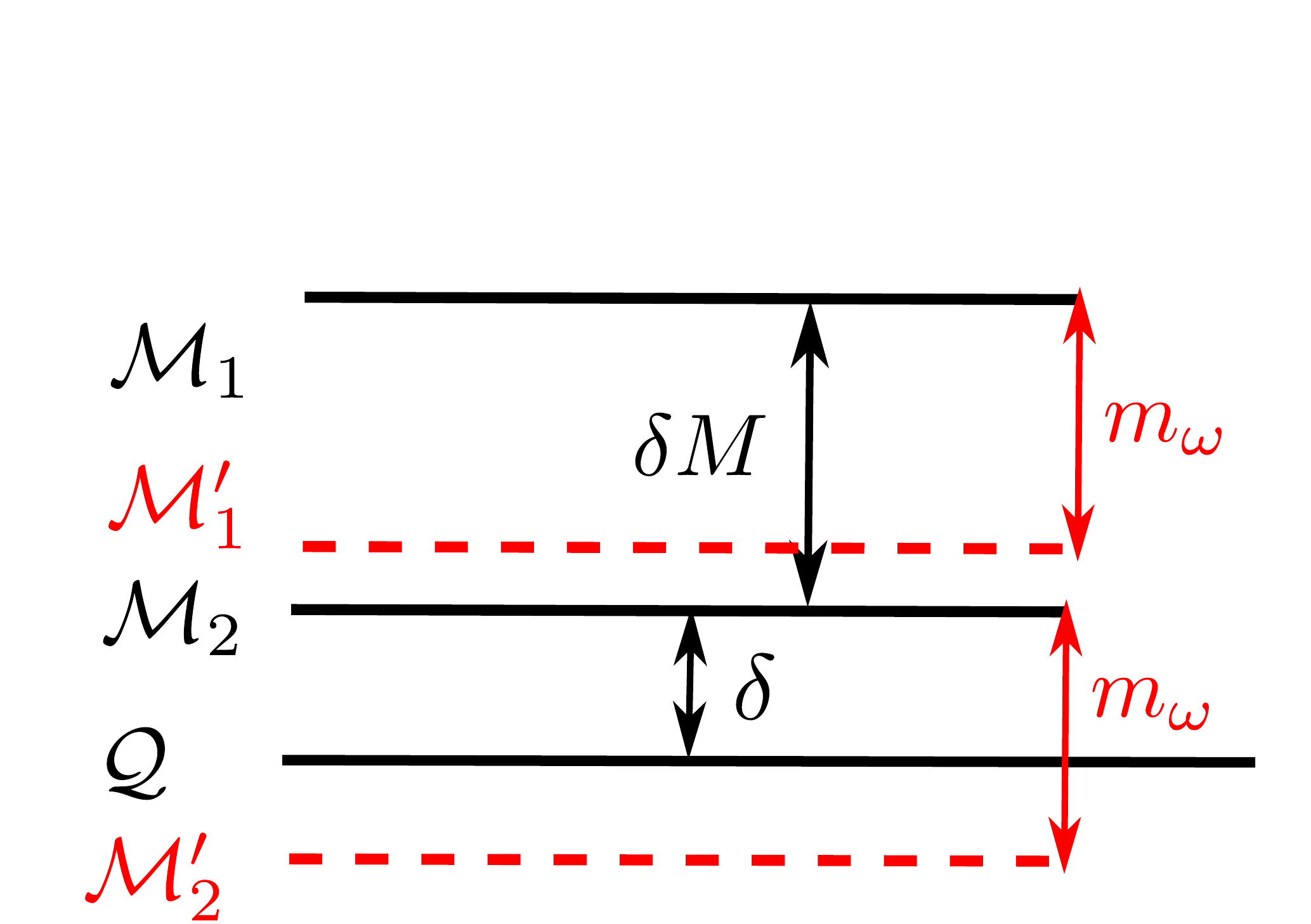}
    
    \vspace{0.5cm}
    
    \caption[``Two-level system'' schematic representation of the CCC]{A ``two-level system'' schematic representation of the CCC for the cases given above. In the left panel, the emission of a mass $m_w$ imposes no danger to the CCC once $\delta > m_w$. On the right panel there may be a violation of the CCC for the second black hole, since $\delta < m_w$.}\label{fig:twolevel}
\end{figure}

Due to the mapping from $r$ to $r^\star$, the transmitted packet is identified as the emitted packet (shown in figure~\ref{fig:emission}). This means that the probability of a wave packet to be emitted by the black hole is given by the transmission rate $T$ defined in equation~\eqref{eq:RTwavepacket}. In this way, the first black hole has a probability $T(\mathcal{M}_1)$ to emit a wave packet of mass $m_w$. In the same way, the second black hole has a probability $T(\mathcal{M}_2)$ to also emit a wave packet of mass $m_w$. Notice that $\delta$ is the parameter that tells how far the black hole is from extremicity. We now would like to know that, if we fix $\delta M$, by choosing  (a really) small $\delta$, what happens to the transmission rates? Suppose that $T(\mathcal{M}_1) > T(\mathcal{M}_2)$ for any given initial configuration with $\delta=\delta_0$ and $\delta M$ fixed. If in the next configuration we choose $\delta=\delta_1 < \delta_0$ with the same $\delta M$ as before, does the relation between this two transmission rates stay the same as it was for $\delta=\delta_0$, or does it change? If it changes, what is its behavior? And how this reflects on the CCC?

For that matter, we define the ratio between the transmission rates $T(\mathcal{M}_1)$ and $T(\mathcal{M}_2)$ as
\begin{equation}\label{eq:TRatio}
    f_\epsilon := \log\Bigg[\frac{T(\mathcal{M}_1)+\epsilon}{T(\mathcal{M}_2)+\epsilon}\Bigg].
\end{equation}
The cutoff factor $\epsilon$ is necessary to solve the convergence problem  when $T(\mathcal{M}_2)$ goes to zero faster than $T(\mathcal{M}_1)$ and the ratio diverges. Note that $\epsilon$ must be chosen as small as possible so it does not interfere with the values of $T$. Since $T$ ranges from $0$ to $1$, then $\epsilon \leqslant 0.1$ is a reasonable choice for $\epsilon$. It is important to point out here that $f_\epsilon$ does not depend on $\epsilon$ in the given range.

Let us now use the parameter $f_\epsilon$ to analyze a possible violation in the CCC. If $f_\epsilon > 0$, it means that the former configuration $\{M+ \delta M,{\cal Q} \} $ is more likely to emit a particle when compared to the latter $\{M,{\cal Q}\} $. On the other hand, of course, $f_\epsilon < 0$ means the opposite: the latter configuration is more likely to emit a particle when compared to the former. If $f_\epsilon = 0$, the emission probabilities are the same regardless of $\delta M$. Now, given the order of the mass of the packet, we must chose $\delta = 10^{-70}M$ (which is the order of the mass of the packet once converted to GSU) to obtain a violation of the CCC (as seen on the right panel in figure~\ref{fig:twolevel}), but that will lead to use floating point precision for machine number and we will lose information, as shown in figure~\ref{fig:TRatio}.

\begin{figure}[h]
    \centering
    \includegraphics[width=0.95\textwidth]{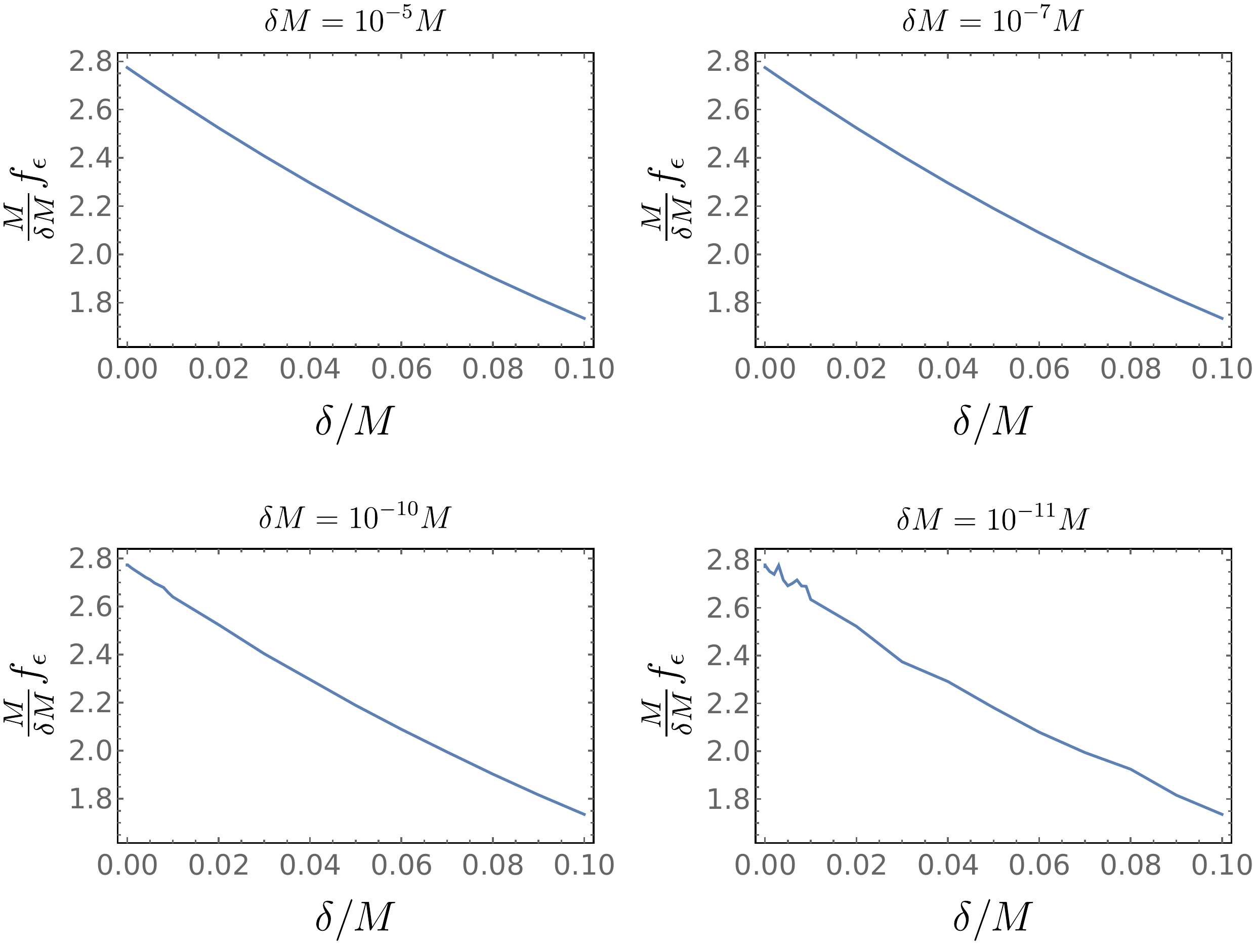}
    \caption[Graphs of $f_\epsilon$ vs. $\delta$]{Behavior of $f_\epsilon$ with respect to $\delta$. Observe that for small $\delta M$ the curve is no longer smooth. Also, notice that for even smaller $\delta$ the graph becomes noisy, meaning that we have lost precision due to floating point precision for machine numbers. In this set of plots, we have $M\omega_0=1.0$, $\ell=5$ and $M\mu = 0.8$.}\label{fig:TRatio}
\end{figure}

Once $\delta$ is chosen to be very small ($\sim10^{-70}M$) it is clear that the highly oscillatory behavior near $\delta=0$ is compatible with a noisy signal, meaning that we have lost precision. Nonetheless, all graphs in figure~\ref{fig:TRatio} show that $f_\epsilon > 0$ in the small$-\delta$ regime, which means that the emission probability of the heavier black hole $M+\delta M$ is greater than the one of the lighter black hole $M$. It is interesting to notice that the numeric value of $f_\epsilon$ does not increase indefinitely as $\delta\to0^+$, but rather approach a fixed value.

Instead of using such small number, we will see if the ratio $f_\epsilon$ shows some kind of tendency along some values of $\delta$ where our calculations ought to be correct, using only machine precision for numbers. We have fitted a model for $f_\epsilon$ in the range $\delta\in[10^{-3},10^{-1}]$ using the linear model,

\begin{equation}\label{eq:dfit}
    f_\epsilon \approx f(\delta) := \frac{\delta M}{M}\Bigg(a\frac{\delta}{M}+b\Bigg)
\end{equation}
finding $a=-10.6\pm0.1$ and $b=2.75\pm0.01$. A plot of $f(\delta)$ is show in figure~\ref{fig:TRatiofit}. 

Thus, in this $\delta-$range, taking the exponential of equation~\eqref{eq:TRatio},
\begin{equation}\label{eq:Tfit}
    \frac{T(M+\delta M)}{T(M)} = e^{f(\delta)} \implies T(M) = T(M+\delta M)e^{-f(\delta)}
\end{equation}
where we write the transmission rate of the lighter black hole as a function of the transmission rate of the heavier black hole by an exponential factor, up to first order. Even though we have $a<0$, in the $\delta-$range we are considering the term $a\delta$ drops once compared to $b$, meaning that, if we were to extrapolate this results and assume they are correct as $\delta$ approaches zero, then the ratio $f_\epsilon$ approaches a positive constant, resulting in a fixed relation that $T(M) = \alpha T(M+\delta M)$, with $0<\alpha<1$ once $f_\epsilon>0$.

\begin{figure}[htbp]
    \centering
    \includegraphics[width=0.95\textwidth]{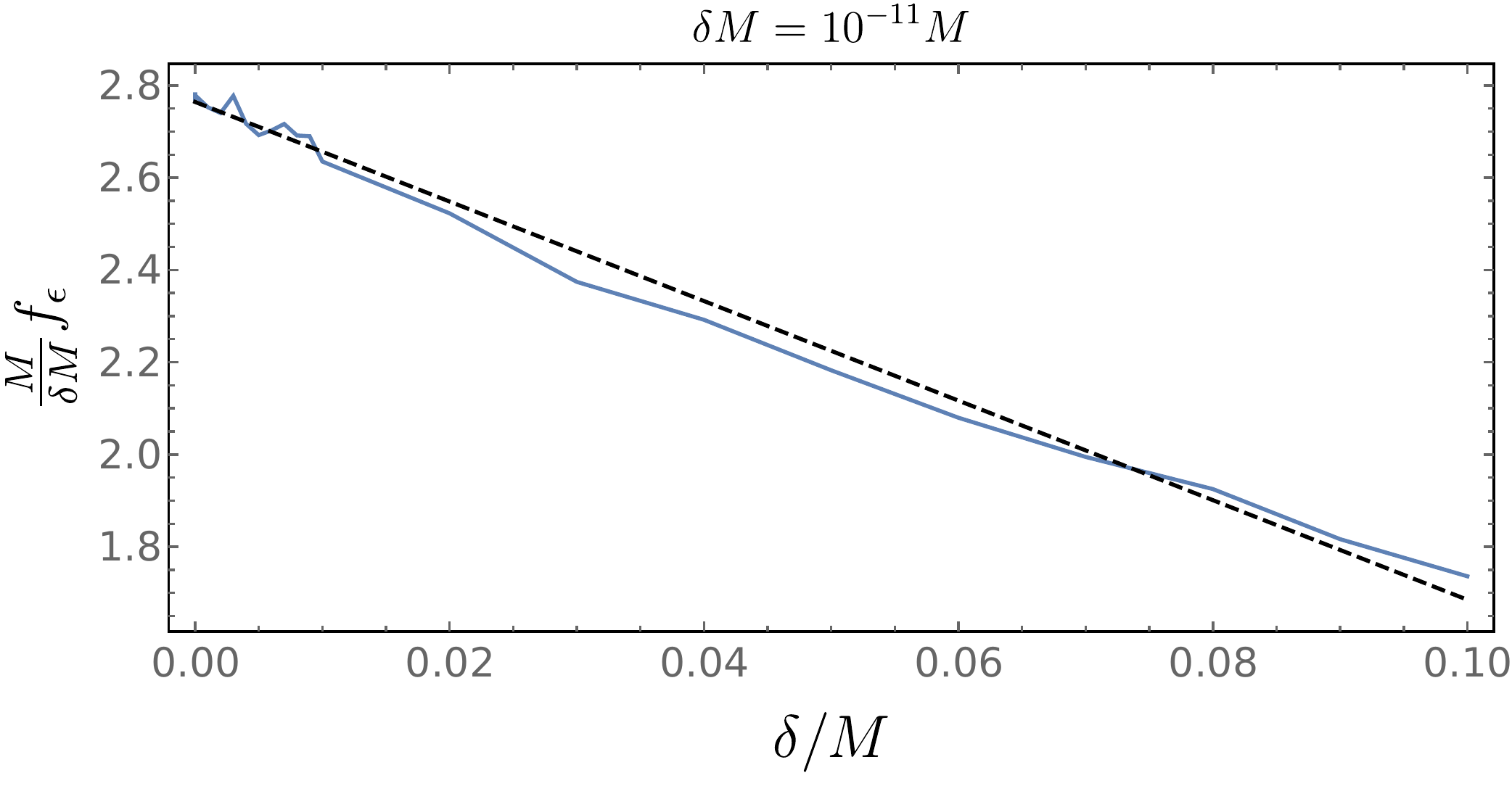}
    \caption[Fit of $f_\epsilon$]{Fit of $f_\epsilon$ (dashed line) obtained in the rage $\delta\in[10^{-3},10^{-1}]$ and extrapolated to be shown along the plot of $f_\epsilon$ with $\delta M = 10^{-11}M$ (solid line).}\label{fig:TRatiofit}
\end{figure}

This results leads to two interesting conclusions: 
\begin{enumerate}
    \item[\textbf{1)}] For two different configurations of a black hole, if the chosen initial conditions are slightly different from each other, the difference between them as they approach extremicity is not zero, but rather a positive constant --- and they are directly proportional.
    \item[\textbf{2)}] Given the results in equations~\eqref{eq:dfit} and~\eqref{eq:Tfit}, once $T(M+\delta M) > T(M)$ in the small$-\delta$ regime, it is always possible to chose $\delta \lesssim m_w$ such that the BH may become overcharged. In other words, there is a violation of the CCC.
\end{enumerate}

Even though there is a tiny, although finite, probability of tunneling and, therefore, overcharging a BH --- or so it seems --- that is not the only necessary ingredient to actually violate the CCC: one should also consider two essential factors. First, the typical discharging time, i.e, the time interval before a BH loses its charge (since it will attract opposite-charged particles from an orbiting cloud). Second, the travel time from the horizon to a typical radius (that may correspond to, say, the peak of the effective potential, eq.~\eqref{eq:VeffRN}, or to the radius circular orbit around the BH), which will provide --- very much like the old-fashioned description of a nuclear alpha decay~\cite{rohlf1994modern} --- the order of magnitude of the time interval between two successive ``attempts'' of such packet to escape.

We can estimate the former considering a massive BH such that the distance $\Delta r$ from $r=2M$ to $r=6M$ (or, as previously mentioned, any typical radius) is  classical, i.e, $\Delta r\gtrsim 10^{-10}\,\mbox{m}$. That requires $M\gtrsim 10^{-13}M_{\odot} \approx 10^{17}\,\mbox{kg}$ and, for a quasi-extremal case, $Q\gtrsim 10^7\,\mbox{C}$ --- which corresponds to about $10^{26}$ electrons or $10^{-5}\,\mbox{kg}$, much less than the BH mass. The proper time\footnote{The time-dilation factor that would yield the corresponding time for an observer at infinity is the same for both ingoing and outgoing particles, so it is not necessary to take it into account when comparing both.} of an outgoing packet (assuming it will be at rest at radial infinity) is about $10^{10}$ shorter than the ingoing time for the charged particle to cross the same distance (due to the attractive EM force). Therefore, a charged quasi-extremal BH would not remain so for the tunneling to take place.

We also recall that the ingoing charge does not need to tunnel, as opposed to the outgoing particle. Therefore, even if the discharging process takes, for some reason, the same amount of time of the outgoing flight, the tunneling probability of the outgoing packet --- which is $\sim 0.03$ for such massive BH  --- will decrease even further the rate of successful attempts.



\end{chapter}

\begin{chapter}{Conclusions}
\label{conclusao}

\hspace{5 mm} Throughout the literature, much has been done around waves being absorbed by a black hole and how it can turn a near-extremal black hole into a naked singularity, as seen in references~\cite{Glampedakis:2001cx,PhysRevD.89.104053,PhysRevD.52.1808,Andersson:2000tf,PhysRevD.18.1030,PhysRevD.91.124030,doi:10.1142/S0218271818430125}, but very little was done regarding a wave packet representing a particle being absorbed by the black hole, alas being emitted by one. In particular, Vishveshwara~\cite{vishveshwara1970scattering} studied the gravitational radiation of a Schwarzschild black hole carried out by a wave packet, and became the main reference on this topic.

Even though this is a case study in the sense that we are modeling a neutral and spin$-0$ particle, the method itself is really promising because all quantities brought and calculated here are analytical. The main idea of this work was to propose a feasible and reliable toy model to study the emission of particles by a black hole without having to restrict ourselves to narrow bands of frequency values or specific numerical cases. We have not only recovered all the results from other researches, but we also have shown that this method allows a broader appliance, with all results being analytical and allowing broader (but not restricted) frequency bands.

The asymmetry of the toy model is a key element to approach the problem discussed here once the actual effective potential is asymmetric itself. After we have constructed the toy model with two analytical functions for which the problem is analytically solvable, we were limited only to the choice of the set of parameters $\{M,Q,\ell,\mu\}$ and their physical limits. We could deal with a realistic case of a charged massive packet, but that would imply that the frequency of the field would couple to its own charge in the effective potential, which would no longer be independent of the field's frequency, making it troublesome to build the wave packets. Also, the particle's trajectory would depend on its charge, and not only by the metric.

Dealing with a localized wave packet may be a way of solving the problem of backreaction, where the initial conditions of the problem changes and the feedback for the waves are not as the ones we began with, which leads to numerous critics about given solutions to the violation of the CCC, where the effect of backreaction must be taken into account in order to perceive the problem in its entirety. The plane wave approach corresponds to an eternal and constant flux of particles --- that is in direct opposition to the qualitative change in the metric (the violation of the CCC) which is being searched for. The semiclassical approximation of a particle --- namely, a wave packet --- can tell the probability of absorption each time it is thrown at a black hole, since it is localized in the position space. We also recall that we have worked in the safe side of the metric, i.e., in the presence of an EH and outside it. We have shown that, within the numerical precisions achieved, the CCC holds even when we start from configurations increasingly closer to an extreme one.

We studied the method applied to the Schwarzschild black hole thoroughly its possibilities, including spin contributions. Being the simplest description for a black hole, it allowed us to exploit the toy model extensively while checking its consistency with a known, well tested algorithm: the Pr\"ufer method. This work serves as a compilation of most researches done in the field and presented into a single method, recovering the results done so far.

It is important to point out that we did not consider the superradiance effect where $R>1$. Once the superradiance conditions are $qQ>0$ and $\omega < |qQ|/r_+$, as we can see from reference~\cite{PhysRevD.93.024028}, with $q$ being the particle's charge, in this work we have $q=0$, thus the conditions become null and we have no superradiance.

Equation~\eqref{eq:TRatio} is the main driver to our conclusions regarding CCC violation. As it is said in the previous chapter, $f_\epsilon$ tells us the ratio of probabilities of a particle to be emitted between two given initial configurations. We have shown in figure~\ref{fig:TRatio} that $f_\epsilon > 0$ in the small$-\delta$ regime, meaning that the heavier black hole configuration has a greater probability rate of emitting a particle of mass $m_w$ once compared to the lighter one. The main issue within this analysis is that once $\delta\to0^+$, machine precision is lost and the calculations begin to rely on floating point precision, leading to a noisy signal at this end. For that purpose, we have fitted a linear model for $f_\epsilon$ where we rely only on machine number precision and extrapolated this result for really small $\delta$. Our conclusion was that $f_\epsilon$ approaches a positive constant once $\delta\to0^+$, showing that the ratio of probabilities are directly proportional to each other. This shows that it is possible to achieve a CCC violation due to the fact that we can choose $\delta \lesssim m_w$ for any given $\delta M$ to satisfy the condition on the right panel in figure~\ref{fig:twolevel}.

Even though there is a tiny --- but non-null --- chance for the particle to be emitted and violate the CCC by exposing the singularity, we have calculated the time required for such emission to happen and compared it to the typical time for the BH to capture an external charge and increase the $\delta$ difference. We have found out that the former is greater than the latter. This means that, even though a particle may be emitted and violate the CCC, the BH will capture a charged particle faster than the emission process occurs, which increases the $\delta$ difference and leaves the BH further from extremicity for the time when the emission takes place.

For a Kerr black hole, whose metric is given by equation~\eqref{eq:mKerrNewman} with $Q=0$, it is impossible for us to use the toy model proposed here to describe the given black hole. This is so because the junction set for both functions of the toy model is a null set for this metric. Thus, to describe a Kerr black hole we must choose another approach or a different toy model.

In future works, the consideration of a charged particle may be carried out by starting from where we left here on chapter~\ref{cap4}. Once the charge is considered, the effective potential for the Schr\"odinger-like equation~\eqref{eq:Veff} is~\cite{PhysRevD.93.024028}
\begin{equation}
    V_{eff}(r) = \Bigg(1 - \frac{r_s}{r} + \frac{r_q^2}{r^2}\Bigg)
			\Bigg(\frac{\ell(\ell+1)}{r^2} + \frac{r_s}{r^3} - \frac{2r_q^2}{r^4} + \mu^2\Bigg)
				- \Bigg(\frac{qQ}{r}\Bigg)^2 + \frac{2\omega qQ}{r}
\end{equation}
where the effective potential will now depend explicitly on the field's frequency $\omega$, which will result that the toy model must also depend on $\omega$. This dependency turns the matching (the determination of the junction point $r_0^\star$) of the toy model into a cumbersome task, where it must be done very carefully.

The inclusion of particle spin is a very interesting idea, but its implications are even more complicated. What we need to know is, if a particle with a non-null spin gets absorbed (or emitted) by a static, charged black hole, does it change considerably its angular momentum? If it does, then the metric that started out as static will become a rotating one, then the Reissner-Nordstr\"om black hole will turn into a Kerr-Newman one. Also, how do we describe the toy model in this case, since in the Kerr configuration we discovered that the present toy model cannot describe it? This consideration carries a lot more since the black hole configuration changes during the process and we need to consider another approach for the rotating singularity, but it is a very interesting idea that will be taken into account into future works.

Overall, our method shows promising results regarding particle emission from a black hole via quantum tunneling process. The toy model ended up as an extremely useful tool to avoid frequency approximations and numerical analysis as the only means to obtain the desired answers for a large range of the parameters of the system. The simplicity of the method allows the use of an everyday computer with no particular configuration to calculate complex quantities in matters of minutes with good approximation and achieve frequency bands as broader as the machine can reach. Using wave packets to represent particles is a novelty in the sense that, up to today, no work has applied this approach in the violation of the CCC. Besides, only a few papers have applied toy models to black holes, but in the sense of its properties and stability in the near-extremal case~\cite{Maldacena:2017axo,Maldacena:2016upp,Maldacena:2016hyu}.

\end{chapter}



\newpage


\phantomsection



\appendix
\begin{chapter}{Pr\"ufer Method}
\label{apendicea}

\hspace{5 mm} From the discussion in reference~\cite{Glampedakis:2001cx}, the Pr\"ufer method may be applied to any ordinary differential equation (ODE) of the form
\begin{equation}\label{eq:edoprufer}
    \dtot{}{x}\Bigg(P(x)\dtot{u}{x}\Bigg) + Q(x)u=0,
\end{equation}
defined in the interval $x\in(a,b)$ where $P'(x)>0$, and both $P'(x)$ and $Q(x)$ are continuous. We will omit the dependency on the independent variable for the next calculations as to keep the notation slim. 

Let us apply the Pr\"ufer substitution,
\begin{equation}
    Pu' = r\cos\theta, \quad
    u = r\sin\theta
\end{equation}
to equation~\eqref{eq:edoprufer}, where $r:=r(x)$, $\theta:=\theta(x)$ and the primes denotes the derivative with respect to the independent variable. Note that
\begin{equation}\label{eq:ptsquared}
    (Pu')^2 = r^2\cos^2\theta, \quad
    u^2 = r^2\sin^2\theta
\end{equation}
such that
\begin{equation}\label{eq:prufer}
    (Pu')^2 + u^2 = r^2, \quad \tan\theta = \frac{u}{Pu'}.
\end{equation}

Let us suppose that $u(x)$ is real. This function may be represented in a Ponicar\'e plane as a parametrized curve by the independent variable $x$. It is interesting to note that the transformation
\begin{equation}
    (Pu',u) \leftrightarrow (r,\theta)
\end{equation}
is non-singular for all $r\neq 0$. Besides, $r>0$ for all non-trivial solutions. In fact, if for any given $x$, $r(x)=0$, we have from equation~\eqref{eq:ptsquared} that $u(x)=0$ and $u'(x)=0$. Thus, by the existence and uniqueness theorem for a second order differential equation that $u(x)=0$ for any $x$.

We then obtain a system of first order ODE equivalent to equation~\eqref{eq:edoprufer}. Differentiating the reciprocal of the second equation in~\eqref{eq:prufer}
\begin{equation}
    \cot\theta = \frac{Pu'}{u}
\end{equation}
we have
\begin{equation}
    -\csc^2\theta\dtot{\theta}{x} = \frac{(Pu')'}{u} - \frac{Pu'^2}{u^2} = -Q - \frac{1}{P}\cot^2\theta
\end{equation}
where we defined
\begin{equation}
    Q:= -\frac{(Pu')'}{u}.
\end{equation}
Thus,
\begin{equation}\label{eq:p1}
    \dtot{\theta}{x} = Q\sin^2\theta + \frac{1}{P}\cos^2\theta
\end{equation}
is the \emph{Pr\"ufer's differential equation for the phase}.

To find the Pr\"ufer's differential equation  for the amplitude, we differentiate the first equation in~\eqref{eq:prufer}
\begin{equation}
    r^2 = u^2 + (Pu')^2
\end{equation}
to obtain
\begin{align}
    r\dtot{r}{x} &= uu' + (Pu')(Pu')' 
    =\frac{u}{P}(Pu') - (Pu')Qu \nonumber\\[.5em]
    &= \frac{1}{P}r^2\sin\theta\cos\theta-Q r^2\cos\theta\sin\theta
\end{align}
or
\begin{equation}\label{eq:p2}
    \dtot{r}{x} = \Bigg(\frac{1}{P}-Q\Bigg)\frac{r\sin 2\theta}{2}.
\end{equation}

We then proceed on solving the system of equations~\eqref{eq:p1} and~\eqref{eq:p2}, which is equivalent to equation~\eqref{eq:edoprufer}.

From both Pr\"ufer's equations, the one for the phase $\theta(x)$ is the most important, because it defines the qualitative behavior of $u(x)$. The reason that makes the phase equation more attractive is that it is a first order ODE independent of the amplitude $r(x)$ --- that is, $r(x)$ has no influence whatsoever in the phase $\theta(x)$.

For any given initial value $(a,\gamma)$ there is only one solution that satisfies
\begin{equation}
    \dtot{\theta}{x} = F(x,\theta)
\end{equation}
and $\theta(a) = \gamma$ every since $P$ and $Q$ are continuous in $a$.

Once $\theta(x)$ is known, Pr\"ufer's amplitude function is found integrating equation~\eqref{eq:p2},
\begin{equation}
    r(x) = K\exp\Bigg[\int_a^x\Bigg(\frac{1}{P}-Q\Bigg)\frac{\sin 2\theta}{2}\,\diff{x}\Bigg]
\end{equation}
with $K=r(a)$. Therefore, each solution to Pr\"ufer's system depends on the initial amplitude $K = r(a)$ and initial phase $\gamma=\theta(a)$. We notice that a change in the constant $K$ simply multiplies the solution $u(x)$ by a constant factor. Thus, the zeros and the oscillatory behavior of $u(x)$ are specified studying the phase.

To illustrate the method, we will apply it to the Schr\"odinger-like equation~\eqref{eq:Veff} for the Schwarzschild metric~\eqref{eq:mSchwarzschild}, where
\begin{equation}
    \Bigg[\dtot{^2}{r^{\star^2}} + \omega^2 - V_{eff}(r)\Bigg]rR(r) = 0, \quad
    V_{eff}(r) = \Bigg(1 - \frac{r_s}{r}\Bigg)\Bigg(\frac{\ell(\ell+1)}{r^2} + \frac{r_s}{r^3} + \mu^2\Bigg),
\end{equation}
with the tortoise coordinate $r^\star$ defined in equation~\eqref{eq:tortoise}
\begin{equation}\label{eq:tortoiseA}
    \dtot{r}{r^\star} := 1 - \frac{r_s}{r} \implies r^\star(r) = r + r_s\log\Bigg[\frac{r}{r_s}+1\Bigg].
\end{equation}

Let us introduce the dimensionless variable $x^\star$ and the parameter $x_s$ as
\begin{equation}
    x^\star := \omega r^\star, \quad x_s := \omega r_s,
\end{equation}
then the tortoise coordinate is rewritten as
\begin{equation}
    \dtot{^2}{r^{\star^2}} = \omega^2\dtot{^2}{x^{\star^2}}
\end{equation}
and the effective potential is
\begin{equation}
    V_{eff}(r) = \omega^2V_{eff}(x) 
    := \omega^2\Bigg(1 - \frac{x_s}{x}\Bigg)\Bigg(\frac{\ell(\ell+1)}{x^2} + \frac{x_s}{x^3} + \tilde{\mu}^2\Bigg), \quad
    x := \omega r, \quad \tilde{\mu} := \frac{\mu}{\omega}
\end{equation}
so the Schr\"odinger-like equation~\eqref{eq:Veff} as a function of the dimensionless variable $x^\star$ is
\begin{equation}
    \Bigg[\dtot{^2}{x^{\star^2}} + 1 - V_{eff}(x)\Bigg]u(x) = 0, \quad u(x) = xR(x)
\end{equation}
once again, we point out the mixed $x$ and $x^\star$ coordinate notation. Defining the potential $V(x) := 1 - V_{eff}(x)$, then
\begin{equation}\label{eq:Veffdimensionless}
    \Bigg[\dtot{^2}{x^{\star^2}} + V(x)\Bigg]u(x) = 0
\end{equation}
and we see that equation~\eqref{eq:Veffdimensionless} is of the form of equation~\eqref{eq:edoprufer} with $P(x)=1/\tilde{\varpi}$ and $Q(x)=V(x)/\tilde{\varpi}$, where the factor $\tilde{\varpi}:=\sqrt{1-\tilde{\mu}^2}$ was introduced for later convenience. We can see that
\begin{equation}
    \lim_{x\to x_s^+} V(x) = 1
\end{equation}
and from equation~\eqref{eq:tortoiseA}, $x\to x_s^+$ implies that $x^\star \to -\infty$. Also, $x \approx x^\star$ when $x^\star \to +\infty$, then we must have the boundary conditions
\begin{equation}\label{eq:asym}
    u(x^\star) \approx
    \begin{cases}
        e^{-ix^\star}, &x^\star\to-\infty \\[.5em]
        B\sin(\tilde{\varpi}x^\star + \xi), &x^\star\to+\infty
    \end{cases}
\end{equation}
where $B$ and $\xi$ are complex constants.

We now define the new variable $G:=G(x^\star)$ such as (primes denotes differentiation with respect to $x^\star$)
\begin{equation}
    G:= \frac{u'}{u} \implies u(x^\star) = \exp\Bigg[\int G(x^\star)\diff{x^\star}\Bigg],
\end{equation}
and $\lim_{x^\star\to-\infty} G(x^\star) = -i$. Notice that
\begin{equation}\label{eq:Gprime}
    G' = \frac{u''}{u} - \Bigg(\frac{u'}{u}\Bigg)^2 
    \implies G' + G^2 + V = 0
\end{equation}
using equation~\eqref{eq:Veffdimensionless} where $u''=-Vu$.

Let us now define a new variable in terms of the phase $\theta:=\theta(x^\star)$
\begin{equation}\label{eq:phase}
    \tilde{G}(x^\star) := \theta(x^\star) - \tilde{\varpi}x^\star
\end{equation}
To find the derivative of $\tilde{G}$, we have the Pr\"ufer's differential equation~\eqref{eq:p1} for the phase
\begin{equation}\label{eq:phase2}
    \theta' = \frac{V}{\tilde{\varpi}}\sin^2\theta 
    + \tilde{\varpi}\cos^2\theta = \Bigg(\frac{1-V_{eff}}{\tilde{\varpi}}\Bigg)\sin^2\theta + \tilde{\varpi}\cos^2\theta
\end{equation}
and use this result in the derivative of equation~\eqref{eq:phase} to obtain
\begin{equation}
    \tilde{G}' + \Bigg(\frac{V_{eff}-\tilde{\mu}^2}{\tilde{\varpi}}\Bigg)\sin^2\Big(\tilde{G}+\tilde{\varpi}x^\star\Big)=0
\end{equation}
using $V_{eff}(x^\star) = 1 - V(x^\star)$. Notice that the above equation guarantees that
\begin{equation}
    \lim_{x^\star\to+\infty}\tilde{G}' = 0.
\end{equation}
Such behavior indulges the success of the method in this case. Notice also that from definition~\eqref{eq:phase} implies $\tilde{G} \to \xi$ when $x^\star\to+\infty$, and we may write
\begin{equation}
    u(x^\star) = B\sin\Big[\tilde{\varpi}x^\star + \tilde{G}(x^\star)\Big] \implies
    G(x^\star) = \frac{u'}{u} = \tilde{\varpi}\cot\theta 
    = \tilde{\varpi}\cot\Big[\tilde{\varpi}x^\star + \tilde{G}(x^\star)\Big]
\end{equation}
with
\begin{equation}
    u'(x^\star) = \tilde{\varpi}B\cos\Big[\tilde{\varpi}x^\star + \tilde{G}(x^\star)\Big].
\end{equation}
which can be used to write $\tilde{G}$ as a function of $G$ and $G'$, that is
\begin{equation}\label{eq:matching}
    \tilde{G} = -\tilde{\varpi}x^\star + \frac{i}{2}\log\Bigg[\frac{G-i}{G+i}\Bigg], \quad
    \tilde{G}' = -\tilde{\varpi} - \frac{G'}{1+G^2},
\end{equation}
where we also wrote down the derivative of $\tilde{G}$ as a function of $G$ and $G'$.

The main idea of the method is based upon numerically integrate equations~\eqref{eq:Gprime} and~\eqref{eq:phase2} instead of equation~\eqref{eq:Veffdimensionless}. The motivation behind it is that while the original solution may rapidly oscillates, the functions $G$ and $\tilde{G}$ vary slowly with respect to $x^\star$.

Such integration of equation~\eqref{eq:Veffdimensionless} is, in general, more stable, specially for high-frequency values. Besides, equations~\eqref{eq:Gprime} and~\eqref{eq:phase2} are well behaved in the classical return points for the particle and also convenient for problems of barrier tunneling.

Nonetheless, we cannot just integrate equation~\eqref{eq:Gprime} from a point just outside the horizon ($x^\star\to-\infty$) up to infinity because of the Stokes effect: small exponential terms, usually neglected, alternate dominance and messes the evolution process up. Such issue is solved by treating the problem in two steps: $G(x^\star)$ is calculated from $x^\star\to-\infty$ up to a junction point $x_m$ set by equation~\eqref{eq:matching}, where the initial value of $\tilde{G}(x^\star)$ is defined. From this junction point up to $x^\star\to+\infty$ the function $\tilde{G}(x^\star)$ is evolved. The final result is stable and reliable if the junction point $x_m$ is on a neighborhood of a maximum of the potential barrier.

The phase-shift is then obtained from equation~\eqref{eq:asym} for $x^\star\to+\infty$, where
\begin{equation}
    u_{\ell m}(x^\star) \approx B\sin(\tilde{\varpi}x^\star + \xi) 
    = B\cos\xi\sin(\tilde{\varpi}x^\star) + B\sin\xi\cos(\tilde{\varpi}x^\star)
\end{equation}
and using Euler's formula,
\begin{align}
    u_{\ell m}(x^\star) 
    &\approx B\cos\xi\Bigg(\frac{e^{i\tilde{\varpi}x^\star} - e^{-i\tilde{\varpi}x^\star}}{2i}\Bigg)
		+ B\sin\xi\Bigg(\frac{e^{i\tilde{\varpi}x^\star}+e^{-i\tilde{\varpi}x^\star}}{2}\Bigg) \\[.5em]
	&= - B\Bigg(\frac{\cos\xi - i\sin\xi}{2i}\Bigg)e^{-i\tilde{\varpi}x^\star} 
		+ B\Bigg(\frac{\cos\xi + i\sin\xi}{2i}\Bigg)e^{i\tilde{\varpi}x^\star}\\[.5em]
	&= -Ae^{-i\xi}e^{-i\tilde{\varpi}x^\star} + Ae^{i\xi}e^{i\tilde{\varpi}x^\star} \\[.5em]
	&= ke^{-i\tilde{\varpi}x^\star} + re^{i\tilde{\varpi}x^\star} 
\end{align}
where we have defined $A:=B/2i$ and
\begin{equation}
    k = -Ae^{-i\xi}, \quad r = Ae^{i\xi}.
\end{equation}
From the definition of the phase-shift in equation~\eqref{eq:phaseshift}, we have
\begin{equation}
    e^{2i\delta_\ell} := (-1)^{\ell+1}\frac{r}{k} = (-1)^{\ell+1}\frac{Ae^{i\xi}}{(-Ae^{-i\xi})} = (-1)^\ell e^{2i\xi}
\end{equation}
and once $e^{i\pi} = -1$, we obtain the desired relation for the phase-shift
\begin{equation}
    e^{2i\delta_\ell} = e^{i\ell\pi}e^{2i\xi} \implies \delta_\ell = \xi + \frac{\ell\pi}{2}.
\end{equation}

\end{chapter}

\begin{chapter}{Transmission and Reflection rates}
\label{apendiceb}

\hspace{5 mm} Following the discussion in complement $\mbox{N}_{\mbox{\tiny III}}$ of reference~\cite{Cohen-Tannoudji:101367}, consider an arbitrary smooth potential with fixed asymptotic values, as represented in figure~\ref{fig:potential}.

\begin{figure}[htbp]
    \centering
    \includegraphics[width=0.6\linewidth]{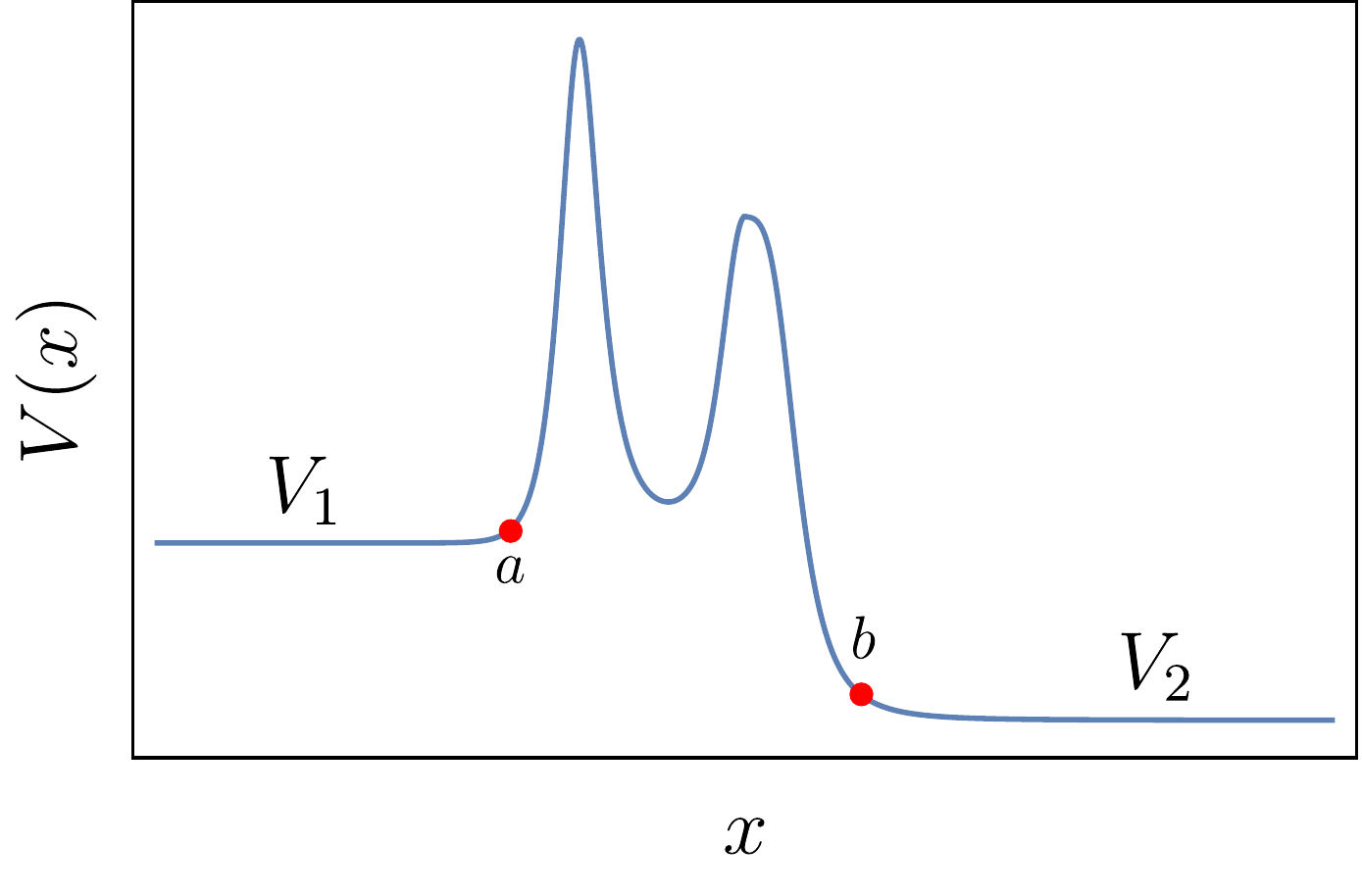}
    \caption[Arbitrary Potential]{Schematic representation of an arbitrary potential $V(x)$. The scale here is unimportant, so it was omitted. The points $a$ and $b$ represent the regions where $x<a$ and $x>b$ the potential is considered to be constant and its values are represented by $V_1$ and $V_2$, respectively.}\label{fig:potential}
\end{figure}

The time-independent Schr\"odinger equation for a massive particle of mass $m$ and energy $E>V_1$ is given by
\begin{equation}\label{eq:schrodinger}
    \psi''(x) + \frac{2m}{\hbar^2}\Big[E - V(x)\Big]\psi(x) = 0.
\end{equation}

Defining the quantities
\begin{equation}
    k_1 := \sqrt{\frac{2m}{\hbar^2}(E-V_1)}, \quad k_2 := \sqrt{\frac{2m}{\hbar^2}(E-V_2)}
\end{equation}
we have that $e^{ik_1x}$ is a solution to equation~\eqref{eq:schrodinger} for $x<a$, while for $x>b$ the solution is a linear combination of $e^{ik_2x}$ and $e^{-ik_2x}$. Let $v_{k_1}(x)$ be the solution for $x<a$, then we have
\begin{equation}
    \begin{cases}
		v_{k_1}(x) = e^{ik_1x}, &x<a \\[.5em]
		v_{k_2}(x) = F(k_1,k_2)e^{ik_2x} + G(k_1,k_2)e^{-ik_2x}, &x>b
    \end{cases}
\end{equation}
where $v_{k_2}(x)$ is the solution for $x>b$, and $F(k_1,k_2)$ and $G(k_1,k_2)$ are coefficients which depend on $k_1$ and $k_2$ as well as the shape of the potential. Similarly, we may introduce the solution $v_{k_1}'(x)$ which is equal to $e^{-ik_1x}$ for $x<a$, that is
\begin{equation}
    \begin{cases}
		v_{k_1}'(x) = e^{-ik_1x}, &x<a \\[.5em]
		v_{k_2}'(x) = F'(k_1,k_2)e^{ik_2x} + G'(k_1,k_2)e^{-ik_2x}, &x>b
    \end{cases}
\end{equation}

The most general solution $\psi_k(x)$ of equation~\eqref{eq:schrodinger} is a linear combination of $v_{k_i}$ and $v_{k_i}'$, $i=1,2$, that is,
\begin{equation}
    \psi_{k_i}(x) = Av_{k_i}(x) + Bv_{k_i}'(x)
\end{equation}
where for $x<a$ we have
\begin{equation}
    \psi_{k_1}(x) = Ae^{ik_1x} + Be^{-ik_1x},
\end{equation}
and for $x>b$ we have
\begin{align}
    \psi_{k_2}(x) &= A\Big[F(k_1,k_2)e^{ik_2x} + G(k_1,k_2)e^{-ik_2x}\Big] 
		+ B\Big[F'(k_1,k_2)e^{ik_2x} + G'(k_1,k_2)e^{-ik_2x}\Big] \nonumber\\[.5em]
    &= \tilde{A}e^{ik_2x} + \tilde{B}e^{-ik_2x}
\end{align}
with
\begin{equation}\label{eq:system}
    \begin{cases}
        \tilde{A} &= AF(k_1,k_2) + BF'(k_1,k_2) \\[.5em]
        \tilde{B} &= AG(k_1,k_2) + BG'(k_1,k_2)
    \end{cases}
\end{equation}

Defining the \emph{transmission matrix} $M(k_1,k_2)$ as
\begin{equation}
    M(k_1,k_2) := \begin{pmatrix} F(k_1,k_2) & F'(k_1,k_2) \\[.5em] G(k_1,k_2) & G'(k_1,k_2) \end{pmatrix}
\end{equation}
then the system of equations~\eqref{eq:system} may be rewritten as
\begin{equation}
    \binom{\tilde{A}}{\tilde{B}} = M(k_1,k_2)\binom{A}{B}
\end{equation}
This tells us that if we happen to know the behavior of the wave function to the left of the potential, then $M(k_1,k_2)$ let us know its behavior to the right of the potential.

It is not hard to see that, once $V(x)$ is real, then the complex conjugated of $\psi_{k_i}(x)$, that is, $\psi_{k_i}^\star(x)$ is also a solution of equation~\eqref{eq:schrodinger}. In this way, we also realize that $v_{k_i}^\star(x) = v_{k_i}'(x)$, which allows us to conclude that
\begin{equation}\label{eq:coefs}
    F^\star(k_1,k_2) = G'(k_1,k_2), \quad G^\star(k_1,k_2) = F'(k_1,k_2),
\end{equation}
then $M(k_1,k_2)$ is written as
\begin{equation}
    M(k_1,k_2) := \begin{pmatrix} F(k_1,k_2) & G^\star(k_1,k_2) \\[.5em] G(k_1,k_2) & F^\star(k_1,k_2) \end{pmatrix}
\end{equation}
and we only need to know two of the four coefficients in system~\eqref{eq:system} to determine the transmission matrix. Also, due to probability conservation, we have
\begin{equation}\label{eq:consprob}
    \det[M(k_1,k_2)] = |F(k_1,k_2)|^2	 - |G(k_1,k_2)|^2 = 1.
\end{equation}

We point out that no particular assumptions were made about the shape of the potential $V(x)$, only that it is smooth in $x\in(-\infty,+\infty)$ and that it has well defined asymptotic values for $x\to\pm\infty$.

If the particle comes from $x\to-\infty$ to $x\to+\infty$, then the coefficients $A$ and $\tilde{B}$ are related to the incoming waves, while the coefficients $B$ and $\tilde{A}$ are related to the outgoing waves. In this description, it is useful to introduce the \emph{scattering matrix} $S$, which allows us to calculate the amplitude of the outgoing waves in terms of the incoming waves,
\begin{equation}\label{eq:scattering}
    \binom{\tilde{A}}{B} = S(k_1,k_2)\binom{A}{\tilde{B}}
\end{equation}
where the system of equations~\eqref{eq:system} may be written now as, given the results in equation~\eqref{eq:coefs},
\begin{equation}
    \begin{cases}
        \tilde{A} &= AF(k_1,k_2) + BG^\star(k_1,k_2) \\[.5em]
        \tilde{B} &= AG(k_1,k_2) + BF^\star(k_1,k_2)
    \end{cases}
\end{equation}
where, from $\tilde{B}$, we have
\begin{equation}
    B = \frac{1}{F^\star(k_1,k_2)}\Big[\tilde{B} - AG(k_1,k_2)\Big],
\end{equation}
and for $\tilde{A}$ we may write
\begin{align}
    \tilde{A} &= AF(k_1,k_2) + \frac{G^\star(k_1,k_2)}{F^\star(k_1,k_2)}\Big[\tilde{B} - AG(k_1,k_2)\Big] \\[.5em]
	&= \frac{1}{F^\star(k_1,k_2)}\Big\{A\Big[F(k_1,k_2)F^\star(k_1,k_2) - G(k_1,k_2)G^\star(k_1,k_2)\Big] 
		+ \tilde{B}G^\star(k_1,k_2)\Big\} \\[.5em]
	&= \frac{1}{F^\star(k_1,k_2)}\Big[A + \tilde{B}G^\star(k_1,k_2)\Big]
\end{align}
using equation~\eqref{eq:consprob} on the second line. Thus, $S(k_1,k_2)$ is given by
\begin{equation}
    S(k_1,k_2) = \frac{1}{F^\star(k_1,k_2)}
    \begin{pmatrix} 1 & G^\star(k_1,k_2) \\[.5em] -G(k_1,k_2) & 1 \end{pmatrix}
\end{equation}
Due to conservation of probability once again, it is very easy to see that $S(k_1,k_2)S^\dagger(k_1,k_2)=S^\dagger(k_1,k_2)S(k_1,k_2)=1$, meaning that $S(k_1,k_2)$ is unitary.

Once we have been dealing with a particle traveling from $x\to-\infty$ to $x\to+\infty$, the definition of the transmission and reflections coefficients are
\begin{equation}
    R_1:=\Bigg|\frac{B}{A}\Bigg|^2, \quad
    T_1:=\Bigg|\frac{\tilde{A}}{A}\Bigg|^2
\end{equation}
and for that case, we have no incoming waves from $x\to+\infty$, meaning that $\tilde{B}=0$. Then, from equation~\eqref{eq:scattering} we have
\begin{equation}
    \tilde{A} = \frac{A}{F^\star(k_1,k_2)}, \quad
    B = -\frac{AG(k_1,k_2)}{F^\star(k_1,k_2)}
\end{equation}

The reflection and transmission coefficients then
\begin{equation}
    R_1 = \Bigg|\frac{G(k_1,k_2)}{F^\star(k_1,k_2)}\Bigg|^2, \quad
    T_1 = \frac{1}{|F^\star(k_1,k_2)|^2}
\end{equation}

If we consider the other way around, a particle coming from $x\to+\infty$ to $x\to-\infty$, then we have that $A$ and $\tilde{B}$ are the coefficients of the incoming wave while $B$ and $\tilde{A}$ are the coefficients of the outcoming wave, where the reflection and transmission coefficients are now
\begin{equation}
    R_2:=\Bigg|\frac{\tilde{A}}{\tilde{B}}\Bigg|^2, \quad
    T_2:=\Bigg|\frac{B}{\tilde{B}}\Bigg|^2,
\end{equation}
and with $A=0$, then
\begin{equation}
    \tilde{A} = \frac{\tilde{B}G^\star(k_1,k_2)}{F^\star(k_1,k_2)}, \quad
    B = \frac{\tilde{B}}{F^\star(k_1,k_2)}
\end{equation}
and then
\begin{equation}
    R_2 = \Bigg|\frac{G^\star(k_1,k_2)}{F^\star(k_1,k_2)}\Bigg|^2, \quad
    T_2 = \frac{1}{|F^\star(k_1,k_2)|^2}.
\end{equation}
Once $|F(k_1,k_2)|=|F^\star(k_1,k_2)|$ and $|G(k_1,k_2)|=|G^\star(k_1,k_2)|$, this shows us that $T_1=T_2$ and $R_1=R_2$, and the problem that considered the particle coming from $x\to+\infty$ may be considered as the particle coming from $x\to-\infty$, and the results would be just the same for the transmission and reflection coefficients.

Considering now the other way around, from the top, we have that for $x>b$, the function $e^{ik_2x}$ is a solution to equation~\eqref{eq:schrodinger}, while for $x<a$ it must be a linear combination of $e^{ik_1x}$ and $e^{-ik_1x}$, that is
\begin{equation}
    \begin{cases}
		u_{k_2}(x) = e^{ik_2x}, &x>b \\[.5em]
		u_{k_1}(x) = P(k_1,k_2)e^{ik_1x} + Q(k_1,k_2)e^{-ik_1x}, &x<a
    \end{cases}
\end{equation}
the same way, we have $u'_{k_2}(x) = e^{-ik_2x}$, then
\begin{equation}
    \begin{cases}
		u_{k_2}'(x) = e^{-ik_2x}, &x>b \\[.5em]
		u_{k_1}'(x) = P'(k_1,k_2)e^{ik_1x} + Q'(k_1,k_2)e^{-ik_1x}, &x<a
    \end{cases}
\end{equation}
and for the solutions $\varphi_{k_i}(x)$, we have
\begin{equation}
    \varphi_{k_i}(x) = Cu_{k_i}(x) + Du_{k_i}'(x),
\end{equation}
where for $x<a$
\begin{align}
    \varphi_{k_1}(x) &= C\Big[P(k_1,k_2)e^{ik_1x} + Q(k_1,k_2)e^{-ik_1x}\Big] 
		+ D\Big[P'(k_1,k_2)e^{ik_1x} + Q'(k_1,k_2)e^{-ik_1x}] \nonumber\\[.5em]
    &= \tilde{C}e^{ik_1x} + \tilde{D}e^{-ik_1x},
\end{align}
with
\begin{equation}
    \begin{cases}
        \tilde{C} &= CP(k_1,k_2) + DP'(k_1,k_2) \\[.5em]
        \tilde{D} &= CQ(k_1,k_2) + DQ'(k_1,k_2)
    \end{cases}
\end{equation}
and for $x>b$
\begin{equation}
    \varphi_{k_2}(x) = Ce^{ik_2x} + De^{-ik_2x}.
\end{equation}

From this point on, it should be clear that the procedure is exactly the same as done before and we will have the same results, leading us to conclude that
\begin{equation}
    R = \Bigg|\frac{P^\star(k_1,k_2)}{Q^\star(k_1,k_2)}\Bigg|^2, \quad
    T = \frac{1}{|Q^\star(k_1,k_2)|^2}
\end{equation}
considering the particle coming either from $x\to+\infty$ or from $x\to-\infty$.

\end{chapter}


\end{document}